\documentclass{mn2e}

\usepackage{graphicx} 
\usepackage{amsmath} 
\usepackage{amssymb} 
\usepackage{natbib} 
\usepackage{bibentry} 
\usepackage{subfig} 
\usepackage{float} 
\usepackage{lscape} 
\usepackage{booktabs} 
\usepackage{color} 
\usepackage{textcomp} 
\usepackage[above,below]{placeins} 

\usepackage{rotating} 
\usepackage{titlesec} 
\titleformat{\section}{\normalfont\bfseries\raggedright}{\thesection}{1em}{\uppercase}
\titleformat{\subsection}{\normalfont\bfseries\raggedright}{\thesubsection}{1em}{}
\titleformat{\subsubsection}{\normalfont\itshape\raggedright}{\thesubsubsection}{1em}{}
\titlespacing*{\section}{0pt}{25pt}{5pt}
\titlespacing*{\chapter}{0pt}{50pt}{40pt}
\titlespacing*{\section}{0pt}{3.5ex plus 1ex minus .2ex}{2.3ex plus .2ex}
\titlespacing*{\subsection}{0pt}{3.25ex plus 1ex minus .2ex}{1.5ex plus .2ex}
\titlespacing*{\subsubsection}{0pt}{3.25ex plus 1ex minus .2ex}{1.5ex plus .2ex}
\titlespacing{\paragraph}{0pt}{3.25ex plus 1ex minus .2ex}{1em}
\titlespacing{\subparagraph}{\parindent}{3.25ex plus 1ex minus .2ex}{1em}
\renewcommand{\thefootnote}{\fnsymbol{footnote}}

\usepackage[colorlinks=true]{hyperref} 
\definecolor{teal}{rgb}{0.0,0.6,0.4}
\definecolor{maroon}{rgb}{0.8,0.0,0.0}
\definecolor{purple}{rgb}{0.75,0.0,0.75}
\hypersetup{colorlinks, citecolor=blue, linkcolor=maroon, urlcolor=purple}

\usepackage{tabularx} 
\newcolumntype{Y}{>{\centering\arraybackslash}X} 
\newcolumntype{Z}{>{\raggedleft\arraybackslash}X} 

\newcommand{\hersc}{{\it Herschel}}
\newcommand{\planck}{{\it Planck}}
\newcommand{\HATLAS}{{\it H}-ATLAS}
\newcommand{\Kband}{{\it$K_{S}$}-band}
\newcommand{\fK}{FUV-{\it$K_{S}$}}

\newcommand{\cnp}{{\it Planck} C13N13}
\newcommand{\fg}{$f_{g}^{\it HI}$}

\newcommand{\HI}{H{\sc i}}

\begin{document}

\title[A Blind Local Galaxy Survey with \hersc-ATLAS]{\textbfit{Herschel}-ATLAS: The Surprising Diversity of Dust- Selected Galaxies in the Local Submillimetre Universe}

\author[C.J.R. Clark et\,al.] {\parbox{\textwidth}{C.\,J.\,R.\,Clark$^{1\star}$, 
L.\,Dunne$^{2,3}$,
H.\,L.\,Gomez$^{1}$,
S.\,Maddox$^{2,3}$,
P.\,De\,Vis$^{2}$,
M.\,W.\,L.\,Smith$^{1}$,
S.\,A.\,Eales$^{1}$,
M.\,Baes$^{4}$,
G.\,J.\,Bendo$^{5}$,
N.\,Bourne$^{3}$,
S.\,P.\,Driver$^{6}$,
S.\,Dye$^{7}$,
C.\,Furlanetto$^{7,8}$,
M.\,W.\,Grootes$^{9}$,
R.\,J.\,Ivison$^{3,10}$,
S.\,P.\,Schofield$^{1}$,
A.\,S.\,G.\,Robotham$^{6}$,
K.\,Rowlands$^{11}$,
E.\,Valiante$^{1}$,
C.\,Vlahakis$^{12}$,
P.\,van\,der\,Werf$^{13}$,
A.\,H.\,Wright$^{6}$,
G.\,de\,Zotti$^{14,15}$}\\
\\
{\parbox{\textwidth}{$^{1}$ School of Physics \& Astronomy, Cardiff University, Queens Buildings, The Parade,  Cardiff, CF24 3AA, UK\\
$^{2}$ Department of Physics \& Astronomy, University of Canterbury, Private Bag 4800, Christchurch, 8140, New Zealand\\
$^{3}$ Institute for Astronomy, University of Edinburgh, Royal Observatory, Blackford Hill, Edinburgh, EH9 3HJ, UK\\
$^{4}$ Sterrenkundig Observatorium, Krijgslaan 281 S9, B-9000 Gent, Belgium\\ 
$^{5}$ UK ALMA Regional Centre Node, Jodrell Bank Centre for Astrophysics, University of Manchester, Manchester, M13 9PL, UK\\
$^{6}$ International Centre for Radio Astronomy Research, The University of Western Australia, Crawley, Perth, 6009, Australia\\
$^{7}$ School of Physics \& Astronomy, University of Nottingham, University Park, Nottingham, NG7 2RD, UK\\
$^{8}$ CAPES Foundation, Ministry of Education of Brazil, Bras\'ilia/DF, 70040-020, Brazil\\
$^{9}$ Max-Planck-Institut f\"ur Kernphysik, Saupfercheckweg 1, 69117 Heidelberg, Germany\\
$^{10}$ European Southern Observatory, Karl Schwarzschild Strasse 2, Garching, D85748, Germany\\ 
$^{11}$ School of Physics \& Astronomy, University of St Andrews, North Haugh, St Andrews, KY16 9SS, UK\\
$^{12}$ Joint ALMA Observatory / European Southern Observatory, Alonso de Cordova 3107, Vitacura, Santiago, Chile\\
$^{13}$ Leiden Observatory, PO Box 9513, NL-2300 RA Leiden, The Netherlands\\
$^{14}$ Istituto Nazionale di Astrofisica, Osservatorio Astronomico di Padova, Vicolo dell'Osservatorio 2, 35122 Padova, Italy\\
$^{15}$ SISSA, Via Bonomea 265, I-34136 Trieste, Italy\\
$^{\star}$ {\tt Christopher.Clark@astro.cf.ac.uk}}}}

\date{}

\pagerange{\pageref{firstpage}--\pageref{lastpage}} \pubyear{2015}

\maketitle 

\begin{abstract} We present the properties of the first 250\,\micron\ blind sample of nearby galaxies ($15<D<46\,{\rm Mpc}$) containing 42 objects from the \hersc\ Astrophysical Terahertz Large Area Survey (\HATLAS). {\it Herschel's} sensitivity probes the faint end of the dust luminosity function for the first time, spanning a range of stellar mass ($7.4<{M_{\star}}<11.3\ {\rm log_{10}\,M_{\odot}}$), star formation activity ($-11.8<{\it SSFR}<-8.9\ {\rm log_{10}\, yr^{-1}}$), gas fraction (3--96\,per\,cent), and colour (0.6\,\textless\,\fK\,\textless\,7.0 mag). The median cold dust temperature is 14.6\,K, colder than in the \hersc\ Reference Survey (18.5\,K) and \planck\ Early Release Compact Source Catalogue (17.7\,K). The mean dust-to-stellar mass ratio in our sample is higher than these surveys by factors of 3.7 and 1.8, with a dust mass volume density of $(3.7 \pm 0.7) \times 10^5\,{\rm M_{\odot}\,Mpc^{-3}}$. Counter-intuitively, we find that the more dust rich a galaxy, the lower its UV attenuation. Over half of our dust-selected sample are very blue in \fK\ colour, with irregular and/or highly flocculent morphology; these galaxies account for only 6\,per\,cent of the sample's stellar mass but contain over 35\,per\,cent of the dust mass. They are the most actively star forming galaxies in the sample, with the highest gas fractions and lowest UV attenuation. They also appear to be in an early stage of converting their gas into stars, providing valuable insights into the chemical evolution of young galaxies.\\
\end{abstract}


\section{Introduction} \label{Section:Introduction}

\setcounter{footnote}{0}
\renewcommand{\thefootnote}{\textsuperscript{\arabic{footnote}}}

On average, half of all starlight emitted by galaxies is absorbed by dust and thermally re-emitted in the Far-InfraRed (FIR) and submillimetre (submm) \citep{Fixsen1996B, Driver2007D}. Dust is particularly prevalent in star forming regions, where the high-energy photons emitted by young stars are highly susceptible to absorption by dust grains \citep{Fitzpatrick2004B}. 
The thermal emission from dust in galaxies is normally dominated by the hot component, which is mostly heated by star-forming regions \citep{Kennicutt1998H, Kennicutt2009B}. Thermal emission from dust therefore provides an invaluable avenue for the study of star formation. The cold diffuse dust component dominates the mass of dust in galaxies \citep{Draine2007C,Law2011A, Ford2013, Hughes2014A}, but it is unclear if this cold component is also indirectly heated by star formation through UltraViolet (UV) photons leaking from birth clouds, or if the evolved stellar population is mainly responsible \citep{Bendo2011A, Boquien2011C, Bendo2014}. Ultimately, the ratio of recent/evolved stellar heating is likely to depend on an individual galaxy's dust geometry and star formation activity \citep{Dunne2013A}. Knowledge of how this ratio depends on measurable properties (e.g. morphological type, $L_{\it TIR}$, colour, etc) would make the determination of star formation rates from FIR measurements more reliable.

The InterStellar Medium (ISM) is enriched by evolved stars, which synthesise heavy elements and then introduce them to the galactic environment. Interstellar dust is now understood to be the product of both winds from evolved stars \citep{Ferrarotti2006, Sargent2010B}, and of core-collapse supernovae (SNe), the end-point in the fleeting lives of massive stars \citep{Dunne2003D, Dunne2009A, Barlow2010, Matsuura2011E, Gomez2012B, Indebetouw2014A}. However studies of both local \citep{Matsuura2009B, Dunne2011} and high-redshift \citep{Morgan2003B, Dwek2007B, Michalowski2010, Rowlands2014B} galaxies have shown a disparity between the rate at which dust is removed from the ISM (either by star formation or interstellar destruction), and the rate at which stars replenish it. As such, the origin of dust in galaxies is still very much an open question.

It is difficult to develop a thorough understanding of galaxies without also understanding the properties of their ISM. As FIR and submm astronomy has matured, numerous projects have been undertaken to characterise dust in galaxies. The galaxy dust mass function was first measured for $\sim$\,200 InfraRed (IR) and optically selected galaxies by the Submillimetre Common-User Bolometer Array (SCUBA) Local Universe Galaxy Survey (SLUGS, \citealp{Dunne2000A, Vlahakis2005C}). This is being followed in the era of the \hersc\ Space Observatory\footnote{{\it Herschel} is an ESA space observatory with science instruments provided by European-led Principal Investigator consortia and with important participation from NASA.} \citep{Pilbratt2010D} by the \hersc\ Reference Survey (HRS, \citealp{Boselli2010}) and the Key Insights in Nearby Galaxies Far-Infrared Survey with \hersc\ (KINGFISH, \citealp{Kennicutt2011D}). However, these and other FIR surveys of nearby galaxies may have been hindered by the fact they are not {\it dust selected}, instead they are selected for their properties at other wavelengths. Large-area missions such as with the Infrared Astronomical Satellite (IRAS, \citealt{Neugebauer1984}) and more recently \planck\ \citep{Planck2011I} provide blindly selected FIR/submm samples of local galaxies, including the recent sample by \citet{Clemens2013A}, but lack resolution and sensitivity when compared to the targeted surveys.

Now, however, with the advent of blind, large-area surveys such as the \hersc\ Astrophysical Terahertz Large Area Survey (\HATLAS, \citealp{Eales2010A}) we finally have an unbiased and unrivalled view of the dusty Universe, with resolution and sensitivity hitherto only found in targeted dust surveys.

In this paper, we use \HATLAS\ to select local dusty galaxies, and investigate the properties of sources chosen on the basis of their dust mass. In Section~\ref{Section:Data_and_Sample} we introduce the observations and sample selection. In Section~\ref{Section:Extended-Source_Photometry_and_Uncertainties} we give an account of our extended-source photometry. In Section~\ref{Section:HAPLESS_Properties}, we discuss the key properties of our local \HATLAS\ sample. In Section~\ref{Section:Comparison_to_Other_Samples} we compare the properties of our sample with other samples of nearby dusty galaxies. In Section~\ref{Section:Chemical_Evolution} we examine the gas and dust evolution of the galaxies in our sample. A companion paper on the dust properties of HI-selected galaxies in the local Universe will be presented in De Vis et al. ({\it in prep.}).  We adopt the cosmology of \citet{Planck2013I}, specifically $\rm H_{0} = 67.30\,km\,s^{-1}\,Mpc^{-1}$, $\Omega_{m} = 0.315$, and $\Omega_{\Lambda} = 0.685$.

\section{\textbfit{Herschel} Data and the HAPLESS Sample} \label{Section:Data_and_Sample}

\subsection{Observations} \label{Subsection:Observations}

Observations for \HATLAS\ were carried out in parallel mode at 100 and
160\,\micron\ with the Photodetector Array Camera and Spectrometer
(PACS, \citealp{Pogslitsch2010B}) and at 250, 350 and
500\,\micron\ with the Spectral and Photometric Imaging REceiver
(SPIRE, \citealp{Griffin2010D}) instruments on board
\hersc. Descriptions of the \HATLAS\ data reduction can be found in
\citet{Ibar2010B} for PACS, and \citet{Pascale2011} and Valiante et
al. ({\it in prep.}) for SPIRE. Photometry in the SPIRE bands was
performed upon maps reduced for extended-source measurements. Our \HATLAS\ PACS
maps were reduced using the {\tt Scanamorphos} \citep{Roussel2013A}
pipeline, with appropriate corrections made for the relative areas of
the reference pixels on the focal plane. 

This work makes use of the \HATLAS\ Phase-1 Version-3 internal data
release (Valiante et al., {\it in prep.}, Bourne et al., {\it in
  prep.}), which comprises 161.6 \,deg$^{2}$ coincident with the
Galaxy And Mass Assembly (GAMA, \citealp{Driver2009B}) redshift
survey. GAMA provides spectroscopic redshifts, along with
supplementary reductions and mosaics of ultraviolet (UV) GALEX
(\citealp{Morrissey2007B}; Liske et al., {\it submitted.}; Andrae et al., {\it in prep.}), optical SDSS DR6
\citep{Adelman-McCarthy2008B}, Near-InfraRed (NIR) VISTA VIKING
\citep{Sutherland2012C}, and Mid-InfraRed (MIR) WISE
\citep{Wright2010F,Cluver2014A} data; details of these reprocessed maps can be found in Driver et al., ({\it in prep.}).

\begin{landscape}
\begin{table}
\begin{center}
\caption{Basic properties of the HAPLESS sample. Velocities corrected for bulk deviation from Hubble flow \citep{Baldry2012}. Morphologies from EFIGI \citep{Baillard2011}.}
\label{Table:Sample}
\begin{tabular}{lllrrrrrrrr}
\toprule \toprule
\multicolumn{1}{c}{HAPLESS} &
\multicolumn{1}{c}{Common name} &
\multicolumn{1}{c}{\HATLAS\ IAU ID} &
\multicolumn{1}{c}{SDSS RA} &
\multicolumn{1}{c}{SDSS DEC} &
\multicolumn{1}{c}{$z$} &
\multicolumn{1}{c}{Corrected velocity} &
\multicolumn{1}{c}{Distance} &
\multicolumn{1}{c}{Morphology} &
\multicolumn{1}{c}{Flocculence} \\
\multicolumn{1}{c}{} &
\multicolumn{1}{c}{} &
\multicolumn{1}{c}{} &
\multicolumn{1}{c}{(J2000 deg)} &
\multicolumn{1}{c}{(J2000 deg)} &
\multicolumn{1}{c}{(helio)} &
\multicolumn{1}{c}{(km\,s$^{-1}$)} &
\multicolumn{1}{c}{(Mpc)} &
\multicolumn{1}{c}{(T)} &
\multicolumn{1}{c}{} \\
\midrule
1 & UGC 06877 & HATLAS J115412.1+000812 & 178.55114 & 0.13663 & 0.00379 & 1336 & 18.3$^{b}$ & -1 & 0.25\\
2 & PGC 037392 & HATLAS J115504.7+014310 & 178.77044 & 1.71981 & 0.00421 & 1796 & 26.7 & 8 & 0.75\\
3 & UGC 09215 & HATLAS J142327.2+014335 & 215.86297 & 1.72630 & 0.00457 & 1726 & 25.6 & 6 & 0.75\\
4 & UM 452 & HATLAS J114700.5-001737 & 176.75303 & -0.29422 & 0.00470 & 1970 & 29.3 & 11 & 0.25\\
5$^{a}$ & PGC 052652 & HATLAS J144430.6+013120 & 221.12828 & 1.52201 & 0.00475 & 1728 & 25.7 & 10 & 0.25\\
6 & NGC 4030 & HATLAS J120023.7-010553 & 180.09843 & -1.10008 & 0.00477 & 1978 & 29.4 & 3 & 0.75\\
7 & NGC 5496 & HATLAS J141137.7-010928 & 212.90774 & -1.15908 & 0.00488 & 1840 & 27.4 & 6 & 0.75\\
8 & UGC 07000 & HATLAS J120110.4-011750 & 180.29502 & -1.29751 & 0.00489 & 2016 & 30.0 & 9 & 0.50\\
9 & UGC 09299 & HATLAS J142934.8-000105 & 217.39416 & -0.01823 & 0.00516 & 1904 & 28.3 & 9 & 1.00\\
10 & NGC 5740 & HATLAS J144424.3+014046 & 221.10186 & 1.67977 & 0.00520 & 1890 & 28.0 & 3 & 0.50\\
11 & UGC 07394 & HATLAS J122027.6+012812 & 185.11526 & 1.46974 & 0.00526 & 2197 & 32.7 & 7 & 0.25\\
12 & PGC 051719 & HATLAS J142837.8+003311 & 217.15652 & 0.55280 & 0.00527 & 1952 & 29.0 & 7 & 0.50\\
13$^{a}$ & LEDA 1241857 & HATLAS J145022.9+025729 & 222.59524 & 2.95853 & 0.00533 & 1928 & 28.6 & 10 & 0.50\\
14 & NGC 5584 & HATLAS J142223.4-002313 & 215.59903 & -0.38766 & 0.00548 & 2033 & 22.1$^{b}$ & 6 & 1.00\\
15$^{a}$ & MGC 0068525 & HATLAS J144515.7-000936 & 221.31587 & -0.15953 & 0.00548 & 1964 & 29.2 & 10 & 0.25\\
16 & UGC 09348  & HATLAS J143228.6+001739 & 218.11878 & 0.29402 & 0.00558 & 2044 & 30.4 & 8 & 0.50\\
17 & UM 456 & HATLAS J115036.2-003406 & 177.65119 & -0.56866 & 0.00561 & 2250 & 33.4 & 10 & 0.75\\
18 & NGC 5733 & HATLAS J144245.8-002104 & 220.69130 & -0.35108 & 0.00565 & 2028 & 30.1 & 9 & 0.75\\
19 & UGC 06780 & HATLAS J114850.4-020156 & 177.21002 & -2.03224 & 0.00569 & 2261 & 33.6 & 8 & 0.75\\
20 & NGC 5719 & HATLAS J144056.2-001906 & 220.23484 & -0.31821 & 0.00575 & 2067 & 30.7 & 1 & 0.25\\
21 & NGC 5746 & HATLAS J144455.9+015719 & 221.23300 & 1.95495 & 0.00575 & 2077 & 30.9 & 1 & 0.25\\
22$^{a}$ & NGC 5738 & HATLAS J144356.1+013615 & 220.98488 & 1.60418 & 0.00582 & 2100 & 31.2 & -2 & 0.00\\
23 & NGC 5690 & HATLAS J143740.9+021729 & 219.42114 & 2.29082 & 0.00583 & 2130 & 31.6 & 3 & 0.75\\
24$^{a}$ & UM 456A & HATLAS J115033.8-003213 & 177.64179 & -0.53782 & 0.00585 & 2391 & 31.6 & 10 & 0.50\\
25 & NGC 5750 & HATLAS J144611.2-001324 & 221.54635 & -0.22294 & 0.00588 & 2094 & 31.1 & 1 & 0.25\\
26 & NGC 5705 & HATLAS J143949.5-004305 & 219.95704 & -0.71846 & 0.00591 & 2097 & 31.2 & 9 & 0.75\\
27 & UGC 09482 & HATLAS J144247.1+003942 & 220.69560 & 0.66173 & 0.00607 & 2177 & 32.3 & 8 & 0.50\\
28 & NGC 5691 & HATLAS J143753.3-002354 & 219.47225 & -0.39888 & 0.00626 & 2244 & 33.4 & 3 & 0.50\\
29 & NGC 5713 & HATLAS J144011.1-001725 & 220.04794 & -0.28897 & 0.00633 & 2261 & 33.6 & 3 & 0.50\\
30 & UGC 09470 & HATLAS J144148.7+004121 & 220.45287 & 0.68697 & 0.00633 & 2265 & 33.6 & 9 & 0.75\\
31 & UGC 06903 & HATLAS J115536.9+011417 & 178.90395 & 1.23717 & 0.00635 & 2535 & 37.7 & 6 & 0.75\\
32 & CGCG 019-084 & HATLAS J144229.4+013006 & 220.62338 & 1.50040 & 0.00652 & 2330 & 34.6 & 10 & 0.75\\
33 & UM 491 & HATLAS J121953.0+014623 & 184.97165 & 1.77347 & 0.00671 & 2673 & 39.7 & 10 & 0.50\\
34 & UGC 07531 & HATLAS J122611.1-011813 & 186.54927 & -1.30475 & 0.00675 & 2654 & 39.4 & 9 & 0.75\\
35 & UGC 07396 & HATLAS J122033.9+004719 & 185.14066 & 0.78806 & 0.00706 & 2779 & 41.3 & 8 & 0.50\\
36 & CGCG 014-014 & HATLAS J122106.0+003306 & 185.27385 & 0.55283 & 0.00719 & 2820 & 41.9 & 8 & 0.25\\
37 & UGC 6879 & HATLAS J115425.2-021910 & 178.60434 & -2.31955 & 0.00803 & 2774 & 45.6 & 4 & 0.75\\
38 & CGCG 019-003 & HATLAS J141919.9+010952 & 214.83417 & 1.16516 & 0.00806 & 2893 & 43.0 & 9 & 0.50\\
39 & UGC 04684 & HATLAS J085640.5+002229 & 134.16946 & 0.37500 & 0.00859 & 2796 & 41.5 & 7 & 1.00\\
40 & NGC 5725 & HATLAS J144058.3+021110 & 220.24298 & 2.18626 & 0.00543 & 2035 & 30.2 & 9 & 0.75\\ 
41$^{a}$ & UGC 06578 & HATLAS J113636.7+004901 & 174.15315 & 0.81543 & 0.00375 & 1164 & 17.3 & 10 & 1.00\\
42$^{a}$ & MGC 0066574 & HATLASJ143959.9-001113 & 219.99950 & -0.18609 & 0.00620 & 2246 & 33.4 & 11 & 0.25\\
\bottomrule
\end{tabular}
\end{center}
\begin{list}{}{}
\item[$^{a}$] HAPLESS 5, 13, 15, 22, 24, 41, and 42 are not included in the luminosity-limited sub-sample.
\item[$^{b}$] Redshift-independent distances used for HAPLESS 1 (UGC 06877, \citealp{Tonry2001B}), and HAPLESS 14 (NGC 5584, \citealp{Riess2011A}).
\end{list}
\end{table}
\end{landscape}

The source extraction algorithm used in \HATLAS\ (MADX, Maddox et
  al., {\it in prep.}, Valiante et al., {\it in prep.}) isolates
  \textgreater\,2.5\,$\sigma$ peaks in the SPIRE 250\,\micron\ maps
  and then measures the fluxes in all three SPIRE bands at the
  position determined by the 250\,\micron\ fit. For our catalogue we
  further select only those sources which have a \textgreater\,5\,$\sigma$
  detection at 250\,\micron.

Optical counterparts to \HATLAS\ sources were found by matching
\HATLAS\ sources to SDSS DR7 objects \citep{Abazajian2009B} within a
10\arcsec\ radius using a likelihood ratio technique
\citep{DJBSmith2011}. This method uses the optical-submm separation,
SPIRE positional errors, and $r$-band magnitudes of potential
counterparts, to derive the probability that a given optical galaxy is
genuinely associated with the SPIRE source in question (see
\citealp{DJBSmith2011} and Bourne et al., {\it in prep.} for details of
the method). Sources with a probability of association $R > 0.8$ are
deemed to be `reliable' IDs. 

\subsection{The Sample} \label{Subsection:Sample_Selection}

A sample of 42 galaxies was assembled from the \HATLAS\ Phase-1
Version-3 catalogue in the distance range $15 < D < 46\,{\rm Mpc}$. We
wished to sample a volume local enough that we retained sensitivity to
the lowest-mass and coldest sources, populations not previously well studied, and
our upper distance limit of 46\,Mpc serves this purpose well. We do
not include galaxies at $D < 15\,{\rm Mpc}$, where recessional velocity is no longer
a reliable indicator of distance.  These galaxies form the \hersc-ATLAS
Phase-1 Limited-Extent Spatial Survey, hereafter referred to as
HAPLESS. Multiwavelength imagery of the full sample can be found in
Appendix \ref{AppendixSection:HAPLESS}, Figure~\ref{AppendixFig:HAPLESS_Imagery}.


We require all sources to have reliable SDSS counterparts ($R \ge
  0.8$, \citealp{DJBSmith2011}) and to have been assessed as having
  science quality redshifts (nQ\,$\geq$\,3, \citealp{Driver2011}) by GAMA. We eyeballed the \HATLAS\ maps at the location of all optical sources within the redshift range and found no other candidates which may have been missed by our ID process.
The total number of false IDs expected in our
  sample can be estimated by summing $(1-R)$ (where $R$ is the
  reliability assigned in the likelihood ratio analysis), which gives
  a false ID rate of 0.7\,per\,cent.  

\begin{figure}
\begin{center}
\includegraphics[width=0.5\textwidth]{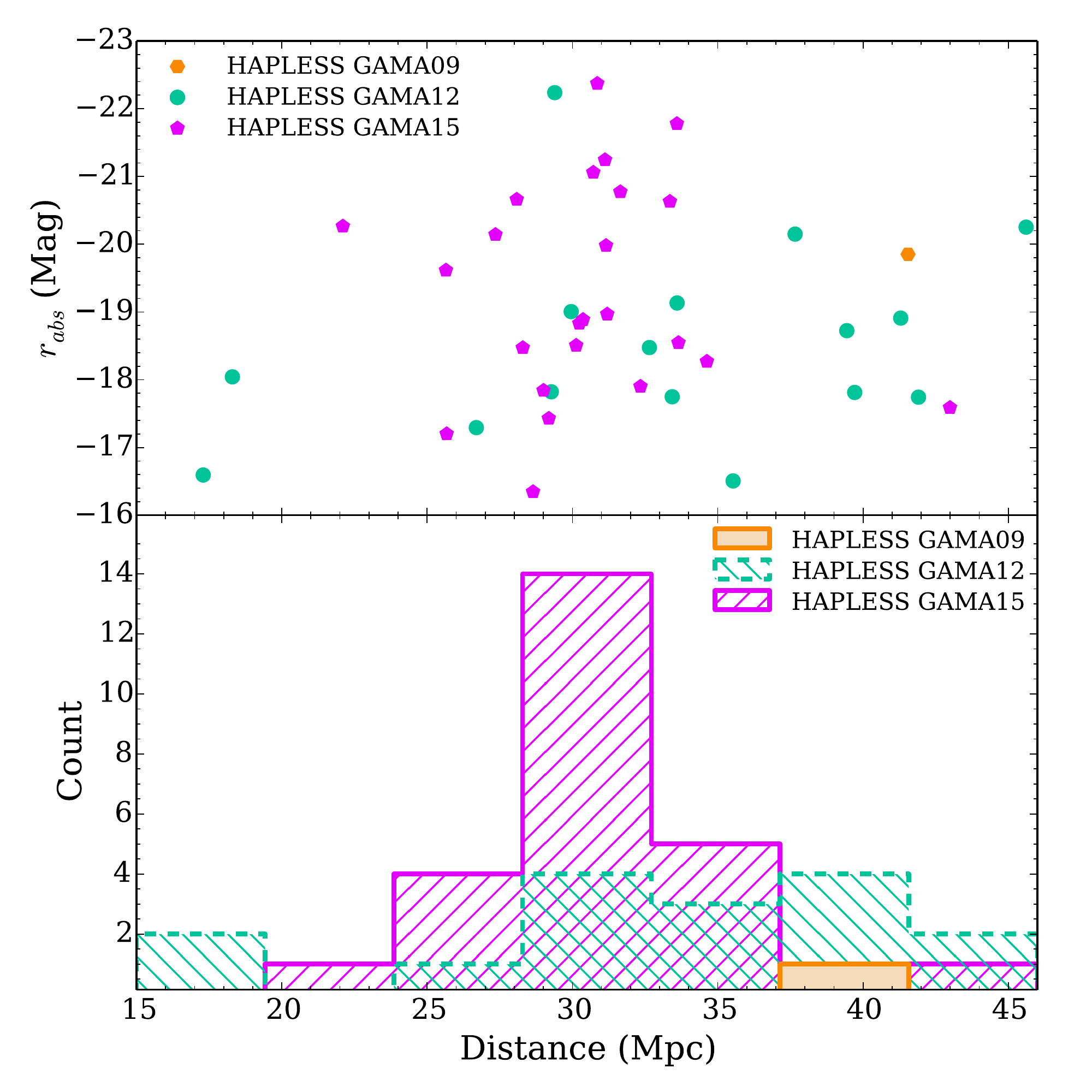}
\caption{{\it Upper:} Absolute {\it r}-band magnitude against distance, for the 42 galaxies of the HAPLESS sample. The different colours and shapes denote whether the galaxy lies in the GAMA09 (orange hexagons), GAMA12 (green circles), or GAMA15 (purple pentagons) fields sampled as part of the \HATLAS\ Phase 1 data release. {\it Lower:} The distance distribution of HAPLESS sources in the different fields. }
\label{Fig:HAPLESS_Distance_Grid}
\end{center}
\end{figure}

Distances were calculated using spectroscopic redshifts, velocity
corrected by GAMA \citep{Baldry2012} to account for bulk deviations
from Hubble flow \citep{Tonry2000C}. For ${\rm H_{0} =
  67.30\,km\,s^{-1}\,Mpc^{-1}}$, the distance limits we impose
correspond to a (flow corrected) redshift range of $0.0035 \lesssim z
\lesssim 0.01$. Reliable redshift-independent distances were used for
the two sources for which they were available; the distance to UGC
06877 has been determined using surface brightness fluctuations
\citep{Tonry2001B}, and the distance to NGC 5584 is known from
measurements of Cepheid variables \citep{Riess2011A}. 

\begin{figure}
\begin{center}
\includegraphics[width=0.45\textwidth]{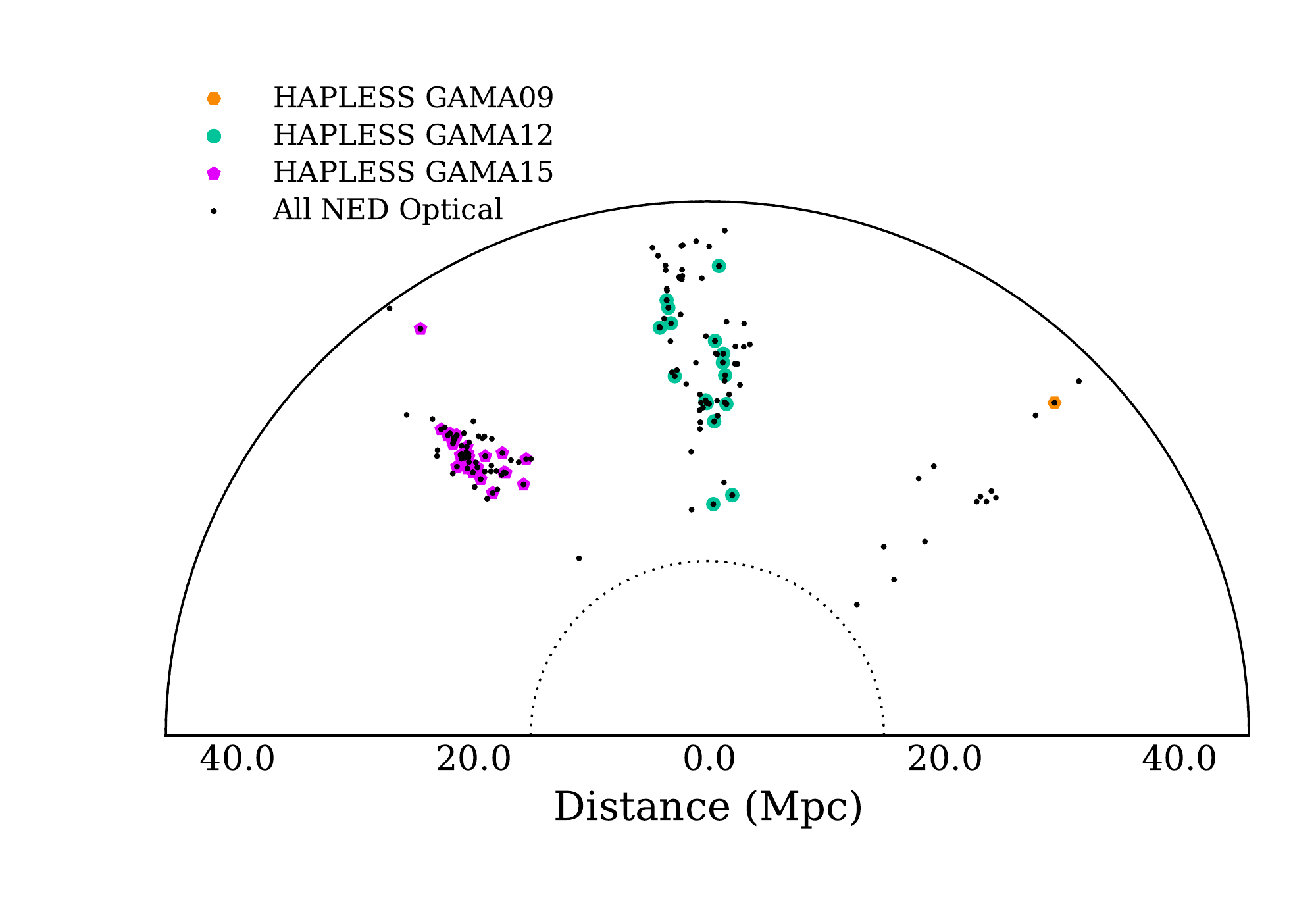}
\caption{Polar plot of the volume sampled by HAPLESS in the GAMA09, GAMA12, and GAMA15 fields (right-to-left). The positions of the HAPLESS galaxies are shown (same symbols as Figure~\ref{Fig:HAPLESS_Distance_Grid}). Also shown are all the optical sources with redshifts that place them in the volume. The inner distance limit of $D = 15\,{\rm Mpc}$ is demarked by the dotted black line.}
\label{Fig:HAPLESS_Polar_Plot}
\end{center}
\end{figure}
  
Comparing {\it r}-band absolute magnitude
(Table~\ref{Table:Misc_Properties}) to distance, as shown in the upper
panel of Figure~\ref{Fig:HAPLESS_Distance_Grid}, shows that there
appear to be fewer galaxies at greater distances, where larger
volumes are being sampled. This is likely to be due to large scale structure (Figure~\ref{Fig:HAPLESS_Polar_Plot}), since the percentage cosmic variance on the number counts in the volume sampled by HAPLESS is $\sim 166\,per\,cent$ \citep{Driver2010B}. The total number of
  sources listed in the NASA/IPAC Extragalactic Database
  (NED\footnote{\url{http://ned.ipac.caltech.edu/}}) in the same
  volume as our sample is 141; we therefore detect 30\,per\,cent of this
  population. Note that the three \HATLAS\ fields (GAMA09, GAMA12, and
  GAMA15; see Figure~\ref{Fig:HAPLESS_Distance_Grid}) contain 1, 16,
  and 25 HAPLESS sources respectively, representing detection rates of
  7\,per\,cent, 24\,per\,cent, and 42\,per\,cent.

We identified the portion of our sample which is limited by
intrinsic 250\,\micron\ luminosity; this gives us a volume limited sample
above $L_{250} = 8.9 \times 10^{21}\,{\rm W\,Hz^{-1}}$ (corresponding
to a 250\,\micron\ flux of 35\,mJy at a distance of 46\,Mpc). Of the
42 HAPLESS galaxies, 35 would still be detected were they located at
the furthest distance of the volume sampled. Following the assumptions
detailed in Section~\ref{Subsection:SED_Fitting}, this is equivalent
to a dust mass limit of $7.4 \times 10^{5}\,{\rm M_{\odot}}$ for a
dust temperature of 14.6\,K (the average dust temperature of the
sample, see Section~\ref{Subsection:SED_Fitting}). The 7 sources fainter than this limit are HAPLESS 5, 13, 15,
22, 24, 41, and 42. These objects are included when describing
the properties of our sample in
Section~\ref{Section:HAPLESS_Properties} and comparing to other
surveys in Section~\ref{Section:Comparison_to_Other_Samples} but are
plotted as hollow circles. We correct for the accessible volume of these sources when considering dust mass volume
densities in
Section~\ref{Section:Dust_Mass_Volume_Density}.

Finally, UGC 06877 (HAPLESS 1) hosts an AGN \citep{Osterbrock1983E}, with
a significant contribution from non-thermal continuum emission in the
UV \citep{Markaryan1979D}. This contaminates our star formation rate
estimate for this galaxy, rendering it unreliable. We therefore omit
HAPLESS 1 from discussions of star formation. The key characteristics
of the HAPLESS sample, such as their common names, redshifts,
distances and morphologies, can be found in Table~\ref{Table:Sample}. We note that 12 of our sources are also part of the smaller nearby sample of \HATLAS\ galaxies presented in \citet{Bourne2013}.

\subsection{Curious Blue Galaxies} \label{Subsection:Curious_Blue_Galaxies}

\begin{figure}
\begin{center}
\includegraphics[width=0.475\textwidth]{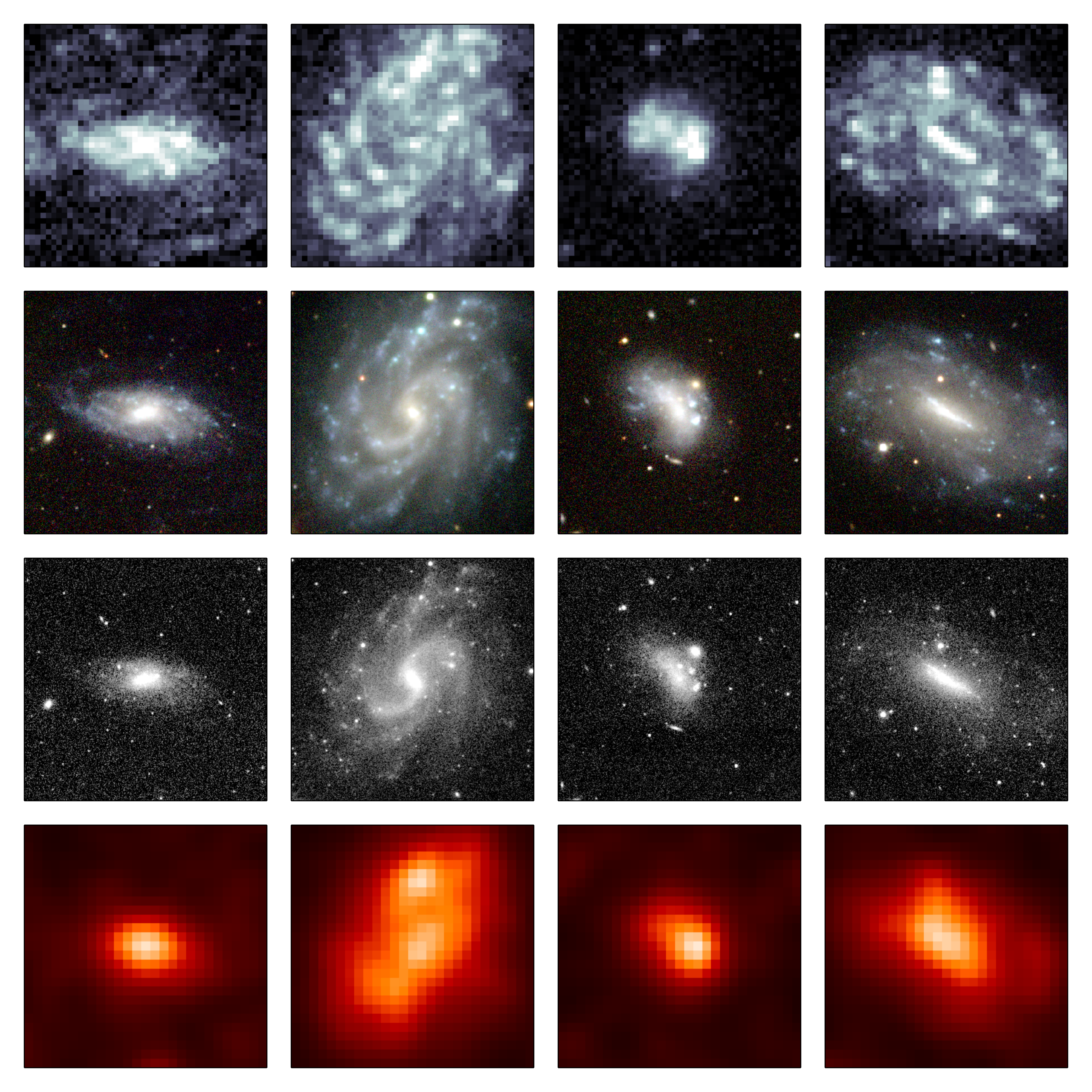}
\caption{Multiwavelength imagery of four examples of the curious very blue galaxies found in the HAPLESS sample. From left-to-right they are, UGC 09299, NGC 5584, NGC 5733, and NGC 5705. The bands displayed, from top-to-bottom, are: GALEX FUV, SDSS {\it gri} three-colour, VIKING \Kband, and PSF-filtered \hersc\ 250 \micron. Each image is 150\arcsec $\times$ 150\arcsec. Note the blue optical colours, flocculent morphologies, NIR faintness, and bright extended UV emission. The whole sample is presented in  Appendix~\ref{AppendixSection:HAPLESS}, Figure~\ref{AppendixFig:HAPLESS_Imagery}.}
\label{Fig:Example_Meanies}
\end{center}
\end{figure}

We obtained morphology information from the EFIGI catalogue of \citet{Baillard2011}, which includes 71\,per\,cent of the HAPLESS galaxies; we visually classified the remainder (all of which were compact dwarf galaxies) using their prescription. The majority of the galaxies in our sample possess very late-type, irregular morphology (Hubble stage $\rm T \ge 8$) though there are two early types (HAPLESS 1 and 22). Furthermore, a large fraction of the sample exhibit a high degree of flocculence (as defined by the EFIGI catalogue). In all, 24 of our sample are classed as irregular, and 19 as highly flocculent; 31 are one or the other, whilst 11 are both (Table~\ref{Table:Sample}). These irregular and flocculent galaxies are bright in the submm and UV, indicating significant dust mass and high specific star formation rates (SSFRs). They exhibit extremely blue UV-NIR colours, arising from the fact that, along with being UV-bright, they are NIR-faint; examples of this can be seen in Figure~\ref{Fig:Example_Meanies}. We find a UV-NIR colour-cut of \fK\,\textless\,3.5\,mag to be an effective criterion for identifying such galaxies. This approach is supported by the work of \citet{GilDePaz2007A}, who found \fK\ colour to be a powerful diagnostic for discriminating morphological type.

These curious blue galaxies with \fK \textless\ 3.5 span a wide range of optical sizes, from 1.3 to 33.3\,kpc, with a median major axis of 9.3\,kpc (derived from $r$-band $R25$, the radius to the 25\textsuperscript{th} magnitude square arcsecond isophote). Whilst many of them, particularly the larger examples, possess disks, they often lack defined spiral structure, and show only a weak bulge contribution. 

Whilst the \fK\ colour of UGC 06780 (HAPLESS 1) is 3.07 mag (which would classify it as a member of the curious blue population), continuum emission from its AGN is contributing to the FUV flux. That said, UGC 06780 clearly emits plentiful UV emission not associated with the AGN (especially for an early-type), as it is more extended in the UV than it is in the optical (Table~\ref{Table:Misc_Properties}). We therefore opt to leave it classed amongst the curious blue population, with this caveat.

GALEX coverage is not available for 2 of the HAPLESS galaxies (HAPLESS 19 and 21); however the colour {\it $u$-$K_{S}$} is well correlated with \fK\ (Spearman rank correlation coefficient of 0.94 for HAPLESS). By comparing the distributions of these colours, we can state with 3\,$\sigma$ confidence that a source with {\it $u$-$K_{S}$}\textless\,1.36 will have \fK\,\textless\,3.5. This indicates that HAPLESS 19 is a member of our curious blue population; visual inspection confirms that it exhibits irregular and extremely flocculent morphology.

The \fK\ colours of the HAPLESS galaxies can be found in Table~\ref{Table:Misc_Properties}. Of the 42 HAPLESS galaxies, 27 (64\,per\,cent) satisfy the very blue \fK\,\textless\,3.5 criterion; 25 (93\,per\,cent) of these exhibit irregular and/or highly flocculent morphology. Of the 15 HAPLESS galaxies with \fK\,\textgreater\,3.5, irregular and/or highly flocculent morphology is exhibited by only 7 (47\,per\,cent); a two-sided Fisher test suggests this difference is significant at the $p<0.01$ level.


\section{Extended-Source Photometry and Uncertainties} \label{Section:Extended-Source_Photometry_and_Uncertainties}

\subsection{Extended-Source Photometry} \label{Subsection:Photometry}

\begin{figure}
\begin{center}
\includegraphics[width=0.23\textwidth]{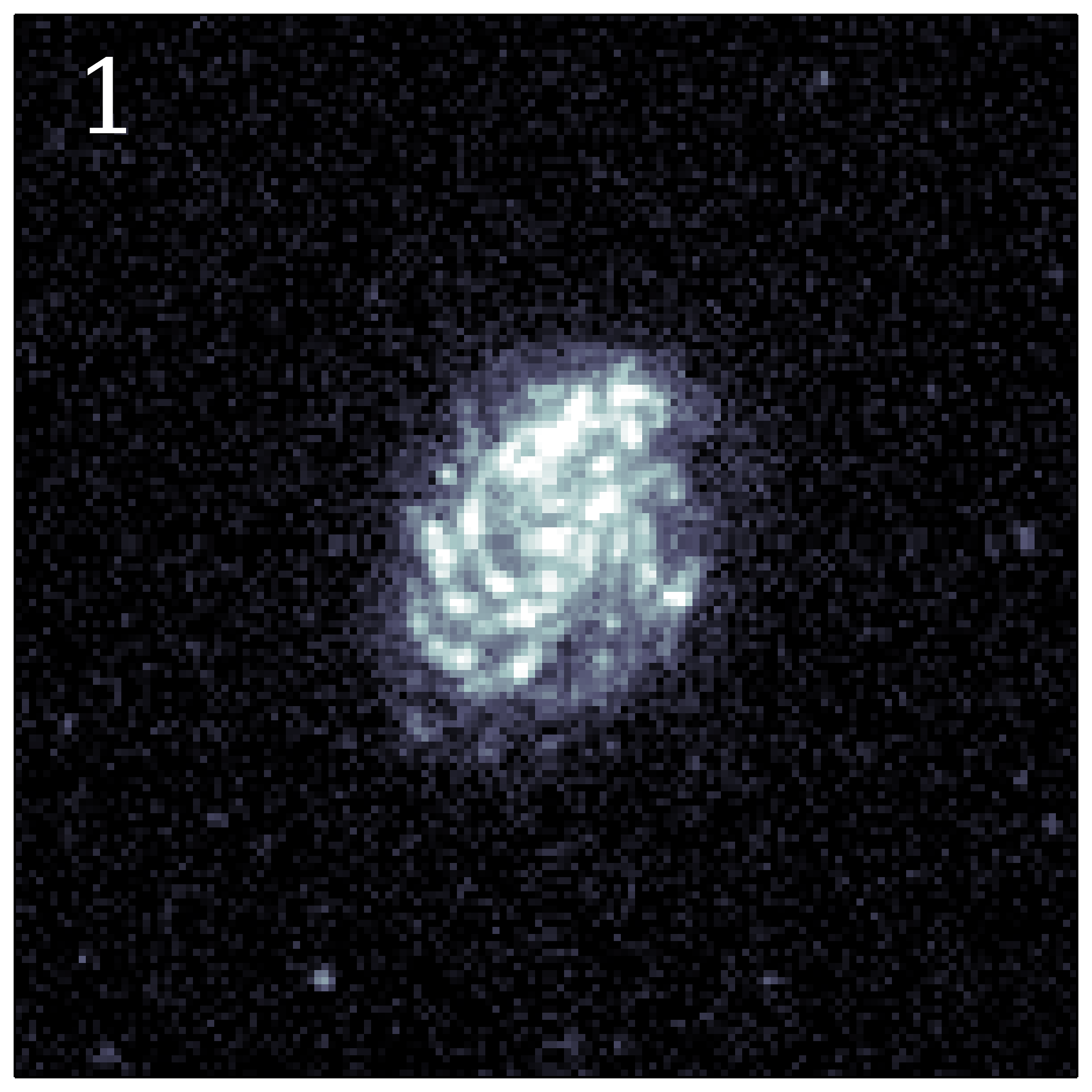}
\includegraphics[width=0.23\textwidth]{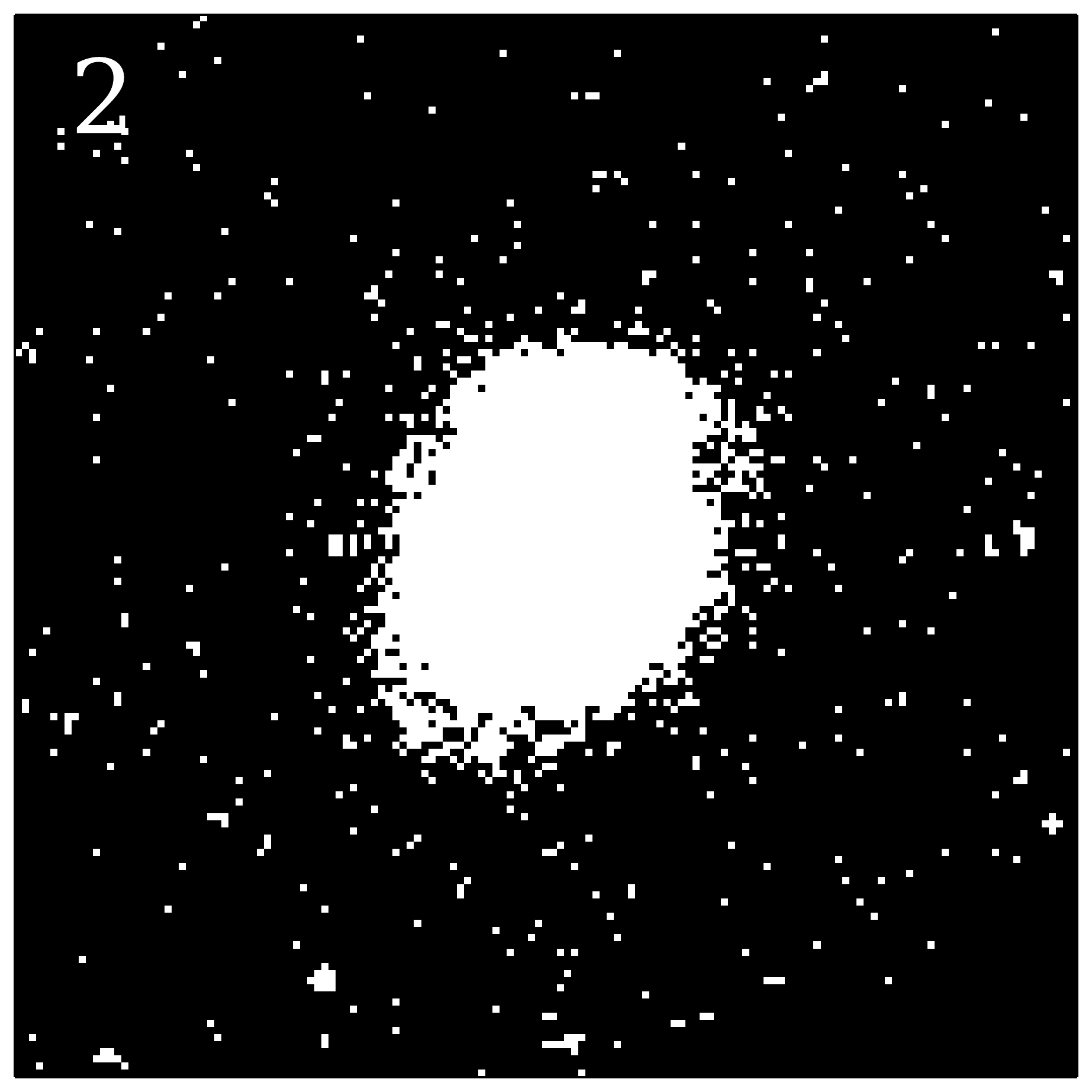}
\includegraphics[width=0.23\textwidth]{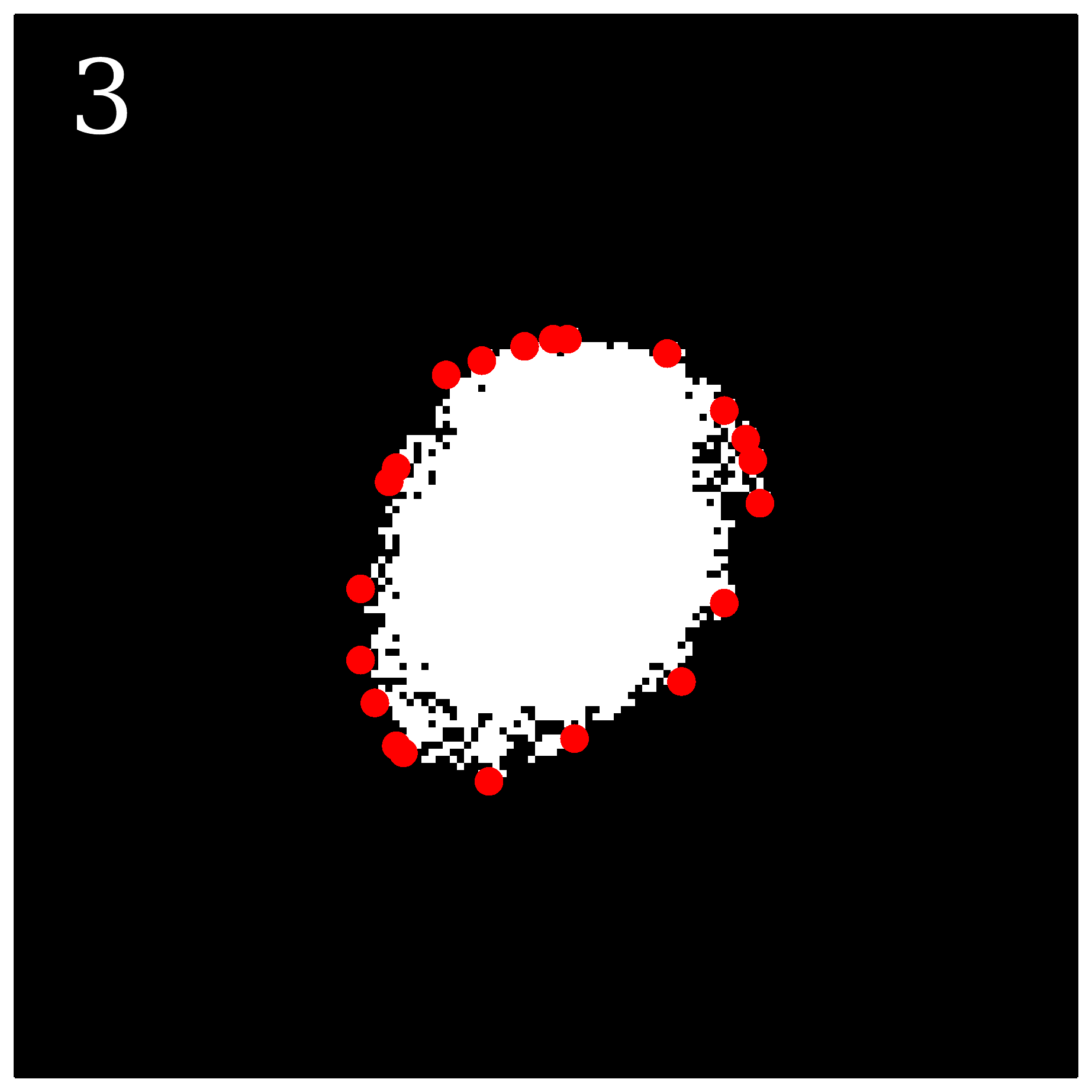}
\includegraphics[width=0.23\textwidth]{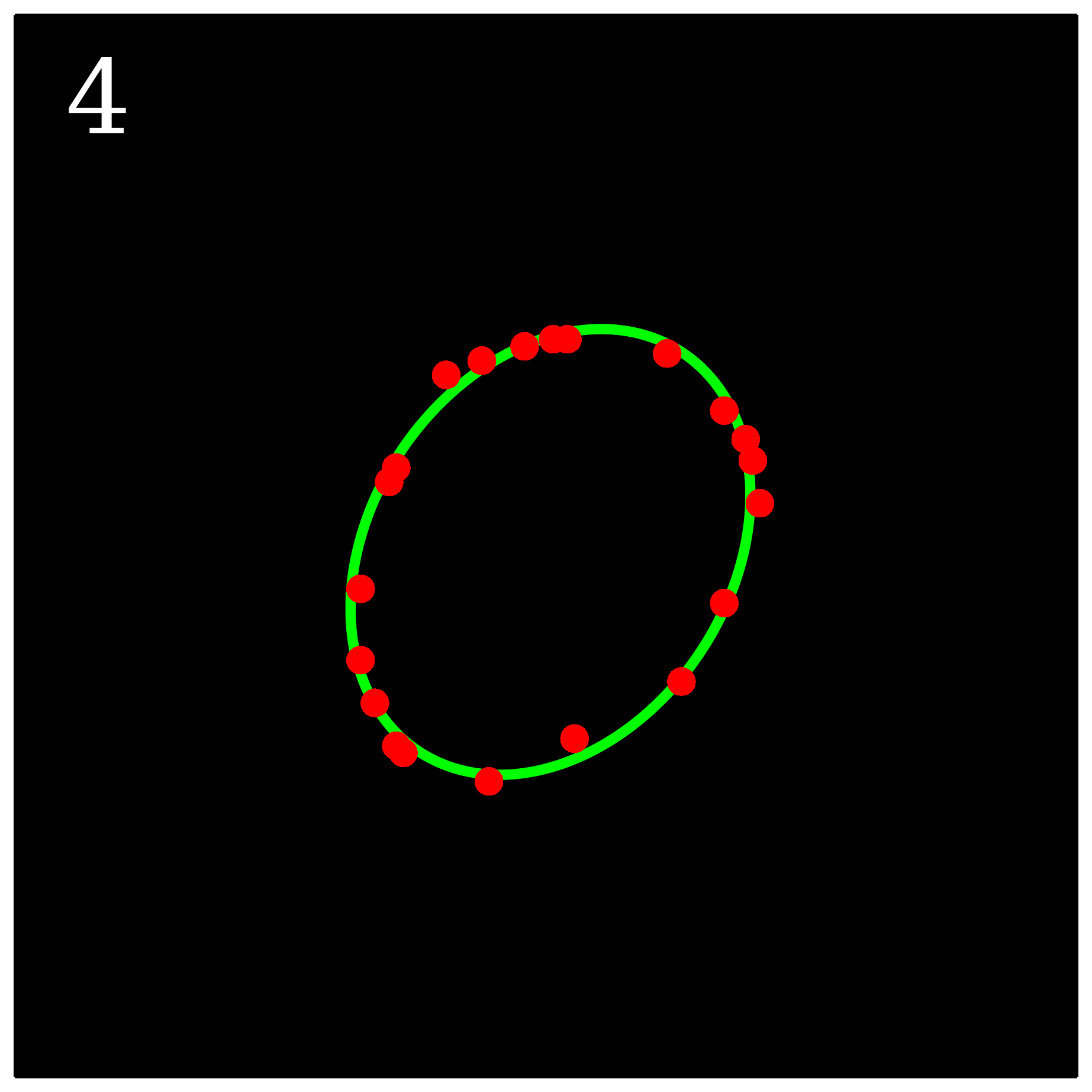}
\includegraphics[width=0.23\textwidth]{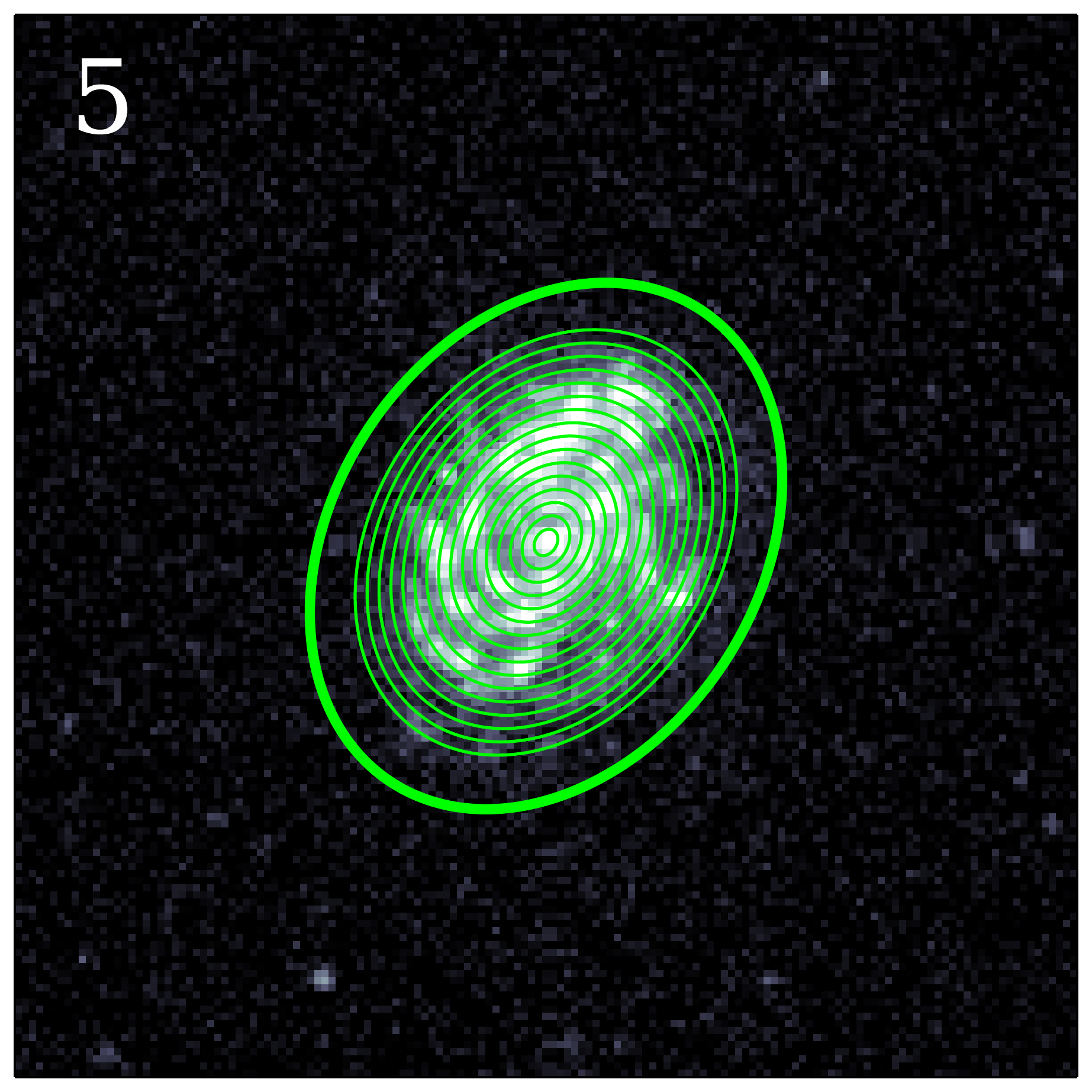}
\includegraphics[width=0.23\textwidth]{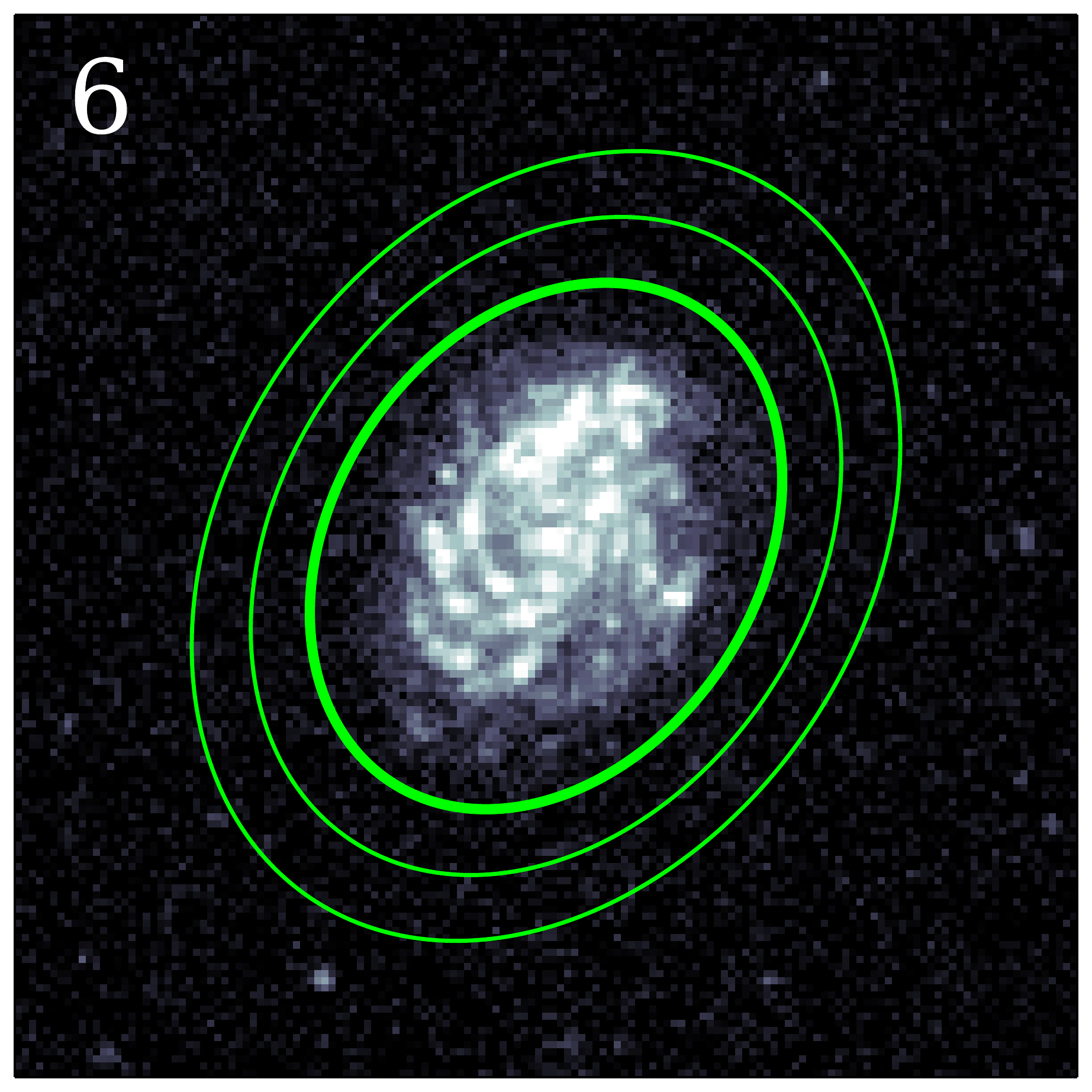}
\caption{Illustration of the stages of our aperture-fitting process, using GALEX FUV imagery of galaxy NGC 5584 (HAPLESS 14) as an example. Panel 1 shows the inner $500\arcsec\times 500\arcsec$ portion of the cutout centred upon the target source. Panel 2 shows all of the pixels in the cutout with SNR \textgreater\ 3. Panel 3 shows the significant pixels of the target source, contained within their convex hull (red points). Panel 4 shows an ellipse fitted to the convex hull; this ellipse provides the position angle and axial ratio of the source aperture. Panel 5 depicts the incremental annuli used to establish the semi-major axis at which annular flux falls to SNR\,\textless\,2 (thin concentric lines); 1.2 times this distance is then used as the semi-major axis of the source aperture (thick line). Panel 6 demonstrates the final source aperture (thick line) and sky annulus (thin lines). The apertures at all bands for a given sources are then compared to select the largest, which is then employed for all bands.}
\label{Fig:CAAPR_Example}
\end{center}
\end{figure}

We conducted our own aperture-matched photometry of the HAPLESS galaxies, across the entire UV-to-submm wavelength range, with exceptions for the IRAS 60 \micron\ measurements, and for the PACS 100 and 160\,\micron\ aperture fitting; these differences are detailed in Sections \ref{Subsubsection:IRAS_Photometry} and \ref{Subsubsection:PACS_Photometry} respectively. At all other wavelengths, we applied a consistent photometric process, tailored to reliably cope with the wide range of sizes and morphologies exhibited by the sample across the 20 photometric bands employed. These bands are: GALEX FUV and NUV; SDSS {\it ugri}, VIKING {\it ZYJHK$_{s}$}, WISE 3.4, 4.6, 12 and 22\,\micron; \hersc-PACS 100, and 160\,\micron; and \hersc-SPIRE 250, 350, and 500\,\micron. In summary, an elliptical aperture was fitted to a given source in the FUV--22\,\micron\ bands\footnote{SPIRE bands were not used to define the aperture size due to the high levels of confusion noise.}.  The sizes of these apertures were compared to identify the largest, which was subsequently then used to perform matched photometry across all bands (see Figure~\ref{Fig:CAAPR_Example}).

In detail, we first cut-out a $2000\arcsec \times 2000\arcsec$ region centred on the target source in each band. In the UV--NIR, bright foreground stars were removed. The SDSS DR9 \citep{Ahn2012} catalogue was used to identify the locations of the brightest $\sim$\,20\,per\,cent of stars in the field. Locations for stars in non-SDSS bands were also taken from the SDSS catalogue, as it was found to provide the most complete and robust identification of the stars present. Each star was profiled using a curve-of-growth technique, to determine the size of the area to be masked. The pixels in the masked region were then replaced by a random sampling of the pixels immediately adjacent to the edge of the mask.

To provide the position angle and axial ratio of the source aperture, we identified all of the pixels in the cutout that had a SNR\,\textgreater\,3 associated with the source, and determined the vertices of their corresponding convex hull\footnote{The convex hull is the tightest polygon that can enclose a given set of points.}. As the vertices of the convex hull trace the outline of the target, least-squares fitting of an ellipse to these points provides the position angle and axial ratio (i.e. the {\it shape}, but not the {\it size}) of the elliptical source aperture for the band in question.

The semi-major axis of the source aperture was determined by placing successive concentric elliptical annuli (with the already-determined postion angle and axial ratio) on the target, centred on the optical SDSS position, with semi-major axes separated by one pixel-width, until a mean per-pixel SNR \textless\ 2 was reached. As flux associated with a source with a Sersic profile will fall beyond the edge of any practical SNR cutoff\footnote{This is true not only for our SNR technique, but also a curve-of-growth approach \citep{Overcast2010} and the SDSS Petrosian method \citep{Blanton2001H}.}, the fitted aperture was multiplied by a factor of 1.2, large enough to be confident of encompassing nearly all the flux, whilst small enough to minimise aperture noise. The effects of using different extension factors, tests upon simulated sources, and visual inspection, all indicate that the factor of 1.2 used here achieves this well. This then defined the size of the source aperture. The semi-major and -minor axes of the generated apertures were compared across wave-bands (after subtracting in quadrature the PSF appropriate to that band), and the largest selected as the definitive photometric aperture, to be employed in every band for a given source. GALEX FUV or NUV served as the defining band for most sources, except in the case of early-type galaxies (and the more early-type spirals), for which it was generally VIKING $Z$-band. We also determined the $r$-band $R25$ and FUV $R28$ (the radius to the 25\textsuperscript{th} and 28\textsuperscript{th} magnitude per square arcsecond isophotes, respectively) of each galaxy, by interpolating between the mean surface density within annuli of one pixel-width; these values are given in Table~\ref{Table:Misc_Properties}.

For the FUV--MIR, we subtracted the background using a sky annulus with inner and outer semi-major axes of 1.25 and 1.5 times that of the source aperture. For the PACS and SPIRE data we used a larger inner and outer annulus of 1.5 and 2 times the source aperture, thus ensuring enough pixels were sampled to make a valid estimation of the value of the background. In both cases, the average background value was calculated by taking the iteratively 3$\sigma$-clipped mean of all pixels within the sky annulus.

The photometry from the FUV to \Kband\ was corrected for Galactic extinction in line with the GAMA method described in \citet{Adelman-McCarthy2008B}.

In the case of NGC 5738 (HAPLESS 22), a dwarf lenticular, emission in the submm and UV is confined to a point source at the centre of the galaxy, as is often seen in early-types \citep{MWLSmith2012B}. The standard aperture, defined by NGC 5738's much larger optical disc, yields poor-quality photometry in the submm bands due to the aperture containing too much background. We therefore opt to utilise \hersc\ point-source photometry in the case of this one object. NGC 5738 is unique amongst our sample -- in all other cases, sources compact in the UV and submm are compact across the spectrum.

\subsubsection{IRAS SCANPI Photometry} \label{Subsubsection:IRAS_Photometry}

For IRAS 60 \micron\ we used the Scan Processing and Integration Tool (SCANPI\footnote{Provided by the NASA/IPAC Infrared Science Archive: \url{http://irsa.ipac.caltech.edu/applications/Scanpi/}}), following the procedure laid out by \citet{Sanders2003A}. The SCANPI tool is unable to process non-detections where the estimated background is greater than the measured flux; in those cases we record a flux of 0, with an uncertainty equal to the IRAS 60 \micron\ 1\,$\sigma$ sensitivity limit of 58\,mJy \citep{Riaz2006C}. 

\subsubsection{\hersc\ PACS Photometry} \label{Subsubsection:PACS_Photometry}

In the standard \HATLAS\ PACS 100 and 160\,\micron\ data reduction (Valiante et al., {\it in prep.}), {\tt Nebuliser} (an algorithm to remove the background emission, \citealp{Irwin2010}) was used to flatten the maps {\it after} they were run through {\tt Scanamorphos} (which deals with $1/f$ noise on the maps, \citealp{Roussel2013A}). For sources with apertures \textgreater\,2.5\arcmin, we used the raw {\tt Scanamorphos} maps instead, as {\tt Nebuliser} removes some emission at these scales. Nonetheless, we still find that using the same apertures for PACS as for the other bands results in poor photometry. Flux at 100 and 160\,\micron\ tends to be concentrated towards the centres of galaxies, often resulting in a small patch of flux at the centre of a much larger aperture; this can drive up the aperture noise enough that a source with clearly-visible flux can count as a `non-detection'. As a result, we define our PACS apertures separately, using the 250\,\micron\ maps for each source, as these are reliable indicators of where dust emission is present. Apart from using a different band to define the apertures, PACS photometry otherwise proceeds in the same manner as described in the main part of Section~\ref{Subsection:Photometry}.

\subsubsection{Comparison with GALEX-GAMA Photometry} \label{Subsubsection:GALEX_GAMA_Comparison}

Given the importance of the UV photometry to this work, and the fact that our apertures in most cases were defined by analysis of surface photometry in the FUV, we have made a detailed comparison of our FUV photometry with the Curve-of-Growth (CoG) FUV photometry provided by the GALEX-GAMA survey (Liske et al., {\it submitted.}; Andrae et al., {\it in prep.}), which has been extensively used in studies of GAMA galaxies. The comparison was conducted for a subset of 17 HAPLESS galaxies relatively unaffected by shredding in the SDSS-based GAMA input catalogue used by the automated GALEX-GAMA CoG analysis. Our FUV apertures were very similar to those derived by the GALEX-GAMA CoG, while our FUV integrated fluxes were initially found to be systematically higher by $\sim$10\,per\,cent, with a similar degree of scatter. This moderate systematic difference in integrated flux was traced to differences in approach to masking foreground stars in the two methods. The only other detectable difference was the additional random uncertainty ($\sim$10\,per\,cent root-mean-square) being introduced by our use of Swarped images in place of the individual tiles used by GALEX-GAMA. We can conclude that both these independent methods are in acceptable agreement.


\subsection{Uncertainties} \label{Subsection:Uncertainties}

To estimate aperture noise for a source, we first 3\,$\sigma$-clipped the pixel values in a given $2000\arcsec\times 2000\arcsec$ cutout (excluding those pixels within the source aperture).   Then random apertures were placed across the cutout (again excluding the location of the source aperture itself). Each random aperture was circular, with the same area as the source aperture, and was background-subtracted in the appropriate manner for each band, as detailed above. The pixel values in each random aperture were inspected; if more than 20\,per\,cent lay beyond the cutout's calculated 3\,$\sigma$ threshold, then that random aperture was rejected. This process was repeated until 100 random apertures had been accepted. We found this clipping technique to be necessary in order to prevent the final aperture noise estimates being too dependant upon the
locations of the random apertures; otherwise the presence of bright background sources in the random apertures could cause the aperture noise estimate to vary wildly between repeat calculations on a given cutout. The WISE 3.4 and 4.6\,\micron\ maps were found to be particularly vulnerable to this effect, due in part to anomalies in the maps (halos, etc) caused by bright foreground stars.

Once 100 random apertures had been accepted, the flux in each was recorded, and the standard deviation of all 100 fluxes was taken to represent the aperture noise. This method of aperture noise estimation includes the contribution from confusion noise in \hersc\ bands.

We wanted the uncertainty values of our flux measurements to include not only the background noise and random photometric uncertainty, but also include the uncertainty in our ability to measure the total flux of a galaxy. To that end, we performed two tests. Firstly, we repeated the photometry with an aperture size 20\,per\,cent larger for each source. Ideally, the fluxes obtained using these larger apertures would be identical to those obtained from the normal apertures; the amount of deviation between the two lets us gauge the effectiveness of both our aperture-fitting and our background-subtraction. Secondly, we repeated the photometry, but instead estimated the background using a sigma-clipped median within the sky annulus, instead of a sigma-clipped mean. These should both be equally valid methods, and so the deviation between the final fluxes returned by them allows us to gauge the limits of our ability to accurately determine the background. The additional uncertainty added by these tests is smaller than the instrumental calibration uncertainties (see below), except in the optical bands, where the instrumental calibration uncertainty is very small.

No systematic difference in measured flux was found for either of these tests. For each of the two tests, the associated error value was determined by calculating the root-median-squared deviation across all 42 sources. For each band, these two error values were then added in quadrature to the band's calibration uncertainty -- as given by \citet{Morrissey2007B} for GALEX, the SDSS DR9 Data Release Supplement\footnote{\url{http://www.sdss3.org/dr9/}} for SDSS, \citet{Edge2013} for VIKING, the WISE All-Sky Data Release Explanatory Supplement for WISE\footnote{\url{http://wise2.ipac.caltech.edu/docs/release/allsky/expsup/}}, the PACS Observers' Manual\footnote{\url{http://herschel.esac.esa.int/Docs/PACS/html/pacs_om.html}} for PACS, and the SPIRE Observer's Manual\footnote{\url{http://herschel.esac.esa.int/Docs/SPIRE/html/spire_om.html}} for SPIRE (see also \citealt{Bendo2013}). This was then added in quadrature to the aperture noise to provide the final photometric uncertainty. 

For the IRAS 60 \micron\ photometry acquired separately using SCANPI, the reported flux uncertainty is added in quadrature to a 20\,per\,cent calibration uncertainty \citep{Sauvage2011A} to provide the total photometric uncertainty for each source.

The final fluxes and uncertainties in all bands can be found in Table~\ref{AppendixTable:Photometry} in Appendix~\ref{AppendixSection:HAPLESS}.

\section{Properties of the HAPLESS Galaxies} \label{Section:HAPLESS_Properties}

\subsection{Modified Blackbody SED Fitting} \label{Subsection:SED_Fitting}
\begin{figure}
\begin{center}
\includegraphics[width=0.5\textwidth]{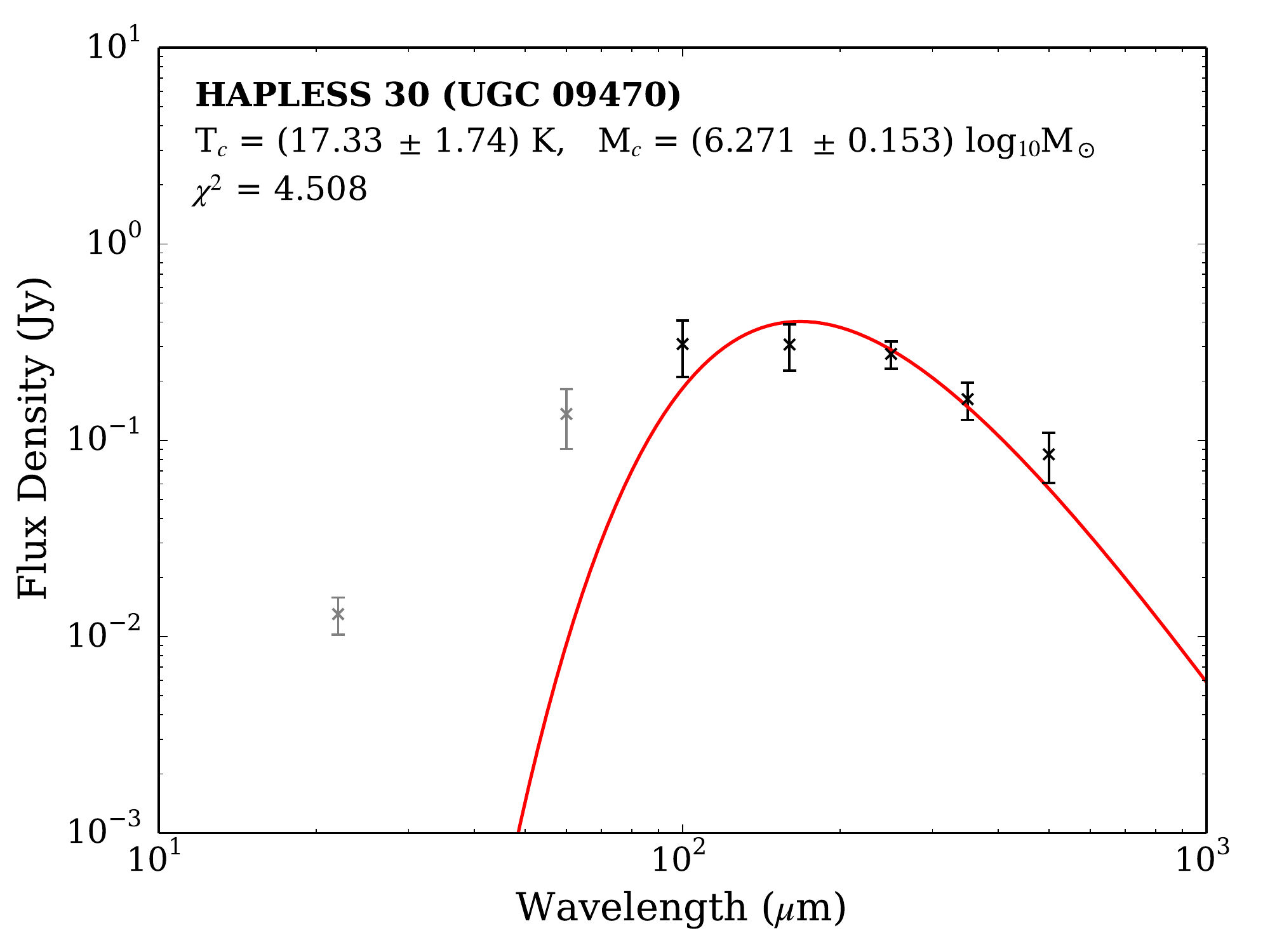}
\includegraphics[width=0.5\textwidth]{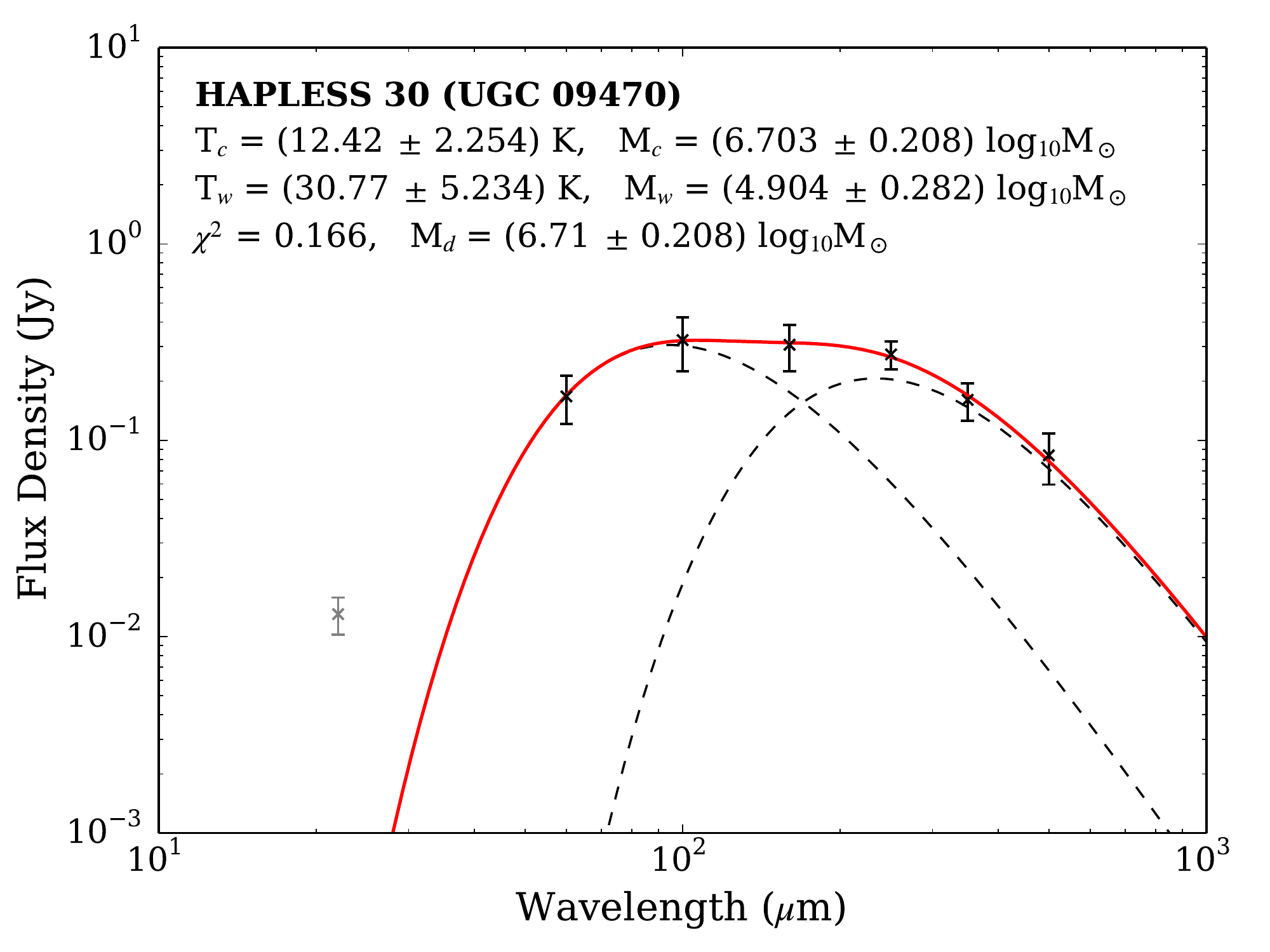}
\caption{Example dust SED of UGC 09470 (HAPLESS 30), with one- and two-temperature component modified blackbody fits attempted (upper and lower panels respectively). Both used a fixed $\beta = 2$. Grey points represent upper limits in the fitting routine. This is an example of a galaxy for which a one-component dust model systematically underestimates the flux at both 100 and 500 \micron, whilst overestimating it at 160\,\micron.}
\label{Fig:SED_Example}
\end{center}
\end{figure}

\begin{figure}
\begin{center}
\includegraphics[width=0.5\textwidth]{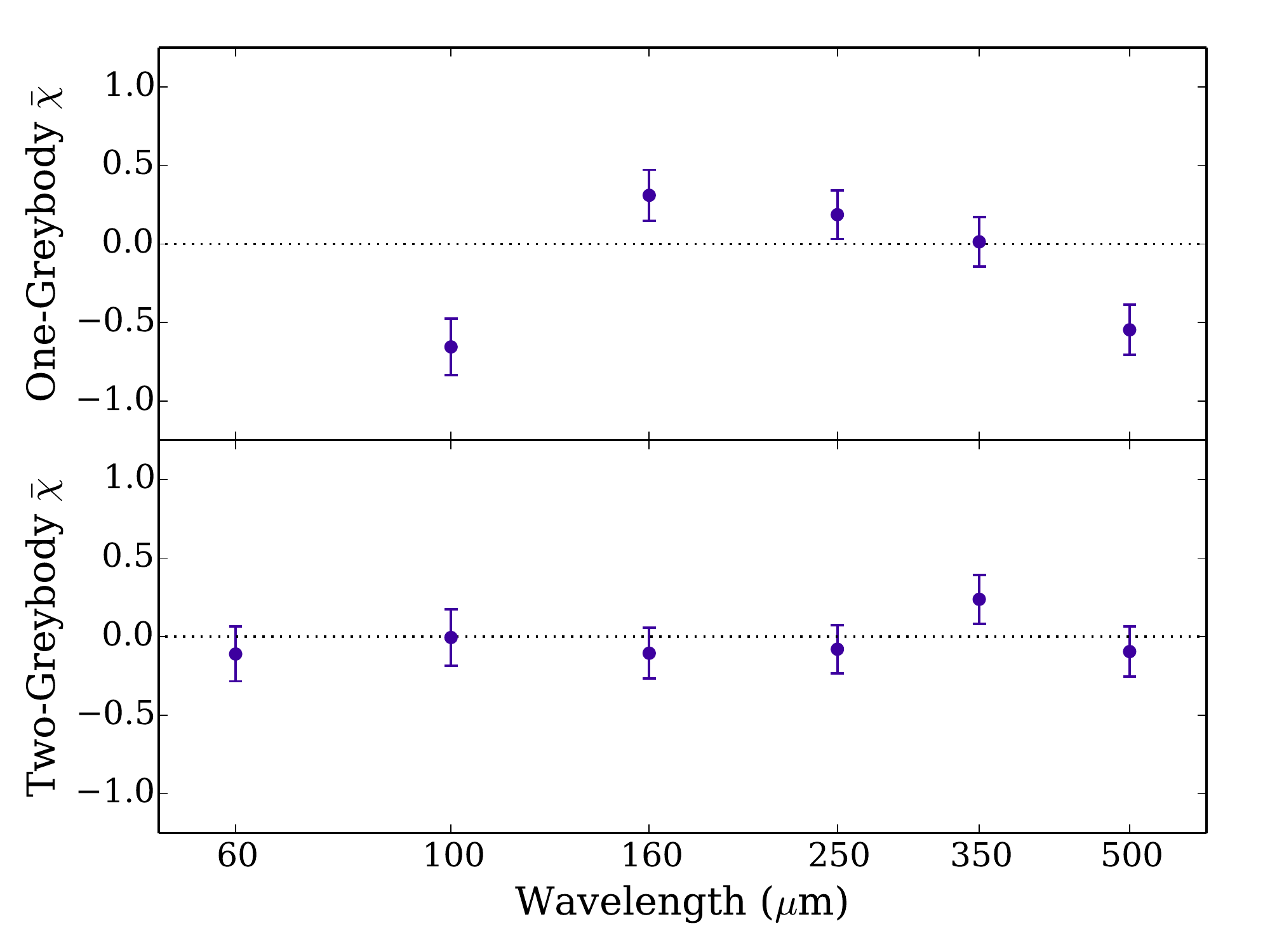}
\caption{The mean residual across the sample between the model and the data as a fraction of the uncertainty, $\bar{\chi}$, in each band, for the one- and two-temperature modified blackbody fits (example in Figure~\ref{Fig:SED_Example}). The single MBB approach systematically overestimates the flux at 160\,\micron\, whilst underestimating it at 100 and 500\,\micron.  The error bars show the uncertainty on the data points, defined by $\sigma = N^{-\frac{1}{2}}$.}
\label{Fig:chi_bar_Comparison_Grid_1}
\end{center}
\end{figure}

To estimate the dust masses and temperatures of the HAPLESS galaxies,
we fit Modified BlackBodies (MBBs) of the form $S_{\nu} \propto
\nu^{\beta}B(\nu,T_d)$ to the FIR and submm Spectral Energy
Distributions (SEDs), where $\beta$ is the dust emissivity index. We
first tried using a single-temperature MBB, keeping $\beta$ fixed at a
value of 2 and fitting only those data points with $\lambda \ge
100\,\mu$m.  This is because the mid-IR part of the SED has
contributions from very small grains which are transiently heated by
single photons, and therefore not in equilibrium with the radiation
field \citep{Boulanger1988,Desert1990}.  This contribution results in
a power-law behaviour for the portion of the SED between 12-70\,\micron\
and including this data in the single-temperature MBB fit would bias
the temperature high.  Figure~\ref{Fig:SED_Example} (upper) shows an
example of a single-temperature MBB; overall we found that this method
systematically underestimated the fluxes at 100 and 500\,\micron,
whilst overestimating them at 160\,\micron. We demonstrate this using
the stacked residuals between the model and the data in
Figure~\ref{Fig:chi_bar_Comparison_Grid_1}.

The residuals suggest that a `flatter' SED, produced either by a lower
value of $\beta$ or by having dust at a range of temperatures
\citep{Dunne2001A, Shetty2009A}, would be more suitable. We next tried
leaving $\beta$ as a free parameter and found a wide range of $\beta$
values (0--4) could adequately fit the HAPLESS sources. Whilst this
greatly reduced the systematic bias, it did not eliminate it.
\citet{Kelly2012B} recently demonstrated that $\chi^{2}$ SED fitting
routines with a given `true' value of $\beta$, can return a wide range
of fitted values for $\beta$ (see also \citealp{DJBSmith2013A}); furthermore \citet{Galametz2012} demonstrated that a
  variable $\beta$ will produce less accurate results than using a
  fixed value. We therefore use a fixed $\beta$ of 2 in this work, as
both observational \citep{Dunne2001A, Clemens2013A, DJBSmith2013A,
  Planck2013XIV} and experimental \citep{Demyk2013A} evidence suggest
values between 1.8--2.0 are appropriate for nearby galaxies.  Using $\beta=2$ also allows us to easily compare our results to other recent {\it Herschel} and \planck\ studies (see Section~\ref{Section:Comparison_to_Other_Samples}). A single MBB only
provides a useful approximation if the large grains have a narrow
range of temperatures (Mattsson et al., {\it in press}), which appears not to be the case for many
galaxies in HAPLESS (and other 
FIR surveys; see Mattsson et al., {\it in press}, \citealp{Bendo2014}). We therefore opt to use an SED model which
incorporates two temperature components:

\begin{equation}
S_{\nu} = \frac{\kappa_{\nu}}{D^{2}} \left[\ M_{w} B(\nu,T_{w}) + M_{c} B(\nu,T_{c})\ \right]
\label{Equation:SED_2}
\end{equation}

\noindent where $S_{\nu}$ is the flux at frequency $\nu$, $\kappa_{\nu}$ is the dust mass absorption coefficient at frequency $\nu$, $M_{w}$ and $M_{c}$ are the hot and cold masses, $B(\nu,T_{h})$ and $B(\nu,T_{c})$ are each the Planck function at frequency $\nu$ and characteristic dust temperatures $T_{h}$ and $T_{c}$, $D$ is the distance to the source. At submm wavelengths, the dust absorption coefficient $\kappa_{\nu}$ varies with frequency as $\kappa_{\nu} \propto \nu^{\beta}$.

We performed the two-temperature MBB fitting from 60--500\,\micron;
the 22\,\micron\ point is used as an upper limit to prevent
unconstrained warm components from being fitted. A
$\chi^{2}$-minimisation routine was used which incorporates
colour-corrections for filter response function and beam
area\footnote{The median colour corrections are 0.957,
    0.995, 0.990, 1.000, 1.004, 0.992 at 60, 100, 160, 250, 350,
    500\,\micron\ across our entire sample.}. Both temperature components were kept within the 5--200\,K range, but were otherwise entirely free. Note that for a galaxy
with an SED that is well-fit by a single-component model, this method
is free to assign negligible mass to one of the dust components, or fit two identical-temperature components. In
keeping with other \HATLAS\ works, we use a value for the dust
absorption coefficient of $\kappa_{850} = 0.077\,{\rm m^{2}\ kg^{-1}}$
from \citet{James2002}, which we extrapolate to other wavelengths
using a $\beta=2$.

Using the two-temperature SED fitting, we no longer encounter any systematic biases in our model fits to the data, as can be seen in the lower panel of Figure~\ref{Fig:chi_bar_Comparison_Grid_1}. Figure~\ref{Fig:SED_Example} shows an example of both one- and two-temperature fits to the SED of HAPLESS 30; the two-temperature fits of all our sources are displayed in Figure~\ref{AppendixFigure:HAPLESS_SED_Grid}.

\begin{table*}
\begin{center}
\caption{Dust properties of the HAPLESS galaxies. Dust masses ($M$) and temperatures ($T$) were derived using a $\chi^{2}$-minimising fit to a two-component modified blackbody SED model, given by Equation~\ref{Equation:SED_2}. Uncertainties were determined by means of a bootstrapping analysis.   }
\label{Table:Dust_Propertires}
\begin{tabular}{lrrrrrrrrr}
\toprule \toprule
\multicolumn{1}{c}{HAPLESS} &
\multicolumn{1}{c}{$T_{c}$} &
\multicolumn{1}{c}{$\Delta T_{c}$} &
\multicolumn{1}{c}{$T_{w}$} &
\multicolumn{1}{c}{$\Delta T_{w}$} &
\multicolumn{1}{c}{$M_{c}/M_{w}$} &
\multicolumn{1}{c}{$\Delta M_{c}/M_{w}$} &
\multicolumn{1}{c}{$M_{d}$} &
\multicolumn{1}{c}{$\Delta M_{d}$} &
\multicolumn{1}{c}{$L_{\it TIR}$} \\
\multicolumn{1}{c}{} &
\multicolumn{1}{c}{(K)} &
\multicolumn{1}{c}{(K)} &
\multicolumn{1}{c}{(K)} &
\multicolumn{1}{c}{(K)} &
\multicolumn{1}{c}{(log)} &
\multicolumn{1}{c}{(dex)} &
\multicolumn{1}{c}{(log$_{10}$ M$_{\odot}$)} &
\multicolumn{1}{c}{(dex)} &
\multicolumn{1}{c}{(log$_{10}$ L$_{\odot}$)} \\
\midrule
1 & 25.6 & 1.9 & 59.7 & 11.4 & 2.1 & 0.8 & 5.4 & 0.1 & 9.0\\ 
2 & 17.2 & 1.6 & 67.2 & 19.3 & 3.6 & 1.8 & 6.0 & 0.2 & 8.5\\ 
3 & 13.5 & 2.4 & 27.7 & 2.9 & 1.3 & 0.5 & 7.2 & 0.2 & 9.5\\ 
4 & 16.7 & 4.9 & 32.7 & 14.2 & 1.4 & 1.3 & 5.7 & 0.4 & 8.4\\ 
5 & 12.9 & 2.6 & 52.5 & 6.8 & 3.1 & 1.8 & 6.0 & 0.3 & 8.1\\ 
6 & 21.7 & 1.0 & 64.2 & 16.4 & 3.2 & 1.4 & 7.9 & 0.1 & 10.9\\ 
7 & 12.2 & 2.8 & 23.2 & 2.4 & 1.1 & 0.6 & 7.4 & 0.2 & 9.5\\ 
8 & 19.6 & 1.3 & 58.0 & 13.8 & 3.0 & 1.0 & 6.4 & 0.1 & 9.1\\ 
9 & 15.0 & 1.6 & 44.6 & 12.1 & 2.9 & 1.0 & 6.7 & 0.2 & 8.8\\ 
10 & 17.5 & 2.6 & 30.1 & 14.0 & 1.2 & 1.0 & 7.3 & 0.1 & 10.0\\ 
11 & 11.3 & 1.4 & 17.7 & 15.6 & 1.5 & 2.1 & 6.9 & 0.2 & 8.4\\ 
12 & 15.7 & 2.4 & 30.7 & 9.8 & 1.5 & 1.0 & 6.4 & 0.2 & 8.8\\ 
13 & 13.2 & 3.1 & 50.6 & 6.5 & 2.9 & 1.6 & 5.7 & 0.3 & 7.9\\ 
14 & 14.6 & 2.4 & 26.5 & 2.9 & 1.1 & 0.6 & 7.4 & 0.1 & 9.8\\ 
15 & 15.5 & 4.3 & 62.6 & 9.0 & 3.4 & 2.3 & 5.5 & 0.5 & 7.9\\ 
16 & 16.1 & 3.2 & 30.0 & 11.2 & 1.4 & 1.1 & 6.7 & 0.2 & 9.2\\ 
17 & 19.6 & 6.1 & 56.8 & 7.9 & 1.9 & 1.0 & 5.3 & 0.6 & 8.8\\ 
18 & 12.8 & 2.5 & 29.9 & 3.1 & 1.5 & 0.5 & 6.7 & 0.2 & 9.0\\ 
19 & 12.3 & 1.6 & 28.8 & 6.3 & 2.2 & 1.1 & 7.0 & 0.2 & 8.8\\ 
20 & 21.2 & 2.4 & 57.9 & 14.2 & 2.7 & 1.0 & 7.5 & 0.1 & 10.5\\ 
21 & 17.4 & 0.9 & 34.0 & 18.6 & 2.5 & 1.4 & 8.0 & 0.1 & 10.3\\ 
22 & 11.5 & 2.1 & 34.1 & 6.8 & 1.9 & 0.7 & 6.0 & 0.4 & 8.1\\ 
23 & 20.7 & 1.9 & 58.1 & 14.8 & 2.8 & 1.1 & 7.6 & 0.1 & 10.5\\ 
24 & 16.3 & 3.9 & 16.4 & 4.5 & 5.7 & 1.5 & 5.7 & 0.3 & 8.1\\ 
25 & 11.7 & 1.1 & 21.5 & 4.2 & 0.2 & 2.2 & 7.2 & 0.1 & 9.6\\ 
26 & 13.0 & 1.8 & 27.2 & 10.2 & 1.7 & 1.1 & 7.4 & 0.2 & 9.4\\ 
27 & 14.2 & 1.6 & 64.2 & 5.3 & 4.3 & 1.8 & 6.2 & 0.2 & 8.2\\ 
28 & 21.5 & 4.1 & 36.6 & 14.8 & 1.3 & 1.1 & 6.9 & 0.1 & 10.1\\ 
29 & 24.4 & 1.5 & 66.6 & 16.8 & 2.8 & 1.5 & 7.6 & 0.1 & 10.9\\ 
30 & 12.4 & 2.4 & 30.8 & 3.9 & 1.8 & 0.8 & 6.7 & 0.3 & 8.8\\ 
31 & 15.8 & 3.3 & 28.5 & 15.8 & 1.7 & 1.5 & 7.2 & 0.2 & 9.5\\ 
32 & 14.1 & 2.9 & 28.7 & 2.5 & 1.0 & 0.5 & 6.6 & 0.2 & 9.2\\ 
33 & 20.8 & 8.2 & 42.8 & 13.8 & 2.1 & 3.4 & 5.7 & 0.8 & 8.7\\ 
34 & 9.2 & 2.7 & 24.3 & 4.6 & 1.9 & 1.0 & 7.2 & 0.5 & 8.6\\ 
35 & 14.5 & 1.5 & 55.9 & 12.3 & 3.4 & 0.8 & 6.7 & 0.2 & 8.8\\ 
36 & 11.2 & 1.1 & 61.2 & 18.6 & 4.1 & 1.3 & 6.8 & 0.2 & 8.3\\ 
37 & 15.3 & 2.8 & 27.3 & 15.5 & 1.6 & 1.4 & 7.3 & 0.2 & 9.5\\ 
38 & 13.3 & 4.0 & 23.4 & 16.0 & 1.5 & 2.7 & 6.2 & 0.6 & 8.2\\ 
39 & 12.5 & 2.9 & 27.0 & 2.5 & 1.2 & 0.6 & 7.1 & 0.3 & 9.4\\ 
40 & 10.8 & 6.3 & 24.5 & 14.5 & 0.5 & 2.3 & 6.7 & 0.6 & 9.2\\ 
41 & 14.6 & 2.5 & 46.1 & 10.9 & 2.3 & 0.8 & 5.8 & 0.3 & 8.4\\ 
42 & 11.7 & 2.5 & 11.7 & 14.8 & 3.2 & 2.4 & 6.3 & 0.4 & 7.4\\ 
\bottomrule
\end{tabular}
\end{center}
\end{table*}
 
Dust masses\footnote{The median dust mass in our sample is higher than that in the overlapping sample of \citet{Bourne2013}, this is due to differences in the distances used and the photometry method.} and temperatures for the HAPLESS galaxies are listed in Table~\ref{Table:Dust_Propertires}. The temperatures of the cold dust components 
range from 9.2 to 25.6 K, with a median temperature of 14.6\,K. The total dust masses range from $2.2 \times 10^{5}$ to $9.5 \times 10^{7}\,{\rm M}_{\odot}$, with a median mass of $5.6 \times 10^{6}\, {\rm M}_{\odot}$. Uncertainties in the derived dust masses and temperatures were estimated by means of a bootstrapping analysis, whereby the fluxes were randomly re-sampled according to a Gaussian distribution defined by the flux uncertainties, and a best fit was made to the re-sampled SED; this was repeated 1,000 times, and the standard deviation in the returned fit parameters was taken to represent their uncertainty. All quoted dust masses are the sum of the cold and warm components, though the cold component significantly dominates the dust mass budget in most of our galaxies (Table~\ref{Table:Dust_Propertires}). 

Some galaxies do have SEDs that would be adequately fit by a one-component MBB; in such cases, there is a risk that using the two-component model could give rise to a spurious low-luminosity cold dust component that would yield an artificially large dust mass, and low cold dust temperature. We gauged the potential impact of this effect by weighting the dust temperatures, according to: 

\begin{equation}
T_{\it weighted} = \frac{M_{c}T_{c}+M_{w}T_{w}}{M_{c}+M_{w}}; 
\label{Equation:Weighted_Temp}
\end{equation}

However, this only causes a significant change in temperature for the two galaxies with the lowest values of $M_{c}/M_{w}$ (HAPLESS 25 and 40). The median $T_{\it weighted}$ is only 0.8\,K greater than the median $T_{c}$, with no significant difference to any of the trends with temperature reported in this work. It is also important to consider that recent work by \citet{Bendo2014} has shown that low-luminosity cold dust components are present in some galaxies; in such cases, a one-component MBB may be an adequate fit to the data, but not reflect the actual nature of the dust in a galaxy.

It is unclear what relationship the systematic 500\,\micron\ excess in our single-temperature MBB fits (Figure~\ref{Fig:chi_bar_Comparison_Grid_1}) bears to the submm excess seen by many other authors \citep{Galliano2003A,Galametz2012A,Remy-Ruyer2013,Ciesla2014A,Grossi2015A} -- as we also see an excess at 100\,\micron, and a deficiency at 160\,\micron. The two-temperature MBB approach is able to account for all of our systematic residuals without the need for extremely cold ($\ll$\,10\,K) dust components.

Total infrared luminosities, $L_{\it TIR}$, from 8--1000\,\micron\ were
estimated using the best-fit SEDs and extrapolating below 60\,\micron\
using a power law to account for the luminosity produced by the
transiently heated small grain population.  This was done by forcing
the SED shape in the mid-IR to a power law, anchored to the WISE
22\,\micron\ flux (or the WISE 12\,\micron\ flux if this was not
available), and the flux at the peak of the best-fit SED (see
\citealt{Ibar2013A} for more details).  This new SED was then
integrated to produce $L_{\it TIR}$; note that the luminosity using
this method was on average 14\,per\,cent higher than simply
integrating the best-fit MBBs from 60--500\,\micron.  The values
determined using this method are in good agreement with those
determined by De Vis et al. ({\it in prep.}) derived from performing
energy-balance modelling of the full UV--submm SED with MAGPHYS
\citep{DaCunha2008}. The resulting $L_{\it TIR}$ values are listed in
Table~\ref{Table:Dust_Propertires}.

\begin{table}
\begin{center}
\caption{Miscellaneous measured and derived properties of the HAPLESS galaxies. Stellar mass is calculated using Equation~\ref{Equation:Stellar_Mass}.}
\label{Table:Misc_Properties}
\begin{tabular}{lrrrrr}
\toprule \toprule
\multicolumn{1}{c}{N\textsuperscript{\underline{o}}} &
\multicolumn{1}{c}{$r_{\it abs}$} &
\multicolumn{1}{c}{$R25_{\it r}$} &
\multicolumn{1}{c}{$R28_{_{\it FUV}}$} &
\multicolumn{1}{c}{\fK} &
\multicolumn{1}{c}{$M_{\star}$} \\
\multicolumn{1}{c}{} &
\multicolumn{1}{c}{(Mag)} &
\multicolumn{1}{c}{(arcsec)} &
\multicolumn{1}{c}{(arcsec)} &
\multicolumn{1}{c}{(mag)} &
\multicolumn{1}{c}{(log$_{10}$ M$_{\odot}$)} \\
\midrule
1 & -18.0 & 32 & 33 & 3.07$^{a}$ & 8.8\\ 
2 & -17.2 & 11 & 17 & 2.03 & 8.1\\ 
3 & -19.6 & 67 & 80 & 2.13 & 9.2\\ 
4 & -17.8 & 21 & 14 & 3.16 & 8.8\\ 
5 & -17.2 & 28 & 13 & 3.58 & 8.5\\ 
6 & -22.2 & 131 & 124 & 4.51 & 10.8\\ 
7 & -20.1 & 124 & 125 & 2.66 & 9.5\\ 
8 & -19.0 & 36 & 37 & 2.41 & 9.0\\ 
9 & -18.4 & 39 & 81 & 1.35 & 8.6\\ 
10 & -20.6 & 89 & 10 & 4.39 & 10.1\\ 
11 & -18.4 & 54 & 56 & 3.74 & 8.9\\ 
12 & -17.8 & 21 & 26 & 3.08 & 8.6\\ 
13 & -16.3 & 13 & 8 & 3.14 & 8.1\\ 
14 & -20.2 & 96 & 92 & 2.72 & 9.5\\ 
15 & -17.4 & 19 & 10 & 3.74 & 8.6\\ 
16 & -18.8 & 43 & 36 & 4.26 & 9.3\\ 
17 & -17.7 & 21 & 21 & 1.55 & 8.1\\ 
18 & -18.5 & 25 & 28 & 2.21 & 8.7\\ 
19 & -19.1 & 102 & - & \textless3.5$^{b}$ & 9.2\\ 
20 & -21.0 & 115 & 34 & 7.00 & 10.8\\ 
21 & -22.3 & 210 & - & \textgreater3.5$^{b}$ & 11.3\\ 
22 & -18.9 & 24 & 34 & 7.12 & 9.7\\ 
23 & -20.7 & 86 & 87 & 4.96 & 10.2\\ 
24 & -16.5 & 10 & 15 & 1.82 & 7.6\\ 
25 & -21.2 & 97 & 51 & 5.85 & 10.6\\ 
26 & -19.9 & 75 & 82 & 2.39 & 9.5\\ 
27 & -17.9 & 37 & 36 & 2.90 & 8.6\\ 
28 & -20.6 & 68 & 39 & 3.99 & 9.8\\ 
29 & -21.7 & 93 & 53 & 4.55 & 10.4\\ 
30 & -18.5 & 33 & 35 & 2.24 & 8.8\\ 
31 & -20.1 & 65 & 74 & 2.94 & 9.6\\ 
32 & -18.2 & 18 & 15 & 3.60 & 8.9\\ 
33 & -17.8 & 13 & 15 & 1.58 & 8.3\\ 
34 & -18.7 & 35 & 15 & 1.16 & 8.6\\ 
35 & -18.9 & 36 & 46 & 2.78 & 9.0\\ 
36 & -17.7 & 21 & 23 & 2.32 & 8.4\\ 
37 & -20.2 & 61 & 56 & 4.09 & 10.0\\ 
38 & -17.5 & 14 & 13 & 2.70 & 8.4\\ 
39 & -19.8 & 36 & 42 & 2.34 & 9.3\\ 
40 & -18.8 & 32 & 29 & 2.60 & 8.9\\ 
41 & -16.5 & 26 & 35 & 0.64 & 7.6\\ 
42 & -15.2 & 4 & 9 & 2.47 & 7.4\\ 
\bottomrule
\end{tabular}
\begin{list}{}{}
\item[$^{a}$] Note that UGC 06877 (HAPLESS 1) is an AGN \citep{Osterbrock1983E}, with a contribution from non-thermal continuum emission in the UV \citep{Markaryan1979D}.
\item[$^{b}$] Sources UGC 06780 (HAPLESS 19) and NGC 5746 (HAPLESS 21) do not have GALEX coverage. We use the $u$-$K_{s}$ colour to infer whether they belong to the curious blue subset.).
\end{list}
\end{center}
\end{table}

\subsection{Stellar Masses} \label{Subsection:Stellar_Masses}

To determine the stellar masses of the HAPLESS galaxies, we follow the
method of \citet{Zibetti2009D}, which assumes a Chabrier
\citep{Chabrier2003C} Initial Mass Function (IMF) and uses $i$-band
luminosity along with a relationship between stellar mass-to-light ratio and $g$-$i$ colour. This method combines stellar population synthesis models \citep{Bruzual2007E} including dust attenuation and compares with a sample of nearby galaxies.  Stellar masses arrived at by this method have a typical uncertainty of 0.1--0.15\,dex \citep{Cortese2012A} modulo uncertainties in the underlying population models. \citet{Zibetti2009D} caution that their approach may not be appropriate where galaxies have very young stellar populations (where $i$ would be overestimated) or significant extinction (where $i$ would be underestimated); ie, sources with obvious dust lanes (only seen in 6 of the HAPLESS galaxies, Figure~\ref{AppendixFig:HAPLESS_Imagery}). As discussed succinctly in \citet{Taylor2011A}, however, variations in extinction (for simple dust geometries), the star formation history, metallicity and age only serve to shift galaxies along the $(g-i)$ vs $M_{\star}/L_i$ relationship, such that uncertainties in these parameters do not produce large errors in the value of stellar mass inferred in this way.

The full formula we employ to calculate stellar mass is:

\begin{equation}
M_{\star} = L_{i} 10^{-0.963 + 1.032(g-i)}
\label{Equation:Stellar_Mass}
\end{equation}

\noindent where $M_{\star}$ is stellar mass and $L_{i}$ is $i$-band luminosity, both in Solar units. Stellar masses are listed in Table~\ref{Table:Misc_Properties}.

The stellar masses of the HAPLESS galaxies range from $2.6 \times 10^{7}$ to $2.2 \times 10^{11}\,{\rm
  M}_{\odot}$, with a median mass of $9.8 \times 10^{8}\,{\rm M}_{\odot}$. The \citet{Zibetti2009D} method yields stellar masses for our sources in excellent agreement with those produced by the more sophisticated MAGPHYS tool which has the ability to model more extincted or highly star-forming systems (De Vis et al., {\it in prep.}), and are also in agreement with the masses derived by GAMA \citep{Taylor2011A}. We continue to use the colour method in this work in order to compare with other nearby FIR surveys (Section~\ref{Section:Comparison_to_Other_Samples}).
   
\subsection{Atomic Gas Masses} \label{Subsection:Gas_Masses}

We searched the literature for the highest-resolution 21\,cm
observations available for each of the HAPLESS galaxies. We found 15
of our sample have observations in the literature; the instrument and
reference for each can be found in
Table~\ref{Table:HI_Properties}. For the remaining sources, we inspected
the \HI\ Parkes All-Sky Survey (HIPASS, \citealp{Meyer2004};
\citealp{Zwaan2004}; \citealp{Wong2006}) catalogue to find HIPASS sources
within the full-width half-maximum (FWHM) of the Parkes beam
(14.3\arcmin) centred on the positions of the HAPLESS galaxies. To avoid the risk of contamination due to confusion, we only accepted
matches for which there were no other known galaxies within
14.3\arcmin\ radius on the sky, nor within 500\,km\,s$^{-1}$ in
velocity.  This ensures that the matches we accept are isolated in
  \HI. From HIPASS we identify 16 additional 21\,cm
  detections associated with HAPLESS galaxies. 

\begin{table*}
\begin{center}
\caption{\HI\ properties of the HAPLESS galaxies. The origin column indicates whether 21\,cm data comes from the HIPASS or ALFALFA catalogues, or published literature values. The \HI\ centroid velocity $V_R$ and linewidth $W_{50}$ are not available for all sources with 21\,cm measurements. \HI\ masses were calculated using Equation~\ref{Equation:HI_Mass}, and gas fraction \fg\ is defined by Equation~\ref{Equation:Gas_Fraction}. Upper limits on the gas mass and gas fraction were derived using Equation~\ref{Equation:HI_Mass_Upper}.}
\label{Table:HI_Properties}
\begin{tabular}{lrrrrrrrr}
\toprule \toprule
\multicolumn{1}{c}{HAPLESS} &
\multicolumn{1}{c}{$S_{\it int}$} &
\multicolumn{1}{c}{$V_R$} &
\multicolumn{1}{c}{$W_{50}$} &
\multicolumn{1}{c}{Telescope} &
\multicolumn{1}{c}{Origin} &
\multicolumn{1}{c}{$M_{\it HI}$} &
\multicolumn{1}{c}{\fg} \\
\multicolumn{1}{c}{} &
\multicolumn{1}{c}{(Jy km s$^{-1}$)} &
\multicolumn{1}{c}{(km s$^{-1}$)} &
\multicolumn{1}{c}{(km s$^{-1}$)} &
\multicolumn{1}{c}{} &
\multicolumn{1}{c}{} &
\multicolumn{1}{c}{(log$_{10}$ M$_{\odot}$)} &
\multicolumn{1}{c}{} \\
\midrule
1 & 1.30 & 1146 & 78 & GBT 91\,m & \citet{Courtois2011A} & 8.08 & 0.13\\ 
2 & 1.39 & 1308 & - & Arecibo & \citet{Salzer1992B} & 8.36 & 0.60\\ 
3 & 23.70 & 1387 & 222 & Parkes & HIPASS & 9.56 & 0.67\\ 
4 & 0.86 & 1439 & 140 & VLA-D & \citet{Taylor1995B} & 8.24 & 0.22\\ 
5 & 0.44 & - & - & Arecibo & \citet{Impey2001H} & 7.83 & 0.18\\ 
6 & 72.00 & 1462 & 306 & Parkes & HIPASS & 10.16 & 0.19\\ 
7 & 60.90 & 1539 & 243 & Parkes & HIPASS & 10.03 & 0.74\\ 
8 & 5.70 & - & - & Arecibo & \citet{Sulentic1983E} & 9.08 & 0.52\\ 
9 & 46.90 & 1537 & 198 & Parkes & HIPASS & 9.94 & 0.96\\ 
10 & 35.80 & 1528 & 287 & WRST & \citet{Popping2011B} & 9.82 & 0.31\\ 
11 & 5.90 & 1624 & 187 & Parkes & HIPASS & 9.17 & 0.62\\ 
12 & 3.97 & 1560 & 176 & Arecibo & ALFALFA & 8.90 & 0.66  \\ 
13 & 0.38 & 1713 & 26 & Arecibo & ALFALFA & 7.87 & 0.37 \\ 
14 & 27.10 & 1638 & 198 & Parkes & HIPASS & 9.76 & 0.47\\ 
15 & 1.08 & 1652 & 72 & Arecibo  & ALFALFA$^a$ & 8.34 &  0.35\\ 
16 & 4.26 & 1673 & 205 & Arecibo  & ALFALFA & 8.97 & 0.32\\ 
17 & 3.50 & 1749 & 120 & VLA-D & \citet{Taylor1995B} & 8.96 & 0.88\\ 
18 & - & - & - & - & - & $<$\,8.67 & $<$\,0.45\\ 
19 & 26.90 & 1729 & 225 & Parkes & HIPASS & 9.85 & 0.82\\ 
20 & 43.50 & 1736 & 431 & GBT 300\,ft & \citet{Davis1983A} & 9.98 & 0.12\\ 
21 & 30.70 & 1724 & 556 & WRST & \citet{Popping2011B} & 9.83 & 0.03\\ 
22 & - & - & - & - & - & $<$\,8.70 & $<$\,0.09\\ 
23 & 25.60 & 1748 & 294 & Parkes & HIPASS & 9.78 & 0.28\\ 
24 & 2.89 & 1859 & 100 & VLA-D & \citet{Taylor1995B} & 8.93 & 0.95\\ 
25 & 5.30 & - & - & Parkes & \citet{Bottinelli199B} & 9.08 & 0.03\\ 
26 & 27.90 & 1760 & 184 & Parkes & HIPASS & 9.80 & 0.66\\ 
27 & 8.40 & 1836 & 224 & Parkes & HIPASS & 9.31 & 0.83\\ 
28 & 5.50 & 1878 & 150 & Parkes & HIPASS & 9.16 & 0.17\\ 
29 & 44.50 & 1897 & 317 & GBT 300\,ft & \citet{Davis1983A} & 10.07 & 0.28\\ 
30 & 3.80 & - & - & NED & NED$^{b}$ & 9.00 & 0.62\\ 
31 & 13.30 & 1891 & 177 & Parkes & HIPASS & 9.64 & 0.51\\ 
32 & 1.81 & 1916 & 113 & Arecibo  & ALFALFA & 8.71  & 0.39\\ 
33 & 6.10 & 1973 & 60 & VLA-D & \citet{Taylor1995B} & 9.35 & 0.91\\ 
34 & 6.50 & 2033 & 99 & Parkes & HIPASS & 9.37 & 0.86\\ 
35 & 3.22 & - & - & Arecibo & \citet{Schneider1990A} & 9.11 & 0.52\\ 
36 & 2.04  & 2143 & 98 & Arecibo  & ALFALFA & 8.93 & 0.77 \\ 
37 & - & - & - & - & - & $<$\,9.03 & $<$\,0.09\\ 
38 & 1.41 & 2433 & 127 & Arecibo  & ALFALFA & 8.79 & 0.71\\ 
39 & 8.80 & 2510 & 148 & Parkes & HIPASS & 9.55 & 0.60\\ 
40 & 3.40 & 1622 & 148 & Parkes & HIPASS & 8.87 & 0.43\\ 
41 & 6.40 & 1098 & 124 & Parkes & HIPASS & 8.65 & 0.90\\ 
42 & - & - & - & - & - & $<$\,8.76 & $<$\,0.96\\ 
\bottomrule
\end{tabular}
\begin{list}{}{}
\item[$^{a}$] Classified by ALFALFA as a low SNR source (SNR = 5).
\item[$^{b}$] A 21\,cm $S_{\it int}$ value for HAPLESS 30 (UGC 09470) is available on NED, but no reference is provided. Despite this, the corresponding \HI\ properties of HAPLESS 30 are typical of the HAPLESS sample, thus we opt to include it. 
\end{list}
\end{center}
\end{table*}

For the 11 sources with neither HIPASS nor literature \HI\ detections available, \HI\ data for 7 were provided by the ALFALFA (Arecibo Legacy Fast ALFA, \citealp{Giovanelli2005D}) survey (Haynes, {\it priv. comm.}). In total we therefore have \HI\
measurements for 38 (90\,per\,cent) of the objects in our sample.

To calculate our \HI\ masses, we used the standard prescription:

\begin{equation}
M_{\it HI} = 2.36 \times 10^{5} S_{\it int} D^{2}
\label{Equation:HI_Mass}
\end{equation}

\noindent where $M_{\it HI}$ is the mass of atomic hydrogen in Solar units, $S_{\it int}$ is the integrated 21\,cm line flux density in\,Jy km s$^{-1}$, and $D$ is the source distance in\,Mpc.

The \HI\ properties for each source are listed in
Table~\ref{Table:HI_Properties}, the atomic gas masses range from $6.8 \times
10^{7}$ to $1.5 \times 10^{10}\,{\rm M}_{\odot}$, with a median mass
of $2.3 \times 10^{9}\,{\rm M}_{\odot}$.

The remaining sources fall below the HIPASS detection
limit, which typically spans the range $1.6 \times 10^{8} < M_{\it HI} <
9.8 \times 10^{8})\,{\rm M_{\odot}}$ for the distance range of our
sample \citep{Haynes2011}. We determine a 3\,$\sigma$ upper limit on the \HI\ mass on our undetected sources using the following prescription from
\citet{Stevens2004}:

\begin{equation}
M_{\it HI} \leq 2.36 \times 10^{5} D^{2} (3 \sigma) \sqrt{18}\sqrt{W_{50}}
\label{Equation:HI_Mass_Upper}
\end{equation}

where $\sigma$ is the RMS noise in a single channel (0.013\,Jy), $D$
is the distance in\,Mpc, the $\sqrt{18}$ accounts for the number of
uncorrelated channels (the velocity resolution of HIPASS is
$\rm 18\,km\,s^{-1}$), and $W_{50}$ is the linewidth measured at 50\,per\,cent peak intensity.  We use the average value of $W_{50}$ observed for the
HIPASS-detected HAPLESS galaxies to estimate the upper limits on the
\HI\ mass (Table~\ref{Table:HI_Properties}). 

To quantify how gas-rich a galaxy is, we calculate the atomic gas
fraction \fg\ for galaxies with detected \HI\ masses (with upper limits
quoted for non-detections); this is defined as:

\begin{equation}
f_{g}^{\it HI} = \frac{ M_{\it HI} }{ M_{\it HI} + M_{\star} }
\label{Equation:Gas_Fraction}
\end{equation}

\noindent where \fg\ provides a lower limit on the fraction of
  the baryonic mass in the gas phase (as molecular gas is not
  considered in this work). 
  
If there is sufficient optical depth in the line of sight for \HI\ clouds, the \HI\ fluxes and masses could be under-estimated due to self-absorption. \citet{Bourne2013} show this correction is on average a factor of 1.08 for the overlapping sample of galaxies between their sources and HAPLESS. As we lack the necessary information to calculate the self-absorption for other nearby galaxy surveys (see Section~\ref{Subsection:Reference_Samples}), we do not consider self-absorption here, but note that our gas masses, particularly for edge on galaxies, could therefore be underestimated by this effect.

Finally, we do not include the molecular gas component in this work due to the lack of uniform measurements for this sample. Since the molecular component only tends to dominate the gas budget in more-massive, earlier-type spirals \citep{Saintonge2011S}, our lack of molecular gas information is unlikely to make a substantial difference to the interpretation in this work.  Using the scaling relations for $\rm H_{2}$/\HI\ and stellar mass from \citet{Bothwell2014}, the molecular-to-atomic gas ratios in our sample are predicted to be negligible ($<0.1$) for all but 10 of our sources, with the remaining galaxies having ratios between 0.1--0.7. The predicted $\rm H_{2}$/\HI\ ratios for our curious blue galaxies range from 0.016--0.14 with a median of 0.06 -- suggesting using the atomic gas only is an appropriate estimate of the total gas component for these sources. Note that adding molecular gas would only serve to increase the gas fractions in Table~\ref{Table:HI_Properties}. The gas masses and gas fractions for the detected galaxies in our sample will be discussed in more detail in Section~\ref{Subsection:Gas}.

\subsection{Star Formation Rates} \label{Subsection:Star_Formation_Rates}

To estimate star formation rate (SFR), we use the
\citet{Hirashita2003D} method of combining UV and IR tracers,
specifically following \citet{Jarrett2013A} to combine GALEX FUV and
WISE 22\,\micron\ measurements to give the total SFR as:

\begin{equation}
{\it SFR} = {\it SFR}_{\it FUV} + {\it SFR}_{22} 
\label{Equation:Total_SFR}
\end{equation}

\noindent where ${\it SFR}_{\it FUV}$ is the FUV-derived unobscured SFR (calculated using Equation~\ref{Equation:UV_SFR}), and ${\it SFR}_{22}$ is the 22\,\micron-derived obscured SFR (calculated using Equation~\ref{Equation:MIR_SFR}). All SFR values are in units of ${\rm M_{\odot}\ yr}^{-1}$.

UV emission traces unobscured high-mass stars, indicating star formation on timescales of $\sim$\,100\,Myr \citep{Kennicutt1998H, Calzetti2005D}. 
For ${\it SFR}_{\it FUV}$, we use the prescription of \citet{Buat2008B, Buat2011C}: 

\begin{equation}
{\it SFR}_{\it FUV} = 10^{-9.69} \nu_{\it FUV} L_{\it FUV} 
\label{Equation:UV_SFR}
\end{equation}

\noindent where $\nu_{\it FUV} L_{\it FUV}$ is the $\nu L_{\nu}$
luminosity in the GALEX FUV waveband\footnote{$\nu_{\it FUV} =
  1.987\,{\rm PHz}$} in units of bolometric Solar
luminosity. \citet{Buat2012C} find the uncertainty in this relation to
be 0.13 dex. It was calibrated using 656 local galaxies (described in
\citealt{Buat2007B}) with stellar masses greater than $10^{10}\,{\rm
  M_{\odot}}$, and extends down to SFRs of $0.07\,{\rm
  M_{\odot}\,yr^{-1}}$; as such it includes a range of actively
star-forming and quiescent systems. The stellar masses of our sample
extend to lower values than the \citet{Buat2007B} sample; however the
\citet{Buat2007B} sample does cover the full luminosity, SSFR, and
colour range (specifically NUV-$r$ against FUV-NUV) exhibited by the
HAPLESS galaxies. Note that their SFR prescription assumes a
\citet{Kroupa2001} IMF; we convert it to the Chabrier IMF (which we
use to derive stellar masses) using a correction factor of 0.94.

MIR emission comes primarily from hot dust, heated by short-wavelength
photons emitted from newborn stars, and traces star formation on time
scales \textless\,100\,Myr \citep{Calzetti2005D,Kennicutt2012A}. The
WISE 22\,\micron\ SFR relation of \citet{Jarrett2013A} was calibrated
by bootstrapping to the {\it Spitzer} 24 \micron\ SFR relation of
\citet{Rieke2009E}, and is given by:

\begin{equation}
{\it SFR}_{22} = (1 - \eta) 10^{-9.125} \nu_{22} L_{22}
\label{Equation:MIR_SFR}
\end{equation}
 
\noindent where $\eta$ is the fraction of MIR emission originating
from dust heated by the evolved stellar population, and $\nu_{22}
L_{22}$ is the $\nu L_{\nu}$ luminosity in the WISE
22\,\micron\ waveband\footnote{$\nu_{22} = 13.64\,{\rm THz}$} in units
of bolometric Solar luminosity. \citet{Rieke2009E} estimate the
uncertainty in their {\it Spitzer} 24 \micron\ SFR relation to be 0.25
dex, and find it to be accurate at gauging the star formation giving
rise to thermal dust emission in IR-selected
galaxies. \citet{Jarrett2013A} find the scatter in their WISE
22\,\micron\ bootstrap to this relation to be negligible ($\sim$1\,per\,cent),
thanks to the close similarity between the {\it Spitzer} 24
\micron\ and WISE 22\,\micron\ passbands.

The value of $\eta$ will vary from galaxy to galaxy depending on its
current star formation activity and dust geometry. $\eta$ may be
calibrated independently if other tracers of dust-corrected SFR are
available, or calculated theoretically; values in the literature for
star forming samples range from $0.17 \leq \eta \leq 0.55$
\citep{Buat2011C,Hao2011B,DJBSmith2012,Kennicutt2012A}. 

\begin{figure}
\begin{center}
\includegraphics[width=0.5\textwidth]{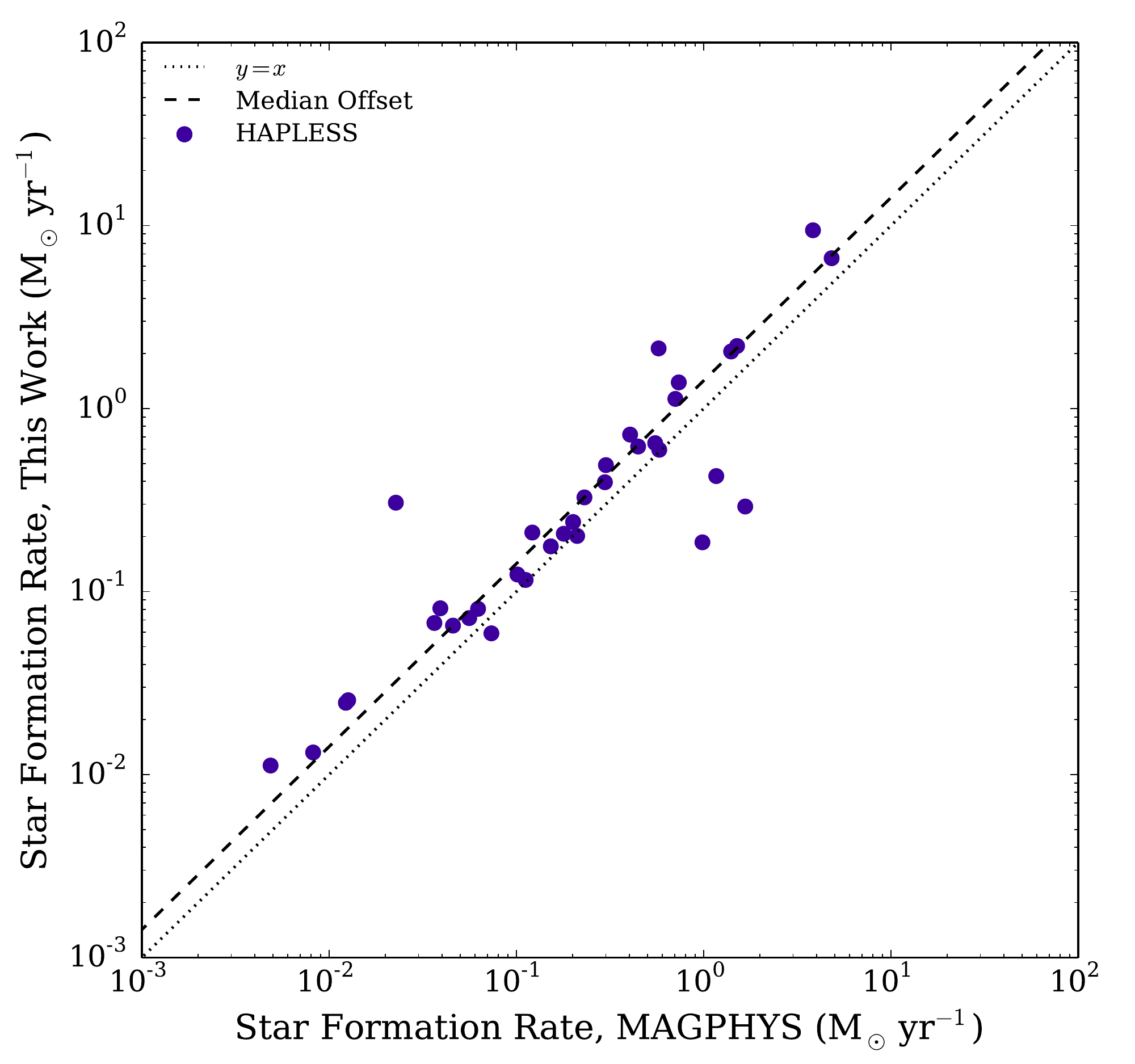}
\caption{The SFRs calculated using Equation~\ref{Equation:Total_SFR} compared to those derived by De Vis et al. ({\it in prep.}) by fitting the full UV-submm SEDs of our sample using MAGPHYS \citet{DaCunha2008}. The offset between the two prescriptions is by a factor of 1.42, as indicated by the dashed line.}
\label{Fig:SFR_Scaling}
\end{center}
\end{figure}

We first set $\eta = 0.17$ to be consistent with
  \citealt{Buat2011C}, and compare our total SFRs using
  Equation~\ref{Equation:Total_SFR} with those derived from SED
  modelling using MAGPHYS (\citealt{DaCunha2008}, De Vis et al. {\it
    in prep.}).  These two techniques produce SFRs offset by a median
  factor of 1.42 (see Figure~\ref{Fig:SFR_Scaling}). The likely cause
is that $\eta=0.17$ is not an accurate measure of the fraction of
22\,\micron\ luminosity powered by the older stellar population for our
sample, whereas MAGPHYS allows this fraction to be determined by the
energy balance between the UV and FIR for each source
individually. There may also be differences in the prescriptions for
${\it SFR}_{\it FUV}$ between \citet{Buat2011C} and the stellar
population models of MAGPHYS (taken from
\citealp{Bruzual2003B,Bruzual2007E}).  Finally, the offset could be explained if a bias existed towards a larger transiently-heated small grain population in our sample compared to the \citet{Rieke2009E} calibration data (indeed there is some evidence that the 22\,\micron\ emission is not correlated with SFR in some \HATLAS\ galaxies, \citealt{Bourne2013}).  Modulo the
  offset, the correlation between the two SFR estimates is tight, with
  the exception of 4 outliers. The 3
    galaxies below the scatter are HAPLESS 9, 33, and 34; these sources
     have extremely blue \fK\ colours (\textless\,2.0), {\it and} SFRs which are significantly dominated by the unobscured, UV component. The outlier well above the line (HAPLESS 25) is at the extreme red end (in terms of \fK) of our sample, and has roughly equal contributions from UV and 22\,\micron\ emission to its SFR using Equation~\ref{Equation:Total_SFR}.  The SFR prescriptions therefore appear to disagree in these extreme regions of the parameter space, though we leave this for a future study (De Vis et al., {\it in prep}).  In order to compare our sources with
  other nearby galaxy studies, including the HRS (for which we do not have
  full multiwavelength data) and the \planck\ sample of
  \citet{Clemens2013A} which uses MAGPHYS (see
  Section~\ref{Section:Comparison_to_Other_Samples}) we therefore
  reduce our SFRs from Equation~\ref{Equation:Total_SFR} by a factor
  of 1.42 to be consistent.  Note that this rescaling factor is well
within the usual variation found between different SFR prescriptions.

\begin{table}
\begin{center}
\caption{Star formation properties of the HAPLESS galaxies. GALEX FUV
  (unobscured) and WISE 22\,\micron\ (obscured) star formation rates,
  ${\it SFR}_{\it FUV}$ and ${\it SFR}_{22}$, are calculated according
  to Equations~\ref{Equation:UV_SFR} and \ref{Equation:MIR_SFR}. Where both GALEX and WISE data exists, we combine this (Equation~\ref{Equation:Total_SFR}) to yield the total
  SFR.}
\label{Table:SFR_Properties}
\begin{tabular}{lrrrrrrrrr}
\toprule \toprule
\multicolumn{1}{c}{N\textsuperscript{\underline{o}}} &
\multicolumn{1}{c}{${\it SFR}_{\it FUV}$} &
\multicolumn{1}{c}{${\it SFR}_{22}$} &
\multicolumn{1}{c}{${\it SFR}$} &
\multicolumn{1}{c}{${\it SSFR}$} \\
\cmidrule(lr){2-4}
\cmidrule(lr){5-5} 
\multicolumn{1}{c}{} &
\multicolumn{3}{c}{(log$_{10}$ M$_{\odot}$  yr$^{-1}$)} &
\multicolumn{1}{c}{(log$_{10}$ yr$^{-1}$)} \\
\midrule
1 & -$^{a}$ & -1.2 & - & -\\ 
2 & -1.3 & - & - & -\\ 
3 & -0.4 & -0.8 & -0.2 & -9.5\\ 
4 & -1.4 & -1.9 & -1.3 & -10.1\\ 
5 & -1.9 & -2.3 & -1.7 & -10.2\\ 
6 & -0.0 & 0.4 & 0.7 & -10.1\\ 
7 & -0.4 & -1.0 & -0.3 & -9.9\\ 
8 & -0.7 & -1.4 & -0.6 & -9.6\\ 
9 & -0.7 & -1.6 & -0.6 & -9.3\\ 
10 & -0.6 & -0.3 & -0.0 & -10.2\\ 
11 & -1.4 & -1.8 & -1.2 & -10.1\\ 
12 & -1.3 & -1.8 & -1.2 & -9.8\\ 
13 & -2.0 & -2.8 & -2.0 & -10.1\\ 
14 & -0.3 & -0.5 & -0.1 & -9.6\\ 
15 & -1.9 & -2.2 & -1.7 & -10.4\\ 
16 & -1.3 & -1.4 & -1.0 & -10.3\\ 
17 & -1.0 & -1.4 & -0.8 & -8.9\\ 
18 & -0.8 & -1.4 & -0.7 & -9.5\\ 
19 & - & -1.4 & - & -\\ 
20 & -1.3 & 0.0 & 0.2 & -10.6\\ 
21 & - & -0.1 & - & -\\ 
22 & -2.3 & -2.5 & -2.0 & -11.8\\ 
23 & -0.7 & 0.0 & 0.2 & -9.9\\ 
24 & -1.4 & -2.0 & -1.3 & -9.0\\ 
25 & -0.9 & -1.0 & -0.6 & -11.2\\ 
26 & -0.4 & -1.2 & -0.3 & -9.8\\ 
27 & -1.4 & -2.0 & -1.3 & -9.9\\ 
28 & -0.6 & -0.2 & 0.0 & -9.8\\ 
29 & -0.3 & 0.6 & 0.8 & -9.6\\ 
30 & -0.9 & -1.6 & -0.8 & -9.6\\ 
31 & -0.4 & -1.1 & -0.3 & -9.9\\ 
32 & -1.4 & -1.4 & -1.0 & -9.9\\ 
33 & -0.9 & -1.8 & -0.8 & -9.2\\ 
34 & -0.5 & -1.9 & -0.5 & -9.1\\ 
35 & -0.9 & -1.8 & -0.9 & -9.9\\ 
36 & -1.3 & - & - & -\\ 
37 & -0.7 & -1.1 & -0.5 & -10.6\\ 
38 & -1.3 & -2.0 & -1.2 & -9.6\\ 
39 & -0.5 & -1.2 & -0.4 & -9.8\\ 
40 & -0.8 & -1.4 & -0.7 & -9.7\\ 
41 & -1.4 & -1.6 & -1.1 & -8.8\\ 
42 & -2.3 & - & - & -\\ 
\bottomrule
\end{tabular}
\begin{list}{}{}
\item[$^{a}$] Note that HAPLESS 1 has contamination from non-thermal continuum emission in the UV; therefore we do not quote a value for $SFR_{\it FUV}$.
\end{list}
\end{center}
\end{table}

Adding in quadrature the uncertainties in the UV (0.13 dex) and MIR
(0.25 dex) relations in Equation~\ref{Equation:Total_SFR} yields an
uncertainty of 0.28 dex in the derived total SFRs (this does not
include the uncertainty in the FUV and 22\,\micron\ luminosities of
individual sources).

We also calculate the specific star formation rate (SSFR), the SFR per stellar mass  
(Table~\ref{Table:SFR_Properties}). The calculated SFRs range from 0.01 to 7.12$\,{\rm M_{\odot} yr}^{-1}$, with a median SFR of $0.18\,{\rm M_{\odot} yr}^{-1}$. Derived SSFRs range from $1.6 \times 10^{-12}$ to $1.4 \times 10^{-9}\,{\rm yr}^{-1}$, with a median SSFR of $1.3 \times 10^{-10}\,{\rm yr}^{-1}$.

\section{Properties in Comparison to Other Dust Surveys of Nearby Galaxies} \label{Section:Comparison_to_Other_Samples} 

We now compare HAPLESS to other surveys of dust in local galaxies. In
this section, we consider our entire sample; however those galaxies
that are not in the luminosity-limited subset of HAPLESS are plotted
in figures as hollow circles. Table~\ref{Table:Parameters} summarises the median
properties of each of the samples, and the results of K-S tests between
them.

\subsection{The Reference Samples} \label{Subsection:Reference_Samples}

\subsubsection{The Herschel Reference Survey} \label{Subsubsection:HRS}

With its stated objective to be the `benchmark study of dust in the
nearby universe', the 323 galaxies of the \hersc\ Reference Survey (HRS,
\citealt{Boselli2010}) have been observed with resolution and sensitivity unrivalled by
any previous FIR survey. The HRS chose \Kband\ brightness as its
selection criteria, because it suffers least from extinction and is
known to be a good proxy for stellar mass. The velocity range of the
HRS ($1050 \le V \le 1750\ \rm km\ s^{-1}$), with corrections made to
account for the velocity dispersion of the galaxies of the Virgo
Cluster, corresponds to a distance range of $15 \le D \le
25\ \rm\,Mpc$ (whereas the HAPLESS distance range is $15 \le D \le
46\ \rm\,Mpc$).

The apparent magnitude limit of the late type galaxies in HRS is
$K_{S} \le 12$, which equates to an absolute magnitude limit between
$K_{S} \le -17.43$ and $K_{S} \le -18.54$, depending on the distance of the
source between the HRS limits\footnote{For early type galaxies, a
  brighter flux limit of $K_S \le 8.7$ is applied.}. From this we can
ascertain that between 4 and 15 of the 42 HAPLESS galaxies would have
been insufficiently luminous in $K_S$ to have been included in the
HRS\footnote{Only 3 HAPLESS galaxies overlap with the distance range
  of HRS and of these, only one would have been bright enough for the
  HRS selection.}.  These faint HAPLESS galaxies are low stellar mass
systems that tend to have very blue \fK\ colours; 13 of the missing 15
satisfy our \fK\,\textless\,3.5 criterion. Galaxies seen by
\HATLAS\ that are faint in $K_S$, but nonetheless dusty, represent an
orthogonal population to the HRS, and reveal selection biases imposed
on targeted dust surveys that \HATLAS, with its blind sample, is not
susceptible to. Another difference between the samples is that the HRS
contains numerous early type galaxies, partly due to the stellar mass
selection, and partly due to the extensive overlap (46\,per\,cent) of the HRS
sample with the Virgo cluster. 

To allow for a direct comparison of HAPLESS to the HRS, we determined
dust masses and temperatures for the HRS galaxies ourselves, using our
own SED-fitting method (as detailed in
Section~\ref{Subsection:SED_Fitting}) and their published
PACS\footnote{We corrected the HRS fluxes to account for a recently-fixed error in the {\tt Scanamorphos} pipeline used to create the HRS PACS maps. The published HRS fluxes at 100 and 160 \micron\ were multiplied by 1.01
  and 0.93 respectively, the average change
  (with scatter $\sim$2\,per\,cent) in extended-source flux in maps produced
  with corrected versions of {\tt Scanamorphos}.}
\citep{Cortese2014A}, SPIRE \citep{Ciesla2012B}, and WISE
\citep{Ciesla2014A} photometry, along with IRAS 60 \micron\ data we
acquired using SCANPI in the same manner as for the HAPLESS galaxies
(described in Section~\ref{Subsubsection:IRAS_Photometry}). We
likewise calculated $L_{\it TIR}$ values for the HRS using the same
method as for HAPLESS.

We note that our dust masses for the HRS galaxies are on average
a factor $\sim 2.2$ lower than in \citet{Ciesla2014A}, consistent with their assumed
lower value for $\kappa_{500} = 0.1\,\rm{m^2\,kg^{-1}}$.

\citet{MWLSmith2012B} also find that the submm emission of two HRS sources, the giant elliptical galaxies M87 and M84, contain significant contamination from their AGN. Therefore, we do not attempt to fit the SEDs of these sources.

For the \HI\ masses of the HRS galaxies, we used the values published in
\citet{Boselli2014B}. The published stellar masses of the HRS
\citep{Cortese2012A} were calculated in the same way as our own. The
UV GALEX and optical SDSS photometry of the HRS is described in
\citet{Cortese2012C}, whilst their NIR \Kband\ photometry
\citep{Boselli2010} was acquired from the 2-Micron All-Sky Survey
(2MASS, \citealt{Jarrett2000A}). To calculate the star formation rates
of the HRS galaxies, we employed the same technique as for the HAPLESS
galaxies (Section~\ref{Subsection:Star_Formation_Rates}), for which we
used the published HRS WISE and GALEX photometry. As for the HAPLESS
galaxies, we obtain morphologies for the HRS from EFIGI
\citep{Baillard2011}.

\subsubsection{{\it Planck}} \label{Subsubsection:Planck}

\begin{figure}
\begin{center}
\includegraphics[width=0.5\textwidth]{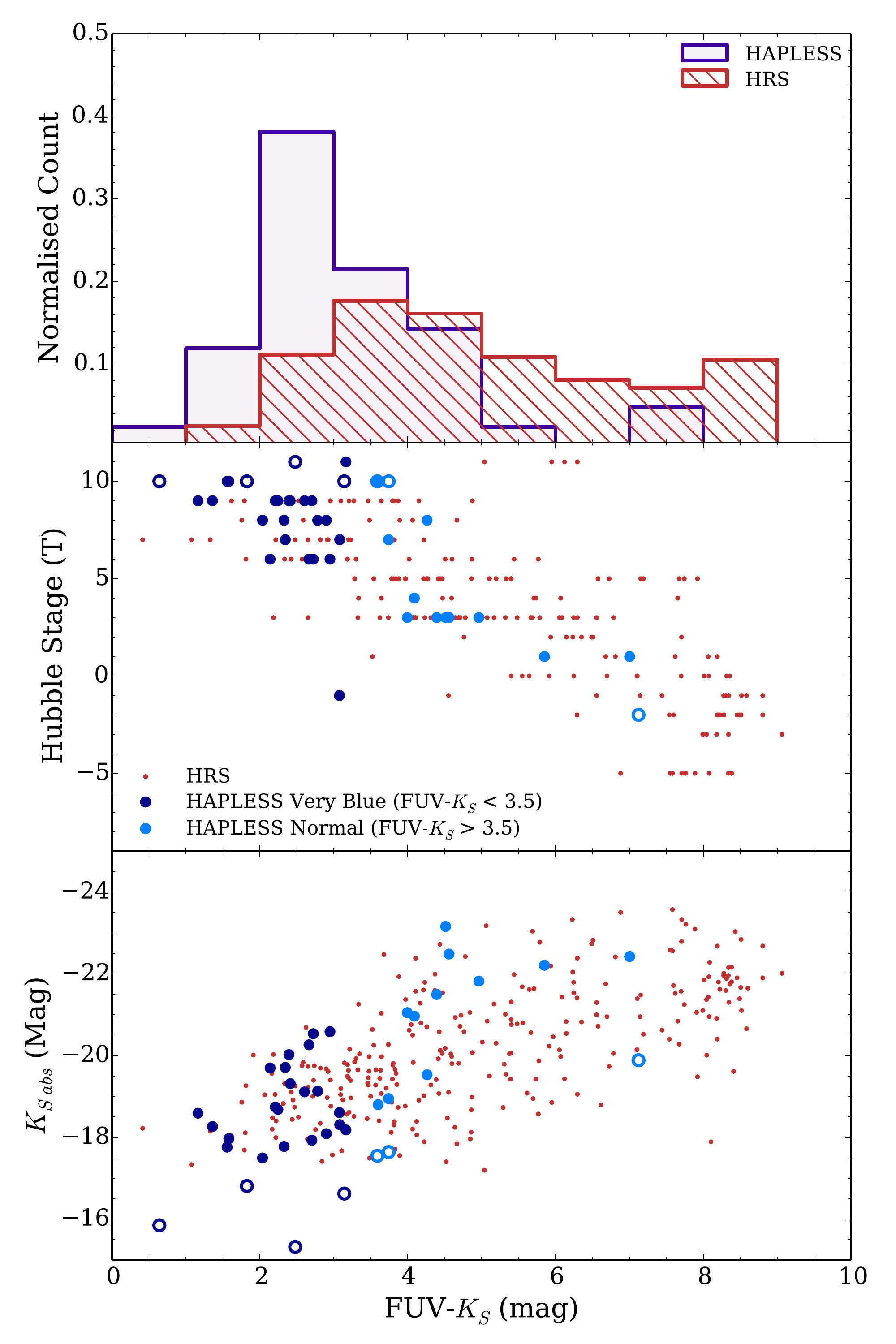}
\caption{{\it Upper:} The distributions of \fK\ colour for the HAPLESS (blue) and the HRS (red) samples. The galaxies of the blind HAPLESS sample tend to be significantly bluer than those of the \Kband\ selected HRS. {\it Central}: Morphology against \fK\ colour for HAPLESS and the HRS. {\it Lower:} Absolute \Kband\ magnitude against \fK\ colour for HAPLESS and the HRS. Hollow circles indicate galaxies that are below the luminosity-limit of the sample.}
\label{Fig:FUV-K_Comparison_Grid_1}
\end{center}
\end{figure}

\begin{figure*}
\begin{center}
\includegraphics[width=1.0\textwidth]{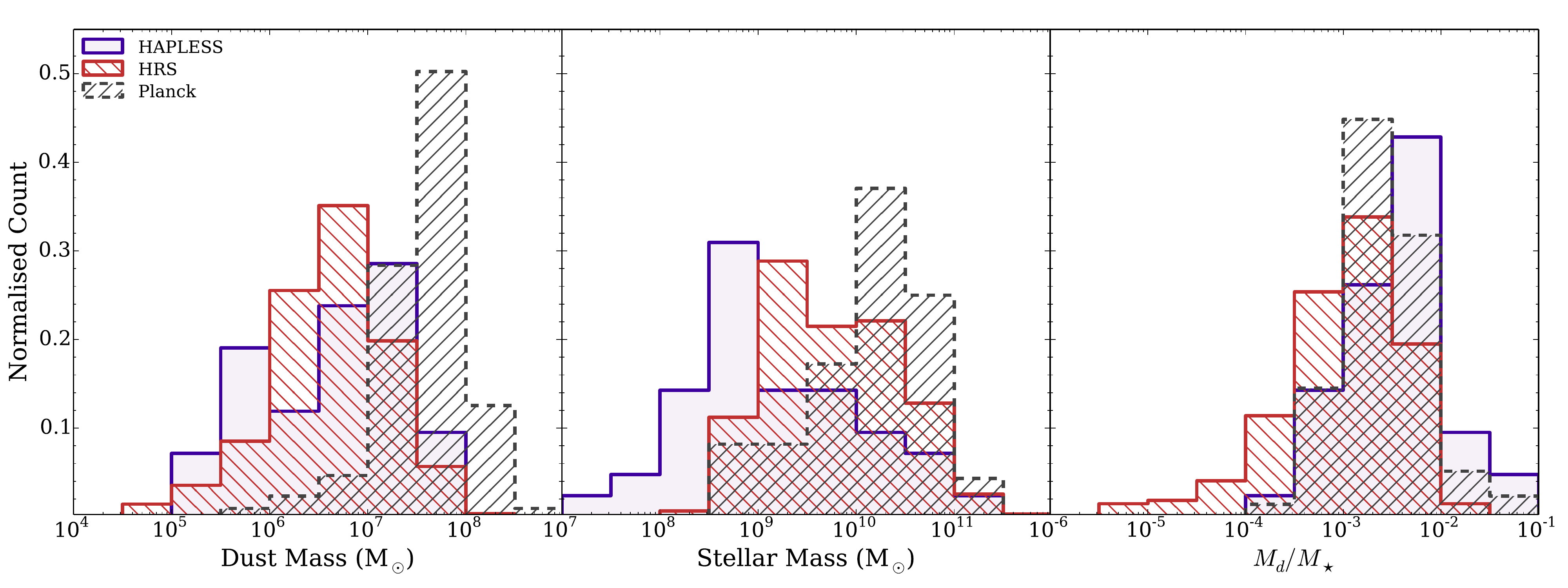}
\caption{The dust and stellar mass properties of the HAPLESS, HRS, and \cnp\ galaxies. {\it Left:} The distribution of dust masses. {\it Centre:} The distributions of stellar masses. Note that \citet{Clemens2013A} derive the stellar masses for the \cnp\ sample using MAGPHYS, whilst for the HAPLESS and HRS samples we use Equation~\ref{Equation:Stellar_Mass}; however the stellar masses produced by both methods are in excellent agreement with each other for the HAPLESS sample (De Vis et al., {\it in prep.}). {\it Right:} The distributions of $M_{d}/M_{\star}$ (ie, specific dust mass). HAPLESS contains a much higher proportion of very dust-rich galaxies than either of the other two samples. }
\label{Fig:Comparison_Dust_Stellar_Hist_Grid}
\end{center}
\end{figure*}

\begin{table*}
\begin{center}
\caption{Median parameters derived for the local-volume surveys compared in this work, including the very blue (\fK\,\textless\,3.5) subset of the HAPLESS sample. Results from Kolmogorov-Smirnov (K-S) tests between HAPLESS and the HRS and \cnp\ surveys are also shown, indicating the likelihood of the null hypothesis that two samples are drawn from the same underlying population.}
\label{Table:Parameters}
\begin{tabular}{lrrrrrrrrrr} 
\toprule \toprule
\multicolumn{1}{c}{Sample} &
\multicolumn{1}{c}{\fK} &
\multicolumn{1}{c}{$T_{c}$ } &
\multicolumn{1}{c}{$M_d$ } &
\multicolumn{1}{c}{$M_{\star}$ } &
\multicolumn{1}{c}{$M_d/M_{\star}$ } &
\multicolumn{1}{c}{SSFR} &
\multicolumn{1}{c}{$M_{\it HI}$} &
\multicolumn{1}{c}{$M_d/M_{\it HI}$ } &
\multicolumn{1}{c}{\fg} &
\multicolumn{1}{c}{$M_{B}$ } \\
\multicolumn{1}{c}{} &
\multicolumn{1}{c}{(mag)} &
\multicolumn{1}{c}{(K) } &
\multicolumn{1}{c}{($10^{6}\,\rm{M_{\odot}}$)} &
\multicolumn{1}{c}{($10^{9}\,\rm{M_{\odot}}$)} &
\multicolumn{1}{c}{($10^{-3}$)} &
\multicolumn{1}{c}{($10^{-11}\,\rm{yr^{-1}}$)} &
\multicolumn{1}{c}{($10^{8}\,\rm{M_{\odot}}$)} &
\multicolumn{1}{c}{($10^{-3}$)} &
\multicolumn{1}{c}{} &
\multicolumn{1}{c}{($10^{9}\,\rm{M_{\odot}}$)} \\
\midrule
HAPLESS & 2.8 & 14.6 & 5.3 & 1.0 & 4.4 & 12.9 & 14.4$^{a}$ & 3.9 & 0.52 & 2.5\\
$\prime\prime$ Very Blue & 2.4 & 14.2 & 4.8 & 0.6 & 6.5 & 20.7 & 12.1$^{a}$ & 2.7 & 0.66 & 2.3 \\
HRS & 4.6 & 18.5 & 4.6 & 4.9 & 1.2 & 4.1 & 8.5$^{a}$ & 6.2 & 0.18 & 5.5\\
\planck & -- & 17.7 & 41.9 & 17.4 & 2.5 & 6.9 & 36.4$^{a}$ & 11.6 & 0.17 & 22.4\\
\midrule
K-S (HRS) & $10^{-8}$ & $10^{-4}$ & 0.15 & $10^{-6}$ & $10^{-6}$ & $10^{-5}$ & $0.03$ & $10^{-2}$ & $10^{-5}$ & $10^{-2}$\\
K-S (\planck) & -- & $10^{-3}$ & $10^{-13}$ & $10^{-11}$ & $10^{-3}$ & 0.01 & $10^{-3}$ & $10^{-10}$ & $10^{-7}$ & $10^{-10}$\\
\bottomrule
\end{tabular}
\begin{list}{}{}
\item[$^{a}$] Gas masses are available for 90\,per\,cent of the HAPLESS sample (93\,per\,cent of the very blue subset), 81\,per\,cent of the HRS, and 90\,per\,cent of the \cnp\ sample.
\end{list}
\end{center}
\end{table*}

\citet{Negrello2013A} used the \planck\ Early Release Compact Source
Catalogue (ERCSC) \citep{Planck2011VII} to assemble a sample of nearby
galaxies. Their flux-limited sample contains 234 dusty galaxies
brighter than 1.8\,Jy at 550 \micron, at distances
$\lesssim\,100\,\rm\,Mpc$ (with the vast majority lying at $z<0.01$);
the authors estimate the sample to be 80\,per\,cent
complete. \citet{Clemens2013A} have used this sample to perform a
study of the properties of nearby dusty galaxies. We hereafter refer
to this as the \planck\ C13N13 sample.

Whilst the \planck-selected sample benefits from being blind and
all-sky (excepting the galactic plane zone of avoidance),
\planck\ suffers from lower sensitivity and resolution compared to
\hersc\ (3.8\arcmin\ in contrast to 18\arcsec). Only 3 of the HAPLESS
galaxies exceed the 1.8\,Jy 550 \micron\ flux limit necessary to
feature in the \cnp\ sample (and none of those are members of the
curious blue subset).

\citet{Clemens2013A} also derived dust masses and temperatures for
their sources by fitting two-component MBB SEDs with
$\beta=2$, which is consistent with our method. For the \cnp\ sample,
the authors adopted a value for the dust absorption coefficient of
$\kappa_{850} = 0.0383 \rm{\ m^{2}\ kg^{-1}}$, in contrast to the
value in this work of $\kappa_{850} = 0.077 \rm{\ m^{2}\ kg^{-1}}$. As
a result, we have divided their dust masses by a factor of 2.01 to
permit comparison.

The \cnp\ stellar masses and star formation rates were estimated using
the MAGPHYS multiwavelength SED-fitting package \citep{DaCunha2008},
which produces stellar masses which agree exceptionally well with the
\citet{Zibetti2009D} method we employ (De Vis et al., {\it in prep.});
both methods also assume the Chabrier IMF.  \HI\ data were available for
220 (94\,per\,cent) of the \cnp\ galaxies \citetext{Clemens, {\it
    priv. comm.}}. Once again, we use EFIGI morphologies
\citep{Baillard2011}.

Whilst almost identical sets of observed and derived properties are
shared by HAPLESS and the HRS, a more limited set of parameters is
available for \cnp; as a result, not all of the following analyses can
include the \planck\ sample.

\subsection{Colour and Magnitude Properties} \label{Subsection:Colour_and_Magnitude} 

As described in Section~\ref{Subsection:Curious_Blue_Galaxies}, we
find \fK\ colour to be an effective way of identifying the subset of
curious blue galaxies in our sample, using a colour cut of
\fK\,\textless\,3.5. We find that 64\,per\,cent (27) of the HAPLESS galaxies
satisfy this criterion, compared to only 27\,per\,cent of the HRS galaxies with
\fK\ colours available. Given that the HRS is \Kband-selected, it is
to be expected that its galaxies will tend to exhibit redder
\fK\ colour. The distributions of \fK\ colours for HAPLESS and the HRS
are shown in the upper panel of
Figure~\ref{Fig:FUV-K_Comparison_Grid_1}. Whilst the HRS more-or-less
equally samples a wide range of \fK\ colours, with a median of 4.6
(Table~\ref{Table:Parameters}), the blindly-selected HAPLESS galaxies
tend to occupy a much narrower range of colours, with a median of
2.8. The distributions are significantly different.

As demonstrated by \citet{GilDePaz2007A}, \fK\ colour is a strong
indicator of morphology, as is also seen in the central panel of
Figure~\ref{Fig:FUV-K_Comparison_Grid_1}. The very blue \fK\ colours
of the HAPLESS galaxies indicate that the dust-selected universe is
dominated by very late type galaxies.

The lower panel of Figure~\ref{Fig:FUV-K_Comparison_Grid_1} is a
colour-magnitude plot constructed using \fK\ colour and
\Kband\ magnitude. Both the blue cloud and red sequence can be seen in the distribution of the HRS, at (3, -19.5) and (8.5, -22); however our HAPLESS sample is skewed towards bluer colours such that the bimodality is not visible in this sample; indeed, many of the HAPLESS galaxies are in fact bluer than the blue cloud peak seen in the HRS distribution.

\subsection{Dust and Stellar Mass} \label{Subsection:Dust_and_Stellar} 

Figure~\ref{Fig:Comparison_Dust_Stellar_Hist_Grid} compares the dust
mass distributions of HAPLESS, HRS, and \cnp. The effect of the
1.8\,Jy flux limit at 550 \micron\ in the \cnp\ sample is immediately
apparent; only galaxies with high dust masses (and a few less massive
but very nearby galaxies) were bright enough to be included in their
sample, which has a median dust mass of $4.2 \times 10^{7}\,{\rm
  M}_{\odot}$. The HAPLESS and the HRS have different selection
effects but ultimately have comparable median dust masses
(Table~\ref{Table:Parameters}).

The three samples also exhibit notably different distributions in
stellar mass (Figure~\ref{Fig:Comparison_Dust_Stellar_Hist_Grid}). The high flux limit of the \cnp\ sample naturally biases it towards more
massive galaxies.  HAPLESS spans the broadest range of stellar
masses, but on average has lower stellar mass systems.  The
median stellar masses of the three samples span over an order of
magnitude, 
and the combination of lower stellar masses, but moderate-to-high dust masses, means the HAPLESS galaxies have the highest median $M_{d}/M_{\star} \sim 4.4\times10^{-3}$ (ie, specific dust mass) out of the three surveys (Figure~\ref{Fig:Comparison_Dust_Stellar_Hist_Grid}, Table~\ref{Table:Parameters}). 
The very blue subset have an even higher median dust-to-stellar mass ratio of $6.5 \times 10^{-3}$; {\it despite accounting for only 6\,per\,cent of the stellar mass in the HAPLESS sample, the curious blue galaxies account for over 35\,per\,cent of the dust mass}.

\subsection{The Dust Mass Volume Density} \label{Section:Dust_Mass_Volume_Density} 

We now measure the dust mass function (DMF) and dust mass volume
density for HAPLESS. In this analysis, we consider
  all 42 galaxies in HAPLESS. For the 7 sources that are fainter than
  the luminosity complete limit, we estimate their accessible volumes
  using the $1/V_{\it max}$ method \citep{Schmidt1968}, while for the
  luminosity complete subset the accessible volume is simply that
  between the 15--46\,Mpc distance limits of the sample
  ($1,540\,\rm{Mpc^{3}}$).



\begin{figure}
\begin{center}
\includegraphics[width=0.5\textwidth]{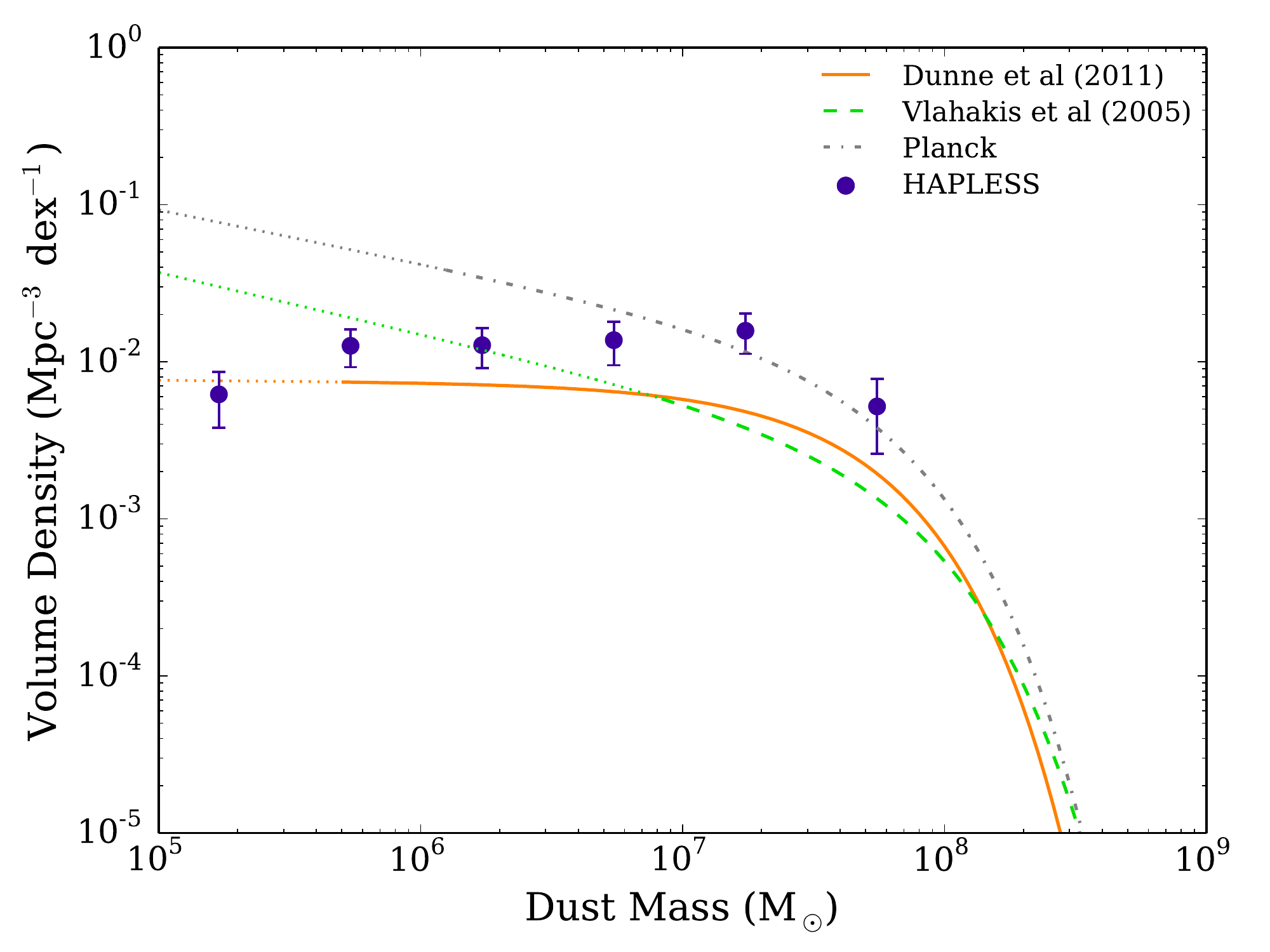}
\includegraphics[width=0.5\textwidth]{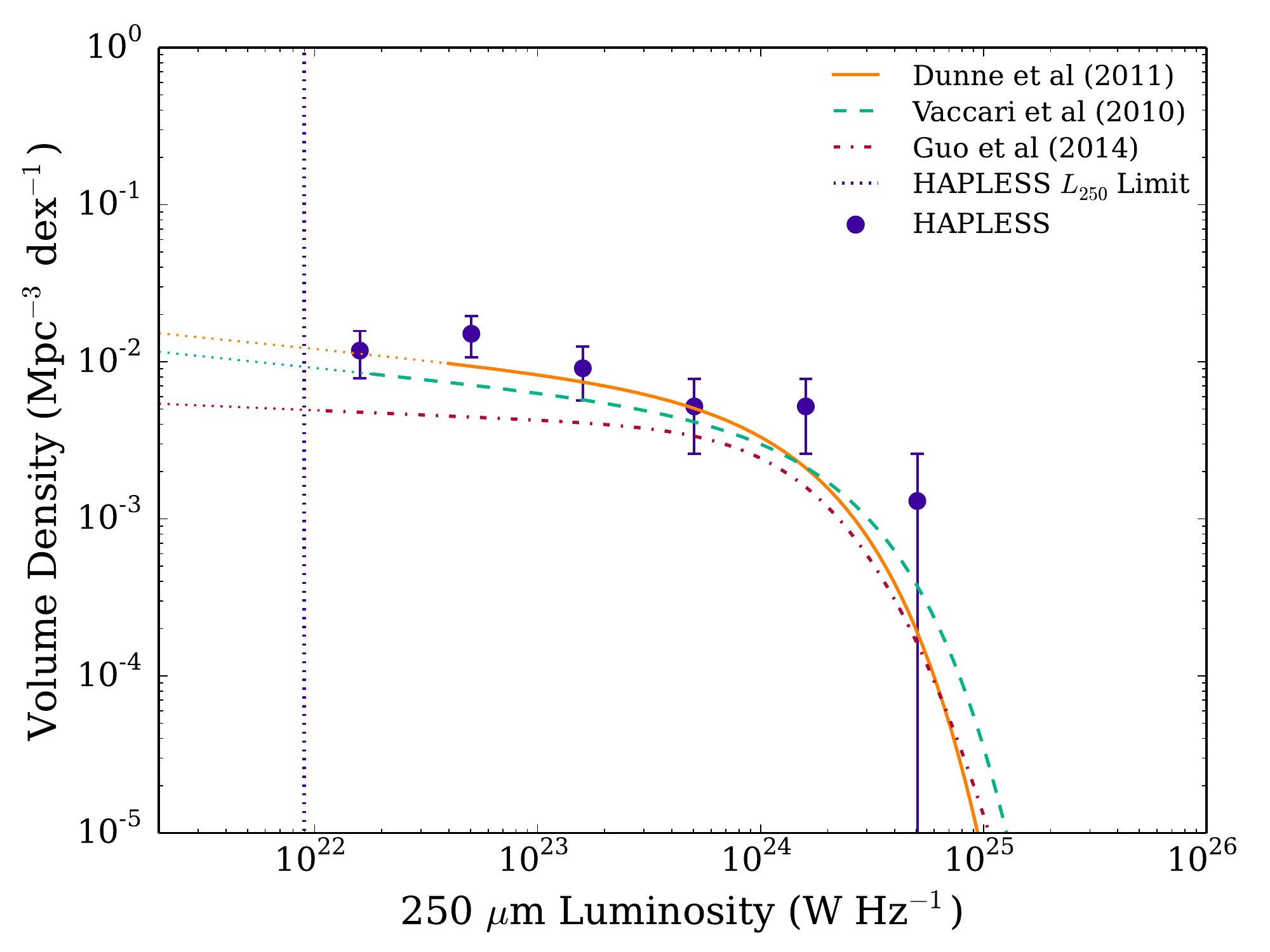}
\caption{{\it Upper:} The HAPLESS dust mass function (with appropriate incompleteness corrections applied) compared with
  those of \cnp, \citet{Vlahakis2005C}, and the \HATLAS\ Science
  Demonstration Phase functions from \citet{Dunne2011}. {\it Lower:} The
  250\,\micron\ luminosity function of HAPLESS compared with the
  $z<0.1$ \HATLAS\ samples from \citet{Dunne2011}, \citet{Guo2014A}
  and the $z<0.2$ sample from HerMES \citep{Vaccari2010B}.  The error bars on points represent Poisson
  uncertainty. The functions are plotted as thin dotted lines in the regions where they are extrapolated. All have been adjusted to our $\kappa_{d}$ and cosmology.}
\label{Fig:Comparison_Volume_Hist_Grid} 
\end{center}
\end{figure}

The upper panel of Figure~\ref{Fig:Comparison_Volume_Hist_Grid}
compares HAPLESS to the dust mass functions of \citet{Dunne2011},
\citet{Vlahakis2005C}, and \citet{Clemens2013A}. The HAPLESS data
points have had the appropriate corrections from \citet{Rigby2011A} applied to
account for the statistical effects of flux boosting and
incompleteness (Section~\ref{Subsection:Sample_Selection}). The
\HATLAS\ Science Demonstration Phase (SDP) result for $0 < z < 0.1$
from \citet{Dunne2011} (orange line in
Figure~\ref{Fig:Comparison_Volume_Hist_Grid}) is based on the first 16
square degree field of H-ATLAS. Their dust mass function shown here
includes a correction factor of 1.4 for the known under-density of the
GAMA09 field at $z$\,\textless\,0.1 relative to the average from SDSS
\citep{Driver2011}. The \citet{Vlahakis2005C} DMF (green line) used
submm/IRAS colour relations from the SLUGS survey to estimate
850\,\micron\ fluxes, and hence dust masses, for all IRAS galaxies in
the PSCz catalogue \citep{Saunders2000J}. In order to translate their
IRAS plus 850\,\micron\ flux estimate to a dust mass they needed to
assume a temperature model for the SED, and their {\it cold} fit
assumes a cold dust temperature of 20\,K, which seemed reasonable at
the time based on submm studies of IRAS galaxies by
\citet{Dunne2001A}. The \planck\ DMF from \citet{Clemens2013A} is
based on the 550\,\micron\ luminosity function from \citet{Negrello2013A}
and uses the same flux limited sample we have described in
Section~\ref{Subsubsection:Planck}. 
Table~\ref{Table:Dust_Mass_Function} lists the parameters for the
different Schechter functions; we have corrected all DMFs to the same
value of $\kappa_d$ and the same cosmology used here. We note that uncertainties in the distance measurements of the different galaxy samples could cause considerable scatter in the shape of the DMF, particularly at the high end, as demonstrated by \citet{Loveday1992B}. This would result in an observed DMF that is effectively a Schechter function convolved with a Gaussian. However, as the distance uncertainties vary both within and between the samples we compare here, we only present the observed mass functions in this work.
    
Above $M_d \sim\,10^{7}\,{\rm M_{\odot}}$, the HAPLESS data points
agree with the \planck\ DMF but are higher than those
from \citet{Dunne2011} and \citet{Vlahakis2005C}. Galaxies with $M_d
\geq 10^7\,\rm{M_{\odot}}$ account for 87\,per\,cent of the total HAPLESS dust
mass. Below this mass, the HAPLESS data points are in closer agreement with the \citet{Dunne2011} DMF and directly probe to lower dust masses than any of the previous works. \citet{Vlahakis2005C} and \cnp\ find a steeper faint-end slope than \citet{Dunne2011} and this work, but their direct sampling of the faint end is 1--2 orders of magnitude less than achieved here. 
With poor statistics in all surveys at the low-mass end, the varying
  estimates of the slope agree with each other within their
  1\,$\sigma$ uncertainties and so we do not consider these differences worrying at present.
  
  \begin{table}
  \begin{center}
  \caption{The best-fit Schecter function parameters of the various dust mass functions and luminosity functions compared in this work. All have been scaled to the same cosmology and value of $\kappa_{d}$ we employ.}
  \label{Table:Dust_Mass_Function}
  \begin{tabular}{lrrr} 
  \toprule \toprule
  \multicolumn{4}{c}{Literature Dust Mass Functions} \\
  \cmidrule(lr){1-4}
  \multicolumn{1}{c}{Reference} &
  \multicolumn{1}{c}{$\alpha$} &
  \multicolumn{1}{c}{$M^{\star}$ } &
  \multicolumn{1}{c}{$\phi^{\star}$ } \\
  \multicolumn{1}{c}{} &
  \multicolumn{1}{c}{} &
  \multicolumn{1}{c}{(${\rm M_{\odot}}$)} &
  \multicolumn{1}{c}{(${\rm Mpc^{-3}\,dex^{-1}}$)} \\
  \midrule
  \citet{Clemens2013A} & -1.34 & $5.27\times 10^{7}$ & $11.0\times 10^{-3}$ \\
  \citet{Dunne2011}$^{a}$ & -1.01 & $4.22\times 10^{7}$ & $7.19\times 10^{-3}$ \\
  \citet{Vlahakis2005C} & -1.39 & $6.49\times 10^{7}$ & $2.97\times 10^{-3}$ \\
  \toprule \toprule
  \multicolumn{4}{c}{Literature Luminosity Functions} \\
  \cmidrule(lr){1-4}
  \multicolumn{1}{c}{Reference} &
  \multicolumn{1}{c}{$\alpha$} &
  \multicolumn{1}{c}{$L^{\star}$ } &
  \multicolumn{1}{c}{$\phi^{\star}$ } \\
  \multicolumn{1}{c}{} &
  \multicolumn{1}{c}{} &
  \multicolumn{1}{c}{(${\rm W\,Hz^{-1}}$)} &
  \multicolumn{1}{c}{(${\rm Mpc^{-3}\,dex^{-1}}$)} \\
  \midrule
  \citet{Dunne2011}$^{a}$ & -1.14 & $1.53\times10^{24}$ & $6.00\times10^{-3}$ \\
  \citet{Guo2014A}$^{b}$ & -1.06 & $1.12\times10^{24}$ & $3.70\times10^{-3}$ \\
  \citet{Vaccari2010B}$^{c}$ & -1.14 & $2.19\times10^{24}$ & $4.22\times10^{-3}$\\
  \bottomrule
  \end{tabular}
  \begin{itemize}{}{}
  \footnotesize
  \item[$^{a}$] Note that this incorporates the 1.42 correction factor applied by \citet{Dunne2011} to account for under density in the GAMA09 field \citep{Driver2011}. 
  \item[$^{b}$] \citet{Guo2014A} use a modified Schechter function to fit their LF, with an additional parameter of $\sigma = 0.30$ (explained in \citealp{Saunders1990B}).
  \item[$^{c}$] \citet{Vaccari2010B} do not provide the parameters to their 250\,\micron\ Schechter fit; these values represent our best fit to their quoted data points.
  \end{itemize}
  \end{center}
  \end{table}

The most significant variation in the dust mass function between HAPLESS and
\citet{Dunne2011} is the excess of HAPLESS galaxies around
$M_d \sim 10^7\,\rm{M_{\odot}}$. This could be due to two possible effects:
  cosmic variance, or incompleteness in the \citet{Dunne2011} DMF.

The volume probed by HAPLESS (and also by the other surveys at the
faint end) is very small and subject to a large uncertainty due to
cosmic variance ($\sim$\,166\,per\,cent,
Section~\ref{Subsection:Sample_Selection}). This effect can be
explored by comparing the 250\,\micron\ luminosity functions since
this removes the complication of relating the 250\,\micron\ emission to
the mass of dust. Any differences in the 250 LF will purely be due to
variations in the space density of sources in the different samples.
We compare HAPLESS to the luminosity functions of previous authors in
Table~\ref{Table:Dust_Mass_Function} and
Figure~\ref{Fig:Comparison_Volume_Hist_Grid} and find good agreement
(within errors) with the $0 < z < 0.1$ \HATLAS\ luminosity function
from \citet{Dunne2011} (from 16\,${\rm deg}^2$ Science Demonstration Phase data, scaled by
their density correction 1.4 factor; this is an
  updated version of the LF presented in \citealt{Dye2010}) and from
HerMES (over 14.7\,${\rm deg}^2$ at $z<0.2$, from
\citealp{Vaccari2010B}). The LF derived from \HATLAS\ Phase-1 data
(161.6\,${\rm deg}^2$) in \citet{Guo2014A} is lower compared to
HAPLESS. This measure has not corrected for the known underdensity of
the GAMA09 field and also uses a brighter optical magnitude threshold for
inclusion of sources than \citet{Dunne2011}. It is not clear how much of a
difference this will make (detailed LFs for the full \HATLAS\ Phase 1
will be presented in future work) but overall, this comparison
indicates that the HAPLESS volume represents a region of fairly
typical 250\,\micron\ luminosity density and certainly is not
significantly overdense relative to the density corrected \citet{Dunne2011}
values. 

\begin{figure}
\begin{center}
\includegraphics[width=0.5\textwidth]{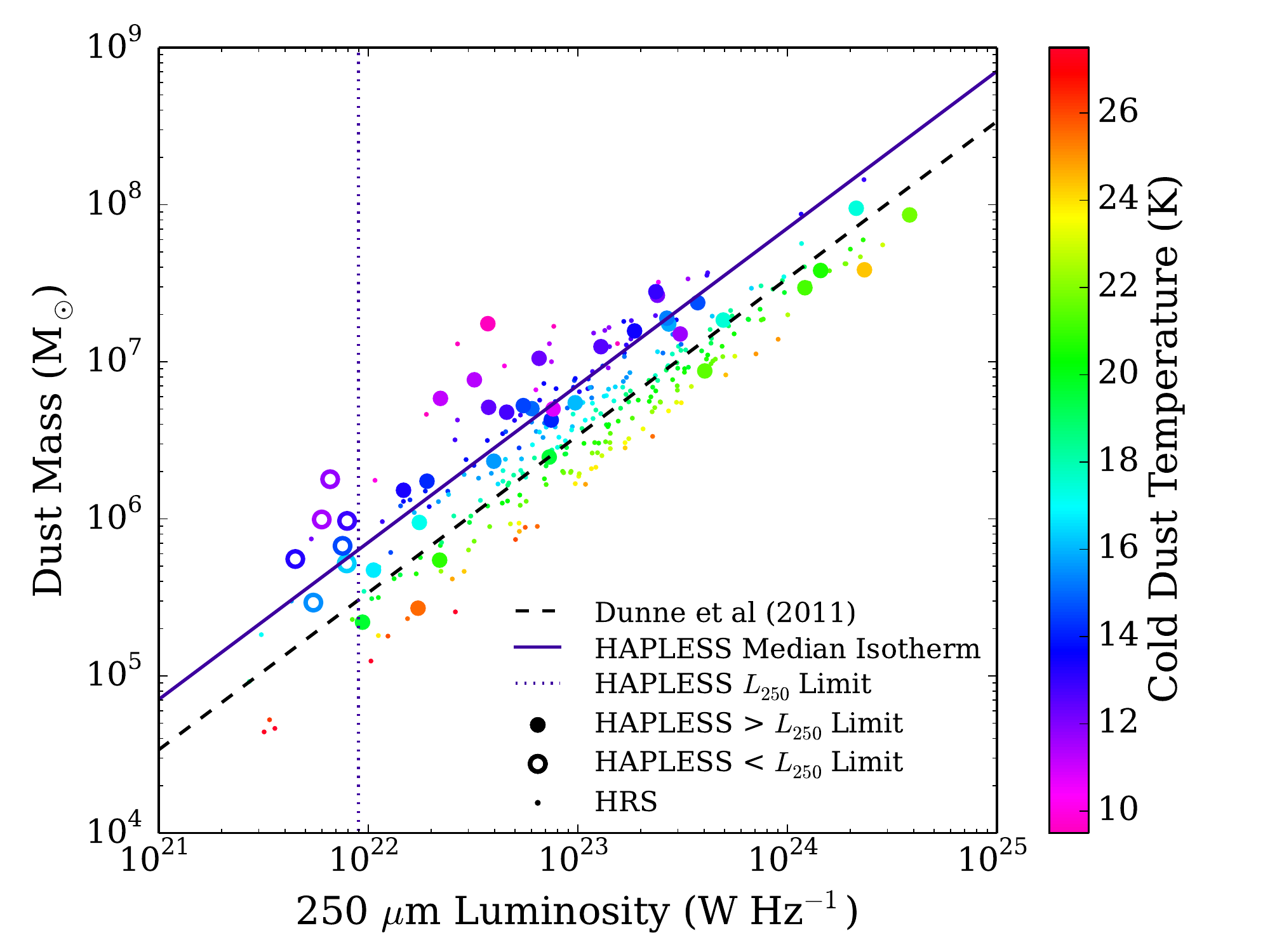}
\caption{Dust mass against 250\,\micron\ luminosity for the HAPLESS and the HRS, colour-coded by cold dust temperature.  Also shown are the median dust temperature (solid) for the HAPLESS sources and the relationship in \citet{Dunne2011} (dashed).  Filled circles show the HAPLESS luminosity complete sub-sample.}
\label{Fig:L250_Td_Scatter} 
\end{center}
\end{figure}

The fact that we find a greater dust mass volume density than
\citet{Dunne2011}, despite having the same 250\,\micron\ luminosity
density detection limit, must therefore be ascribed to a difference in the average
ratio of $L_{250}/M_d$ in the two samples, such that HAPLESS includes dustier objects for a given 250\,\micron\ luminosity threshold. The reason behind this lies in the relationship between our selection parameter,
$L_{250}$, and our parameter of interest, $M_d$. Whether or not we
detect a given mass of dust is strongly dependant upon the temperature
of that dust. This is illustrated by Figure~\ref{Fig:L250_Td_Scatter},
which compares the relation between 250\,\micron\ luminosity and dust
mass for HAPLESS and the HRS. The relationship found by \citet{Dunne2011}
is shown as a dashed line. There is a scatter of $\sim\,1$ dex in this
relationship, due to dust temperature. The HAPLESS galaxies have more
dust mass for a given 250\,\micron\ luminosity than both the HRS and
the \citet{Dunne2011} relation, because they are colder on average
than the galaxies in those samples\footnote{Note that the
  \citet{Dunne2011} best-fit line passes through the $\sim$\,19\,K
  isotherm in their scatter, and does so here also; indicating that
  for a given luminosity and dust temperature, we would find the same
  dust mass.}. The issue is therefore that no surveys have a truly
`dust mass limited sample' but rather, in the case of H-ATLAS, we have
a 250\,\micron\ luminosity limited sample. Our luminosity limit of
$L_{250} = 8.9 \times 10^{21}\,{\rm W\, Hz^{-1}}$ for the HAPLESS
complete sub-sample translates to an approximate dust mass limit of
$7.4 \times 10^{5}\,{\rm M_{\odot}}$, using the average HAPLESS dust
temperature of 14.6\,K.  But if we instead use the warmest and coldest
temperatures in our sample (10--25\,K) this limit becomes `fuzzy' and
ranges from $5 \times 10^6$ to $3 \times 10^5\,\rm{M_{\odot}}$.  We
are assuming that a 250\,\micron\ luminosity limited sample is
equivalent to a dust mass limited sample, when in reality it is
not. If we consider the volume accessible to a source with
$M_d=10^7\,{\rm M_{\odot}}$ at a dust temperature of 14\,K, compared to that for
  a source with the same dust mass but at a temperature of 20\,K, we find that the warmer source {\it
    with the same dust mass} has an accessible volume {\it 8 times
  greater} than the colder one. The small area in the \citet{Dunne2011}
  work, combined with this effect, may have resulted in an
  incompleteness to colder galaxies at the median redshift of sources
  in the $M_d\sim 10^7\,{\rm M_{\odot}}$ bins. At the time of the \citet{Dunne2011} work, it was
  not expected to find many galaxies with such cold dust temperatures.
  Future work in measuring the dust mass function for the full Phase-1
  \HATLAS\ area (Dunne et al., {\it in prep.}) will address this issue
  and aim to correct for it.
  
To determine the dust mass density in the HAPLESS volume, we use the combined dust mass of the individual sources, with our luminosity-incomplete sources weighted to account for the fraction of our volume in which they can be detected, according to:

\begin{equation}
\rho_{d} = \frac{ \sum{ \left( M_{d} \frac{V_{\it tot}}{V_{\it acc}} \right) } }{ V_{\it tot} }
\label{Equation:Volume_Density}
\end{equation}

\noindent where $\rho_{d}$ is the dust mass density, $V_{\it tot}$ is the total sample volume, and $V_{\it acc}$ is the accessible volume for a given source (in the case of the luminosity-complete sources detectable in our entire volume, we treat $V_{\it acc} = V_{\it tot}$). The resulting dust mass density in the HAPLESS volume is $\rho_{d} = (3.7 \pm 0.7) \times 10^{5}\,{\rm M_{\odot}\,Mpc^{-3}}$,
  where the uncertainty includes errors on the dust masses of the
  individual sources and poisson statistics, but not cosmic variance
  (in keeping with the errors from other estimates discussed below).

We integrate the Schechter fits to the dust mass functions of
\citet{Clemens2013A, Dunne2011} and \citet{Vlahakis2005C} down to the
average HAPLESS dust mass limit of $7.4 \times 10^{5}\,{\rm
  M_{\odot}}$ to calculate their values of $\rho_d$. We account for the difference in $\kappa_{d}$ in the
\citet{Clemens2013A} work and also note that the units for
$\phi^{\star}$ in their Table 2 are actually $\rm{Mpc^{-3}}$ for their
own fits and not $\rm{Mpc^{-3} \,dex^{-1}}$ as is written in their
paper. The values they quote for $\phi^{\star}$ for \citet{Dunne2011}
and \citet{Vlahakis2005C} are however in the correct units (Negrello,
{\it priv. comm.}). We also scale all values to
  reflect the cosmology used in this work. The corresponding values of
  the local dust mass volume density are $\rho_d = (3.2 \pm 0.6)
  \times 10^{5}\,{\rm M_{\odot}\,Mpc^{-3}}$ for \citet{Clemens2013A},
  $\rho_d = (1.3 \pm 0.2) \times 10^{5}\,{\rm M_{\odot}\,Mpc^{-3}}$
  for \citet{Dunne2011}, and $\rho_d = 1.1 \times 10^{5}\,{\rm
    M_{\odot}\,Mpc^{-3}}$ for \citet{Vlahakis2005C}. The quoted
uncertainty is estimated by retaining the fractional uncertainty of
the integrated value quoted in the original works (where
applicable). \citet{Driver2007D} also estimate the
  dust mass density from their study of the $B$-band luminosity
  function assuming a constant ratio of $L_B/M_d$. They derive a value
  of $(2.1 - 2.7 \pm 0.8) \times 10^{5}\,{\rm M_{\odot}\,Mpc^{-3}}$,
  after accounting for our choice of cosmology and the difference in
  $\kappa_d$ used by
  \citet{Driver2007D}\footnote{\citet{Clemens2013A}
    did not account for the $h$ scaling in \citet{Driver2007D}; also the value of $\kappa_{d}$ used by \citet{Popescu2002} is
    lower than that used here, not equal to ours as stated in
    \citet{Clemens2013A}. Therefore the \citet{Driver2007D} value is
  lower, not higher, than their estimate, though agrees to within
  their 1$\sigma$ errors.}.

The HAPLESS value of $\rho_d = (3.7 \pm 0.7) \times
  10^{5}\,{\rm M_{\odot}\,Mpc^{-3}}$ is compatible with those
  of \citet{Clemens2013A} and \citet{Driver2007D}, and is significantly
  higher than those of \citet{Dunne2011} and
  \citet{Vlahakis2005C}. The dust mass density is dominated by sources
  at and above the knee in the Schechter function, where HAPLESS
  measures a higher space density than the \citet{Dunne2011} and
  \citet{Vlahakis2005C} surveys. We believe this could be due
  to incompleteness in the other works in accounting for the very coldest dusty
  galaxies we see in HAPLESS.

Extrapolation of the DMFs to zero mass suggest that 2--8 percent of dust mass in the local volume is in galaxies below the approximate HAPLESS mass limit, and thus HAPLESS presents a highly complete census (albeit with small statistics at present) of the dust content of the very local universe.
Future work
  exploiting the full \HATLAS\ survey, which covers 600 square degrees
  of sky (compared to the 161.6 square degrees surveyed in this work),
  will be able to address these matters with far greater statistical
  power.

\FloatBarrier \subsection{Links Between Star Formation, Colour, and
  Dust Temperature} \label{Section:Cold_Blue_Galaxies}

\begin{figure}
\begin{center}
\includegraphics[width=0.5\textwidth]{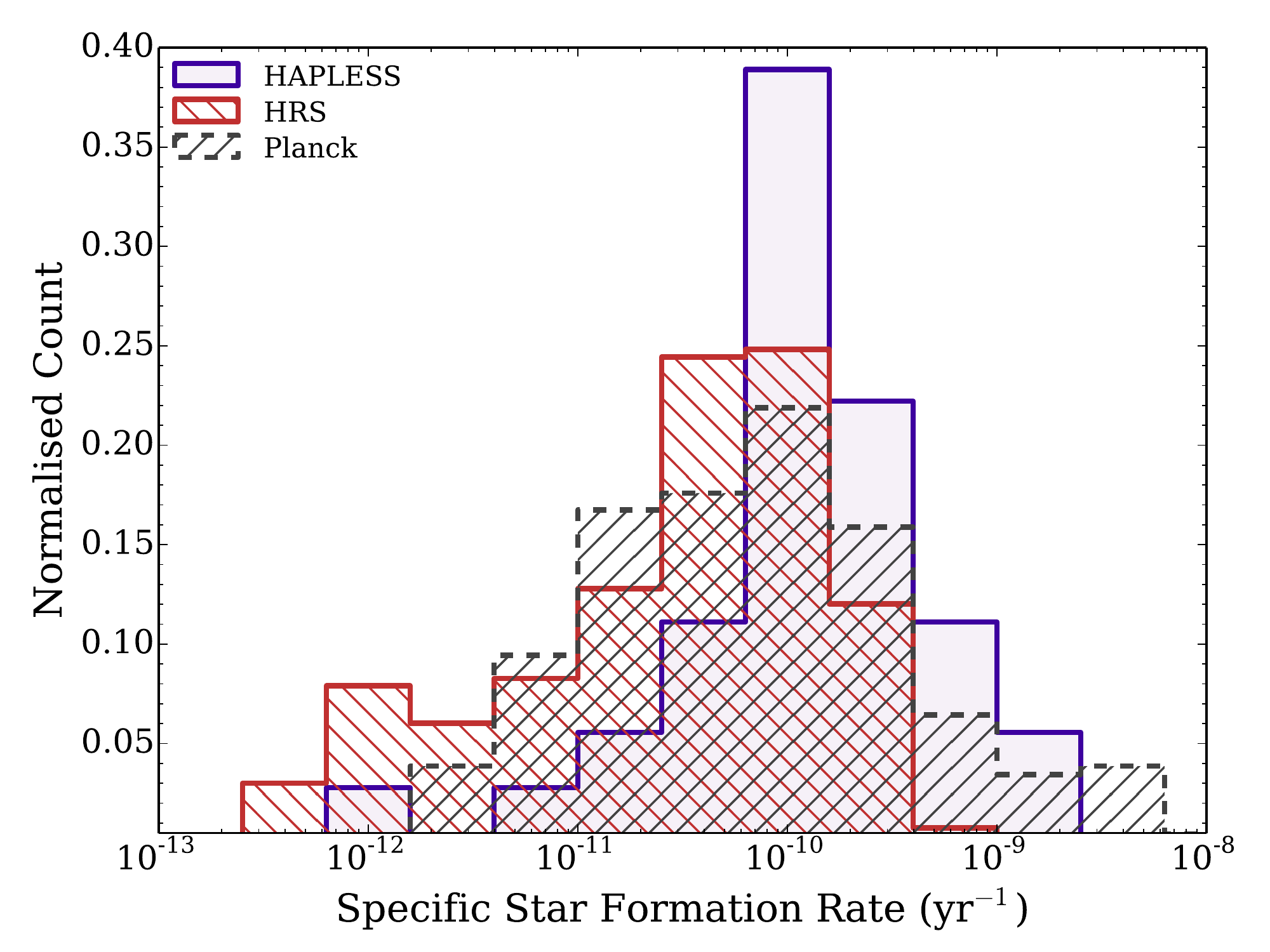}
\caption{The distribution of SSFRs derived for the HAPLESS, HRS, and \cnp\ samples. Whilst the HRS and \cnp\ samples show a broad range of SSFRs, the HAPLESS galaxies generally occupy of much narrow range of values, of relatively high SSFRs.}
\label{Fig:Comparison_SSFR_Hist}
\end{center}
\end{figure}

\begin{figure}
\begin{center}
\includegraphics[width=0.5\textwidth]{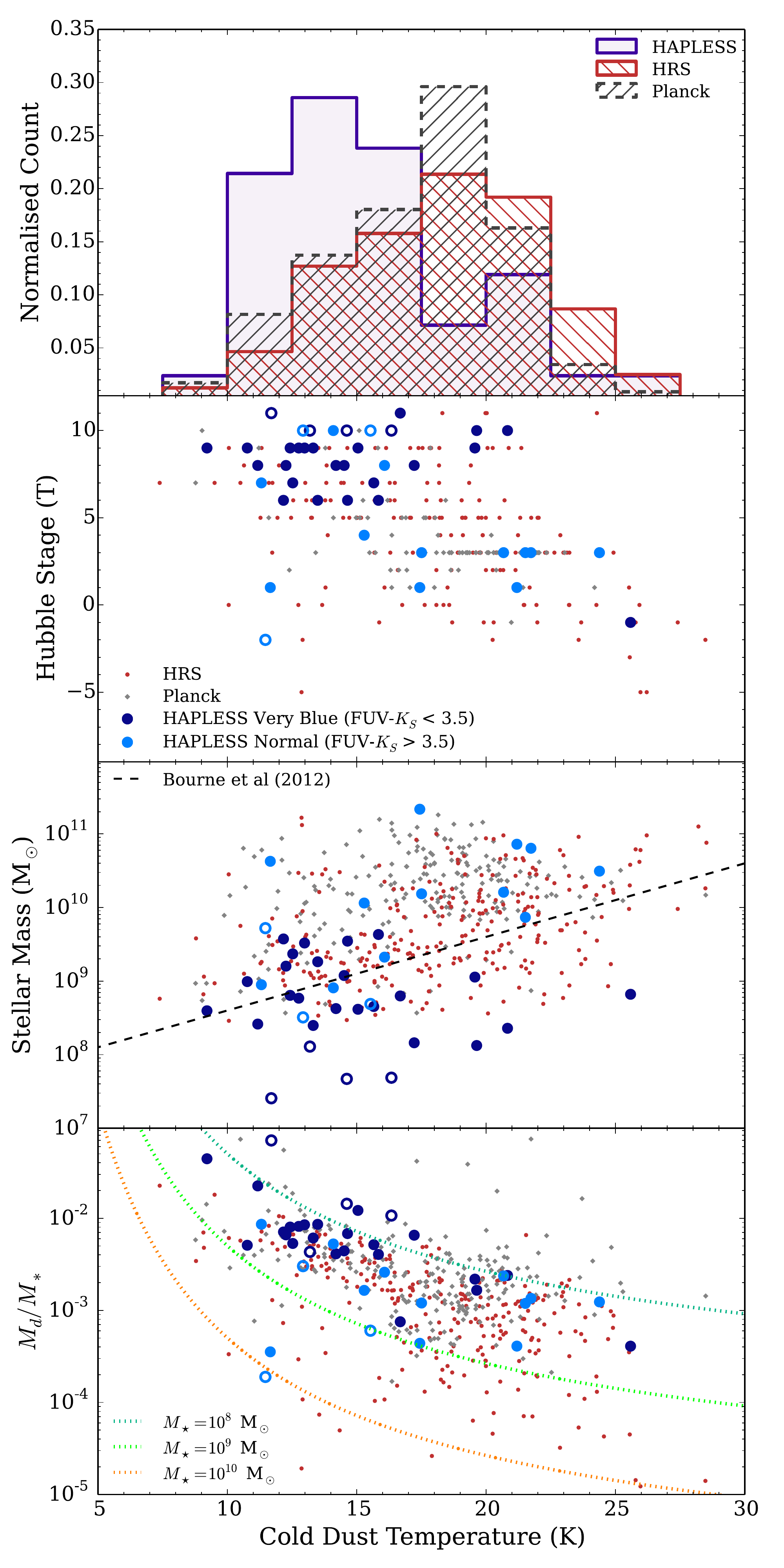}
\caption{Cold dust temperatures relations for the HAPLESS, HRS, and \cnp\ galaxies. {\it From top-to-bottom -- 1\textsuperscript{st}:} The distribution of cold dust temperatures. {\it 2\textsuperscript{nd}:} The relation between morphological type and cold dust temperature; HAPLESS is heavily skewed towards cold late-type galaxies. {\it 3\textsuperscript{rd}:} Stellar mass versus cold dust temperature with the relation from \citet{Bourne2012A}. {\it 4\textsuperscript{th}:} $M_{d}/M_{\star}$ against cold dust temperature. Curves represent different observing limits in $M_{d}/M_{\star}$ due to the cold dust temperature for a given value of $M_{\star}$. Hollow circles indicate galaxies that are beneath the luminosity-limit of the sample.}
\label{Fig:Tc_Comparison_Grid_1}
\end{center}
\end{figure}

\begin{figure}
\begin{center}
\includegraphics[width=0.5\textwidth]{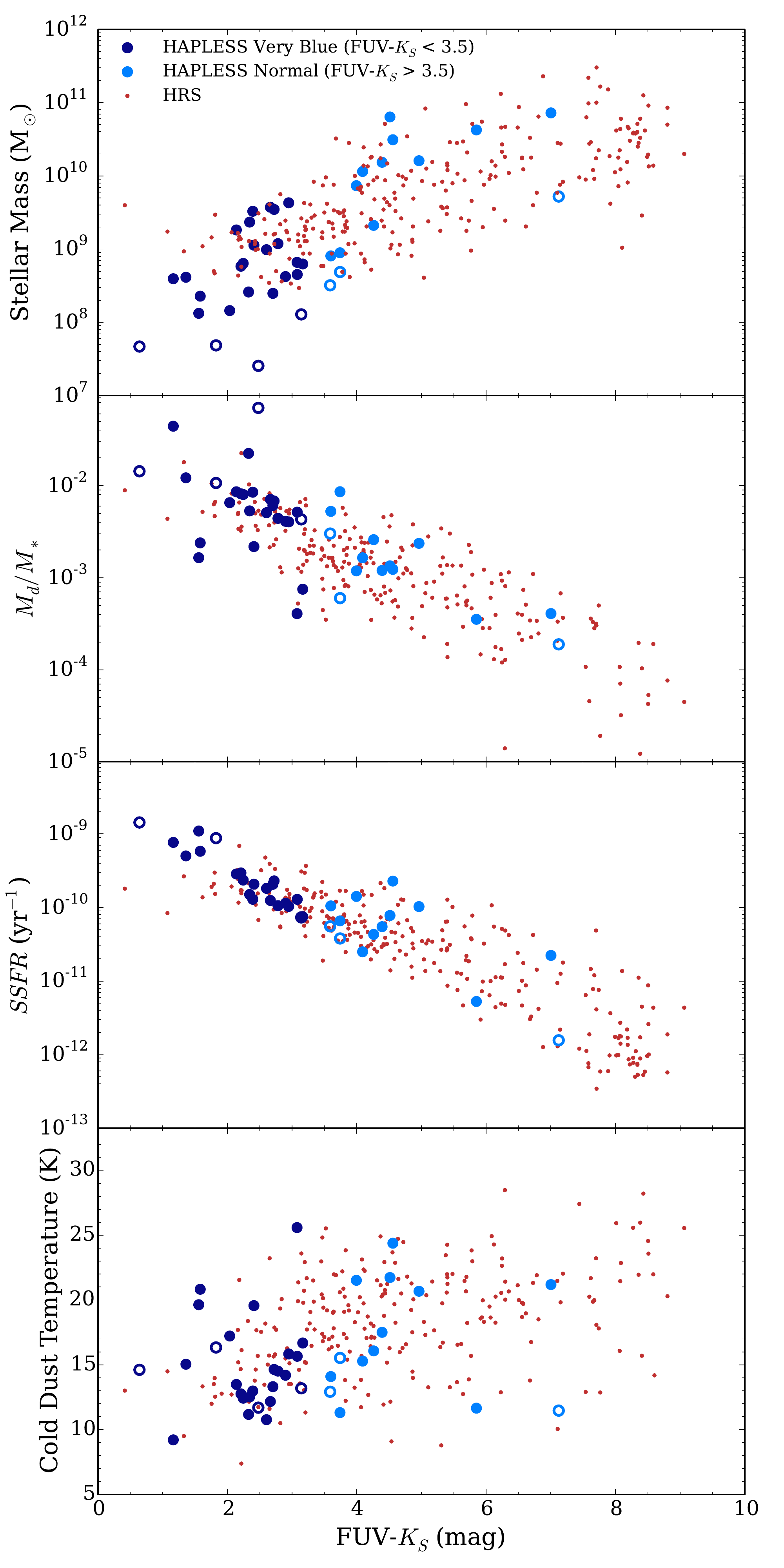}
\caption{Scaling relations with \fK\ colour for HAPLESS and the HRS. {\it From top-to-bottom -- 1\textsuperscript{st}:} Stellar mass against \fK\ colour. {\it 2\textsuperscript{nd}:} $M_{d}/M_{\star}$ (ie, specific dust mass) against \fK\ colour, showing the strong relationship between colour and dust-richness. {\it 3\textsuperscript{rd}:} SSFR against \fK\ colour; the two are tightly related, with our \fK\,\textless\,3.5 colour criterion corresponding to a SSFR $\sim 1.1 \times 10^{-10}\,{\rm yr^{-1}}$.  {\it 4\textsuperscript{th}:} Cold dust temperature against \fK\ colour; HAPLESS shows a preponderance of cold blue galaxies, which only make up a small fraction of the HRS.}
\label{Fig:FUV-K_Comparison_Grid_2}
\end{center}
\end{figure}

\begin{figure}
\begin{center}
\includegraphics[width=0.5\textwidth]{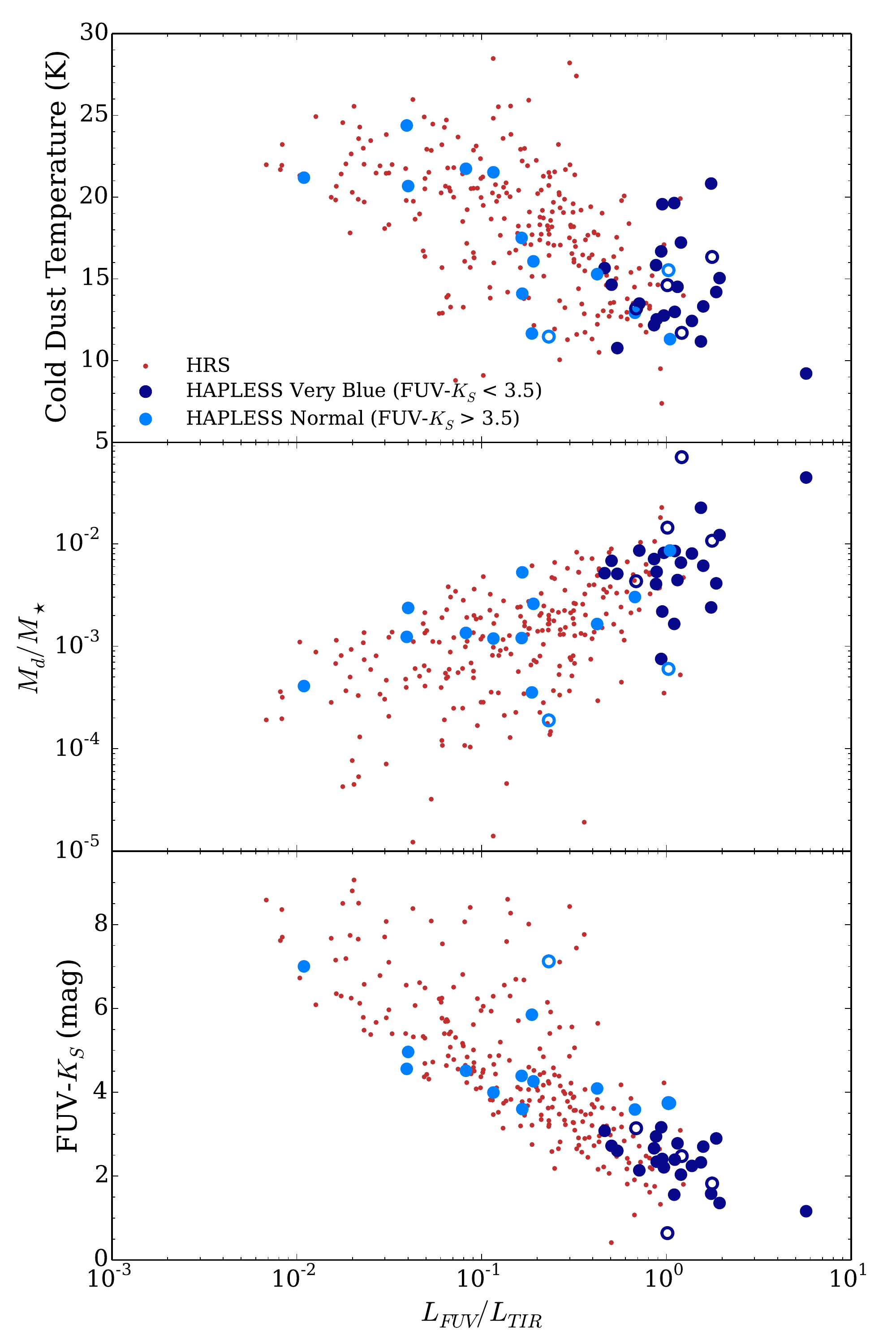}
\caption{Scaling relations with $L_{\it FUV}/L_{\it TIR}$ for the HAPLESS and HRS. {\it Upper:} Cold dust temperature against $L_{\it FUV}/L_{\it TIR}$. {\it Centre:} $M_{d}/M_{\star}$ against $L_{\it FUV}/L_{\it TIR}$; counter-intuitively, the more dust-rich a galaxy, the larger the proportion of FUV photons that go unabsorbed. {\it Lower:} \fK\ colour against $L_{\it FUV}/L_{\it TIR}$; correlation is very tight for bluer galaxies, but far weaker for redder galaxies. }
\label{Fig:FUV_Lum_per_TIR_Lum_Comparison_Grid_1}
\end{center}
\end{figure}

We compare the relative rates of star formation activity between
  the samples using specific star formation rate (SFR/$M_{\star}$), with distributions shown in
  Figure~\ref{Fig:Comparison_SSFR_Hist}. The HAPLESS galaxies tend
towards higher SSFRs, with a median of $1.3 \times 10^{-10}\,{\rm
  yr}^{-1}$. Only 15\,per\,cent (34\,per\,cent) of the HRS (\cnp) galaxies exhibit SSFRs
greater than this. The difference in SSFR distrbutions is
  statistically significant (Table~\ref{Table:Parameters}), again
  highlighting that the HRS stellar mass selection appears to under-sample the regions of the parameter space where the blindly-selected galaxies are found.

Figure~\ref{Fig:Tc_Comparison_Grid_1} (top) shows the cold dust
temperature distributions of the three samples. The HRS and
\cnp\ distributions are similar, with medians of 18.5 and 17.7\,K
respectively (Table~\ref{Table:Parameters}). The HAPLESS distribution,
however, is strikingly different, with a broad peak in the 10--17\,K
range and a median temperature of 14.6\,K; 71\,per\,cent (30) of the HAPLESS
galaxies have dust temperatures colder than both the HRS and
\cnp\ medians.

The relationship between cold dust temperature and galaxy morphology
for the three samples is shown in the 2\textsuperscript{nd} panel of
Figure~\ref{Fig:Tc_Comparison_Grid_1}. A strong correlation is
present; the dust in later galaxy types tends to be much colder than in low
metallicity dwarf galaxies \citep{Remy-Ruyer2013}
and in earlier types \citep{Skibba2011D,MWLSmith2012A}. The HAPLESS
galaxies are heavily skewed towards the late type and very cold end of
the distribution.  The 3\textsuperscript{rd} panel of
  Figure~\ref{Fig:Tc_Comparison_Grid_1} compares stellar mass with
  cold dust temperature. Only a weak correlation is seen (the Spearman
  rank coefficients are 0.23, 0.39 and 0.23 for the HAPLESS, HRS and
  \cnp\ samples respectively; only the latter two samples are
  statistically significant).  We note that \citet{Bourne2012A}
  find a correlation between dust temperature and stellar mass for
  {\em blue cloud} galaxies using a stacking analysis on the H-ATLAS
  data; this trend is plotted as the dashed line.

The last panel of Figure~\ref{Fig:Tc_Comparison_Grid_1} shows a strong
inverse correlation between cold dust temperature and
$M_{d}/M_{\star}$, this is particularly tight for galaxies with cold
dust temperatures below $\sim$\,15\,K. We will explore physical
  connections between stellar heating sources and dust temperature
  next but first we consider whether this trend could be related to
  selection biases. As it is always easier to detect a warm dust
  source at a given mass than a colder one, the lack of galaxies in
  the upper right of this plot cannot be a selection bias. If galaxies
  existed in this part of parameter space (high dust content and warm)
  we would see them. We interpret this as a `heating limit' -- there
  is simply not enough stellar radiation available to heat the dust
  present above the upper temperature envelope. On the
  other hand, the lack of cold galaxies with low $M_d/M_{\star}$
  (lower left) may well be due to the detection bias against low dust
  mass objects with cold temperatures discussed in
  Section~\ref{Section:Dust_Mass_Volume_Density}. Warmer galaxies can still be above the flux limit
  with smaller quantities of dust, and there is indeed more scatter to
  lower $M_d/M_{\star}$ values at higher dust temperatures. The curves in Figure~\ref{Fig:Tc_Comparison_Grid_1} show our observing limits (for inclusion in the luminosity-limited subset of our sample) of $M_d/M_{\star}$ as a function of temperature for different values of $M_{\star}$.  This explains the lack of sources in the lower left region of this plot and the apparent `tightening' of the relationship at cold temperatures.
  
Comparing our parameters with \fK\ colour
    instead of dust temperature in
    Figure~\ref{Fig:FUV-K_Comparison_Grid_2} shows very tight scaling
    relations of colour against stellar mass, $M_{d}/M_{\star}$, and
    SSFR.  The 2\textsuperscript{nd} panel of Figure~\ref{Fig:FUV-K_Comparison_Grid_2} demonstrates that
    bluer galaxies are consistently more {\it dust-rich}. We see that across the
  3.5 orders of magnitude of $M_{d}/M_{\star}$ sampled by HAPLESS and
  the HRS, no galaxies are so dusty that extinction takes over and
  \fK\ colours become redder.  This is in stark contrast to the
  Dust-Obscured Galaxies (DOGs) and SMGs observed at higher
  redshifts, where dust-richness gives rise to severe extinction,
  resulting in red UV-NIR colours \citep{Dey2008A, Calanog2013B,
    Rowlands2014A}.

Despite blue \fK\ colours indicating plentiful ongoing star formation
(Figure~\ref{Fig:FUV-K_Comparison_Grid_2}, 3\textsuperscript{rd}
panel) there is no correlation between `blueness' and dust temperature
(Figure~\ref{Fig:FUV-K_Comparison_Grid_2}, 4\textsuperscript{th}
panel). However, many of the curious blue galaxies are found to have
very cold dust temperatures. A possible explanation for this is that a
large fraction of their UV luminosity escapes unabsorbed by dust. In
Figure~\ref{Fig:FUV_Lum_per_TIR_Lum_Comparison_Grid_1} we examine
$L_{\it FUV}/L_{\it TIR}$ in relation to dust temperature,
$M_{d}/M_{\star}$, and \fK\ colour. $L_{\it FUV}/L_{\it TIR}$
indicates the number UV photons escaping a galaxy (unabsorbed),
relative to the amount of energy which is absorbed by dust and
thermally re-emitted in the IR. In the case where most dust emission
is powered by absorption of UV photons rather than optical photons,
this is equivalent to a measure of the UV
transparency. Overall, both the HAPLESS and HRS
  samples show that the cold dust temperature is anti-correlated with
  $L_{\it FUV}/L_{\it TIR}$, suggesting that the
  higher the factor of UV radiation that is absorbed, the higher the temperature of the cold
  dust. The very bluest HAPLESS galaxies have the highest
  values of $L_{\it FUV}/L_{\it TIR}$, and display a range of cold dust
  temperatures. The central panel of
  Figure~\ref{Fig:FUV_Lum_per_TIR_Lum_Comparison_Grid_1} shows that,
  counter-intuitively, the more dust-rich a galaxy is (as defined by
  $M_{d}/M_{\star}$), the {\it smaller} the fraction of the UV
  luminosity that suffers dust absorption. The combination of dust-richness
  and low attenuation leads to the very cold dust temperatures in the bluest galaxies.
This could be due to some physical difference in the grain
  population, leading to more efficient emission and/or less efficient
  absorption in the UV; or due to a difference in the dust-star geometry
  in the bluest sources. This is beyond the scope of this work but be
  will explored further using the \citet{James2002} method and
  radiative transfer modelling respectively (Dunne et al., {\it in
    prep.}, De Vis et al., {\it in prep.}).

 The lower panel of
Figure~\ref{Fig:FUV_Lum_per_TIR_Lum_Comparison_Grid_1} demonstrates a
tight correlation between $L_{\it FUV}/L_{\it TIR}$ and \fK\ colour --
except for galaxies on the red sequence (\fK\,$\gtrsim$\,6). In these
systems there is a range of $L_{\it FUV}/L_{\it TIR}$ at the same
colour. Recalling that $L_{\it FUV}/L_{\it TIR}$ is really only an
attenuation measure if most IR luminosity is powered by UV photons (as
opposed to optical ones), this wide range of values for $L_{\it
  FUV}/L_{\it TIR}$ on the red sequence may indicate that the dust
heating in this population is not dominated by UV radiation. Dust in
early type galaxies is often acquired during interactions
\citep{Gomez2010B,MWLSmith2012A,Rowlands2012B} which may produce a
range of dust geometries and therefore a wide range of values for
$L_{\it FUV}/L_{\it TIR}$.

In summary, the bluest \fK\ sources exhibit the highest SSFRs,
  the highest specific dust masses, the lowest UV attenuation and
 often display very cold temperatures.

\subsubsection{What is Heating the Cold Dust Component?} \label{Subsection:Cold_Dust_Heating}  

\begin{figure*}
\begin{center}
\includegraphics[width=0.475\textwidth]{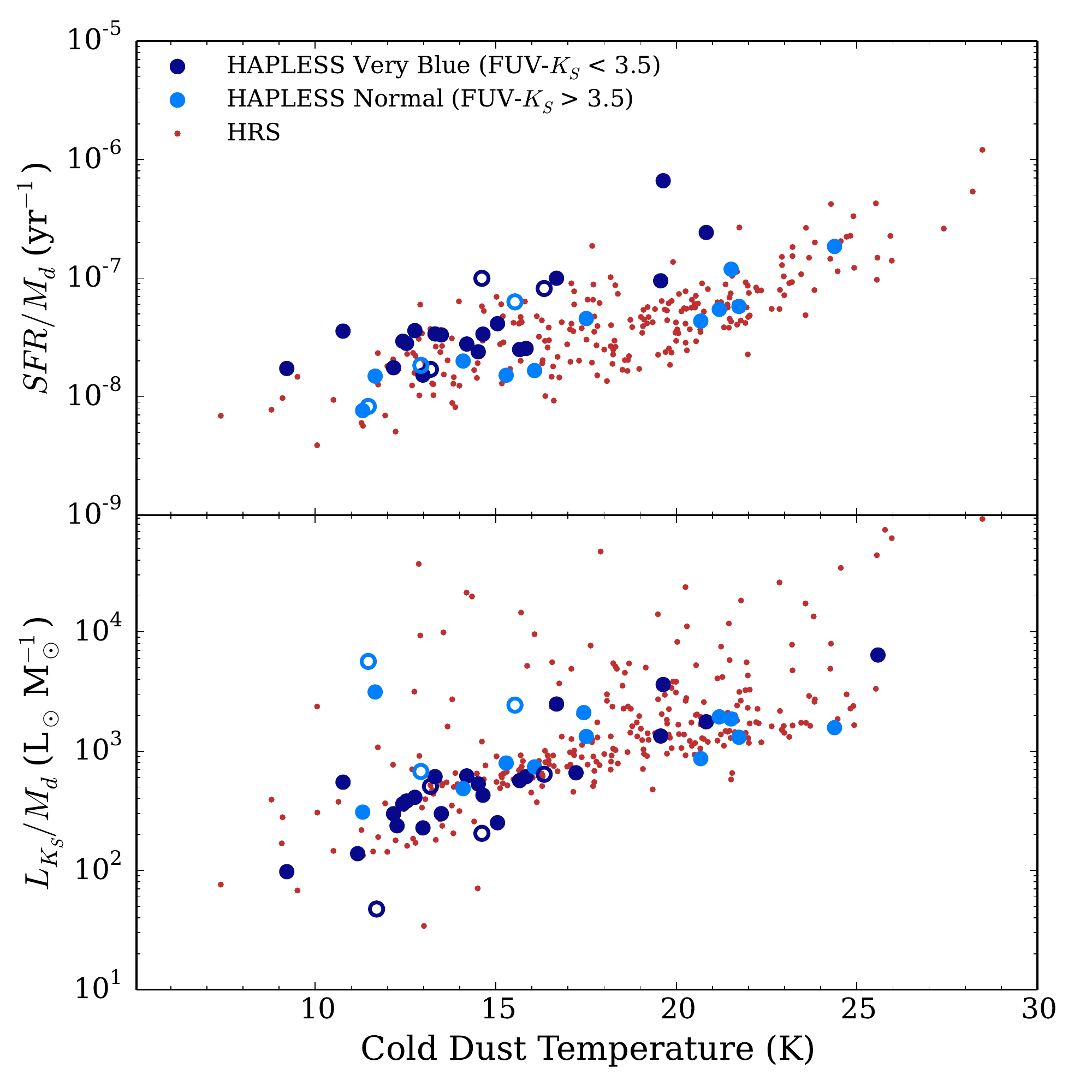}
\includegraphics[width=0.475\textwidth]{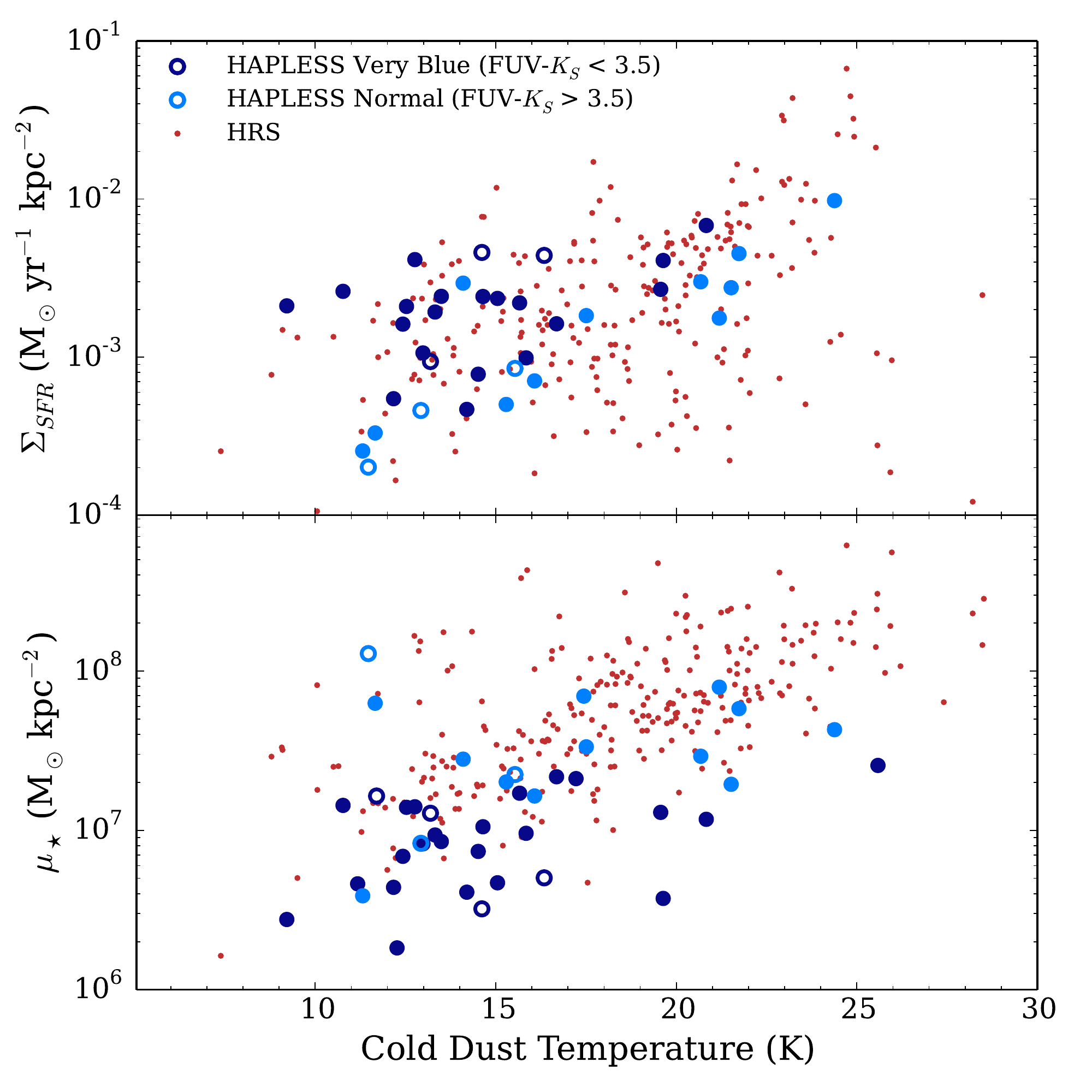}
\caption{The influence of star formation and the older stellar population upon the temperature of the cold dust in the HAPLESS and HRS galaxies. {\it Upper left:} ${\it SFR}/M_{d}$. {\it Lower left:} $L_{K_{S}}/M_{d}$. {\it Upper right:} SFR surface density, $\Sigma_{\it SFR}$. {\it Lower right:} Stellar mass surface density, $\mu_{\star}$. }
\label{Fig:Dust_Heating}
\end{center}
\end{figure*}

Dust heating in galaxies can occur in a variety of ways (see
\citealp{Kennicutt2012A,Dunne2013A}).  Warm dust is thought to be
associated with star-forming dense molecular clouds, with newly formed
stars heating the dust to temperatures \textgreater\,30\,K
\citep{Kennicutt1998H,Kennicutt2009B,Bendo2010C}.  Cold dust is
usually associated with the diffuse ISM
\citep{Rowan-Robinson1989,Boulanger1996,Lagache1998,Tuffs2005,Boquien2011C,Bendo2011A}.
Most dust resides in this diffuse environment
\citep{Dunne2001A,Draine2007C} where it is heated by the general
InterStellar Radiation Field (ISRF), and is often known as a `cirrus'
component \citep{Rowan-Robinson1989}. The ISRF may be largely composed
of photons in the optical produced by the old stellar population; in
this case, the cold dust luminosity would be powered by the old
stellar population and not young newly formed stars
\citep{Tuffs2005,Boquien2011C,Bendo2011A}. However, it is also
possible that high energy UV photons from low optical depth star
forming regions could `leak out' and therefore contribute to heating
the diffuse dust component \citep{Law2011A,
  Popescu2011,Clemens2013A,Hughes2014A}. The dust heating will also
depend on the distribution of dust and stars within a galaxy and the
optical properties of the dust (see
\citealp{Foyle2013}). \citet{Bendo2014} recently used
  a large number of sources from the Very Nearby Galaxy Survey, HRS,
  and KINGFISH to show that the relative contributions of young and
  evolved stars to dust heating varies greatly among nearby spiral
  galaxies. In this section, we wish to investigate the relative
importance of both the young and the old stellar population in heating
the bulk dust mass (ie, the cold component).
  
  Our choice of `heating' parameter is influenced by the study of
  \citet{Foyle2013} who proposed that the amount of star
  formation (or alternatively old stellar luminosity) {\it per
    unit dust mass} should determine the temperature of the bulk
  dust component, not simply the {\it amount} of star formation and/or old
  stars (ie, SFR/stellar mass) or even their surface density. If
  there is more dust to be heated by a particular radiation field then
  its average temperature will be lower. To explore this, Figure~\ref{Fig:Dust_Heating} (left column) compares SFR/$M_d$ (a proxy for energy in star formation per unit dust mass) and $L_{K_{S}}/M_d$ (a proxy for the energy in old stellar photons per unit dust mass) with the cold dust temperature.   

In the upper panel we see that higher values of ${\it SFR}/M_{d}$
correlate with higher values of $T_{c}$ with Spearman $r$ coefficient
of 0.74 for the combined surveys ($r=0.71$ for HAPLESS, $r=0.75$ for
HRS). This was also seen in \cnp. In the lower panel, a positive
correlation is seen between $L_{K_{S}}/M_d$ and $T_{c}$ (Spearman coefficient
$r=0.69$ for the combined surveys -- $r=0.64$ and $r=0.67$ for the
HAPLESS and HRS respectively) with most sources clustered together with a well-defined maximum described by $T_{\it max} \sim (L_{K_{S}}/M_d)^{0.25}$. The galaxies which scatter well above this tight cluster of sources are early-types (E and S0). These relationships suggest that {\it
  both} star formation and the old stellar population are important
contributors to the heating of the diffuse dust component in these
samples, (first noted for the HRS by
  \citealt{Boselli2012}). On average, we also see that for a given
  value of $T_c$, the ${\it SFR}/M_{d}$ is higher in the HAPLESS
  sources compared to the HRS, whereas this is not the case when
  comparing $L_{K_{S}}/M_d$.

\citet{Kirkpatrick2014B} make the same comparison in a 
study of a sample of resolved star forming spirals with average
$M_{\star}=8.2 \times 10^9\, \rm M_{\odot}$ from the KINGFISH
survey. They use 500 \micron\ luminosity as a proxy for dust mass,
3.6 \micron\ luminosity to trace the the older stellar population,
and H$\alpha +24\,\mu$m emission to trace SFR in a study of star-forming spirals. They
find a similar relationship between SFR per unit dust mass and
temperature; however in contrast to this work they find
no significant correlation between the old stellar luminosity per
dust mass and temperature. A key difference in our approach
compared to theirs is that they consider resolved regions within
their galaxies and so photons are required to be absorbed within
the same pixel they were emitted in order to produce a
correlation. It is not clear whether the \citet{Kirkpatrick2014B}
sample would produce the same trends as we see if only the global
integrated values were considered.

\begin{figure}
\begin{center}
\includegraphics[width=0.5\textwidth]{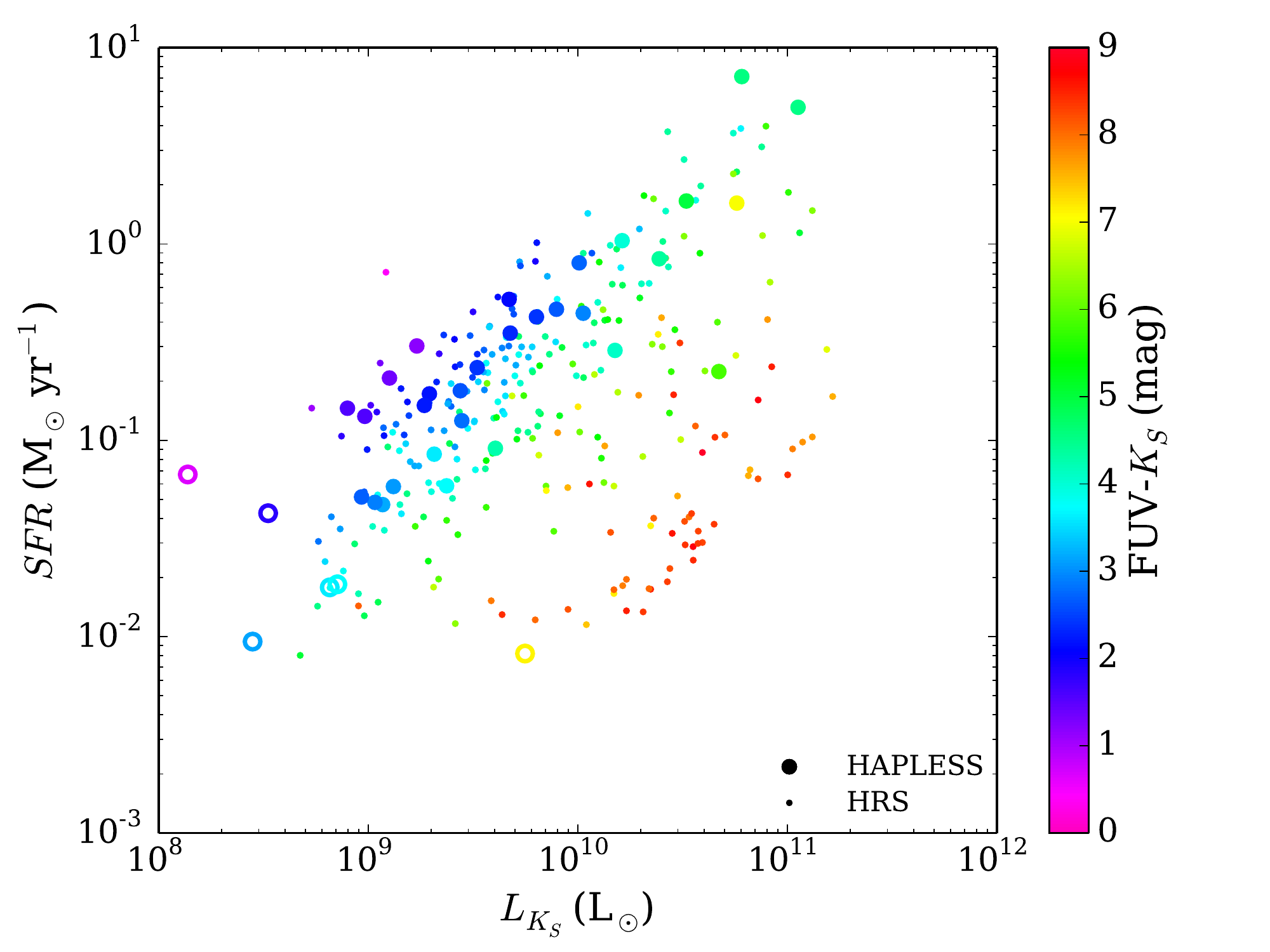}
\caption{The relationship between $L_{K_{S}}$ and ${\it SFR}$, colour-coded by \fK\ colour (indicating morphology), for the HAPLESS and the HRS galaxies. Hollow circles show galaxies beneath the luminosity limit of the sample.}
\label{Fig:Ks_Lum_vs_SFR_vs_FUV-Ks}
\end{center}
\end{figure}

A sufficiently tight, linear correlation between $L_{K_{S}}$ and SFR could make it appear that cold dust temperature is correlated with the heating parameter ${\it SFR}/M_{d}$, even if $L_{K_{S}}$ alone was driving the cold dust temperature, with no contribution from star formation (and vice-a-versa). To test for this, we plot $L_{K_{S}}$ against SFR  in Figure~\ref{Fig:Ks_Lum_vs_SFR_vs_FUV-Ks}. A tight, linear relation in this plot could give rise to spurious correlations with temperature in Figure~\ref{Fig:Dust_Heating}. However, the scatter in this plot is in fact very large -- almost 2 orders of magnitude in SFR are possible for a given value of $L_{K_{S}}$, with distinct sequences of ETGs and LTGs (as indicated by their \fK\ colour) visible. Given how weak the correlation in this plot is (Spearman rank correlation coefficient of 0.28 with both samples taken together), it does not seem possible that it could be artificially driving the tight relations in the left-hand panels of Figure~\ref{Fig:Dust_Heating} (which have Spearman rank correlation coefficients of 0.74 and 0.69 respectively). This was corroborated by using Monte-Carlo simulations of the $L_{K_{S}}$ against SFR relation to generate `spurious' versions of the heating relations in the left half of Figure~\ref{Fig:Dust_Heating}; the `spurious' simulated plots were never able to replicate the degree to which the actual heating relations are tighter than the $L_{K_{S}}$ vs SFR relation.


\begin{figure*}
\begin{center}
\includegraphics[width=1.0\textwidth]{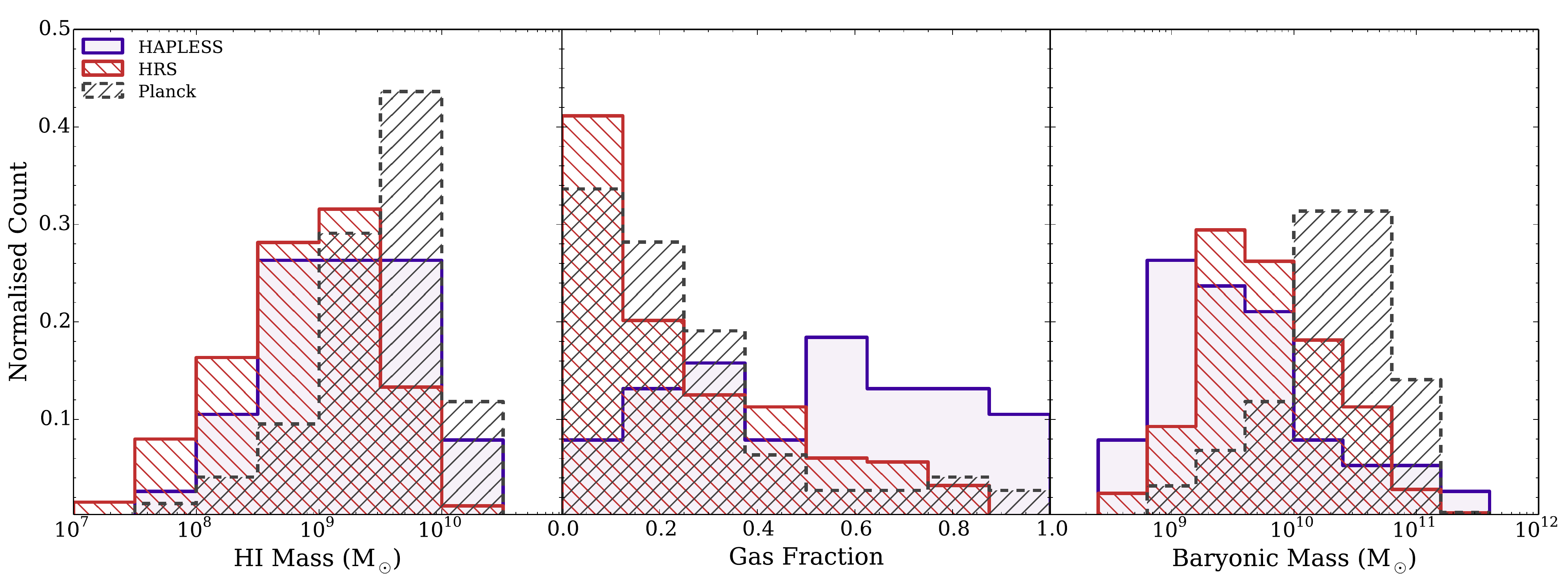}
\caption{The atomic gas properties of the HAPLESS, HRS, and \cnp\ galaxies. {\it Left:} The \HI\ mass distribution. {\it Centre:} The gas fraction (Equation~\ref{Equation:Gas_Fraction}).  The HAPLESS sources have higher gas fractions than seen in other FIR surveys of local galaxies.  {\it Right:} The baryonic mass distribution. Whilst the stellar mass and \HI\ mass distributions of HAPLESS and the HRS are very different, their baryonic mass distributions are rather more similar.}
\label{Fig:Comparison_Gas_Hist_Grid}
\end{center}
\end{figure*}

We also show the relations between the surface densities of star formation ($\Sigma_{\it SFR}$) and stellar
  mass ($\mu_{\star}$) against $T_{c}$ in Figure~\ref{Fig:Dust_Heating} (right column)\footnote{Surface densities were estimated using the $r$-band $R25$ (Table~\ref{Table:Misc_Properties}) to determine the optical radius in kpc, assuming that each galaxy is circular as a first approximation.}.   The surface
density of SFR and the old stellar population should be a first
approximation to the average ISRF contributed by both
populations (the energy per unit area), although this assumption is complicated in the case of the SFR surface density,
as some fraction of the UV radiation will be absorbed locally
by dust in the birth clouds and contribute to heating the warm dust
component rather than the cold.

Both the HAPLESS and HRS (see also \citealt{Boselli2012}) show a clear correlation between $\mu_{\star}$ and $T_{c}$ ($r=0.44$ and $r=0.63$ respectively, with $r=0.65$ when the two samples are combined); the stellar mass surface density is also higher for HRS galaxies at a given dust temperature.   Further evidence for heating by the old stellar population comes from the sample of \HATLAS\ galaxies in \citet{Bourne2013} (of which 12 overlap with our sources) using independent measurements of correlations with dense and diffuse gas components.

Additionally, the HAPLESS galaxies show a weak, but significant correlation between star formation surface density and cold dust temperature ($r=0.32$) while, in contrast, the HRS shows no significant correlation ($r=0.04$). More specifically, the HAPLESS sample shows a range of
  $\Sigma_{\it SFR}$ at the coldest dust temperatures, but requires a higher
  $\Sigma_{\it SFR}$ to reach higher temperatures; the HRS sample instead
  shows a range of $\Sigma_{\it SFR}$ at all temperatures.

This suggests that while both the young and old
  stellar radiation fields play a role in heating the dust in both
  samples, the HRS dust heating is more strongly influenced by the old
  stellar population while HAPLESS sources are on average more
  strongly heated by the young stellar population. As the average
  stellar mass and SSFR of HAPLESS are lower and higher respectively
  than HRS, finding an ISRF dominated by young stars is not
  surprising in the HAPLESS systems.

\subsection{Gas Properties} \label{Subsection:Gas}  

Here we compare the gas properties of the three
samples. Figure~\ref{Fig:Comparison_Gas_Hist_Grid} shows the \HI\ mass
distribution of the three samples; 90\,per\,cent (38) of the HAPLESS galaxies, 81\,per\,cent
(263) of the HRS, and 94\,per\,cent (220) of the \cnp\ galaxies have \HI\ data
available. Interestingly, the median HAPLESS \HI\ mass of $1.4 \times
10^{9}\,{\rm M}_{\odot}$ is greater than the HRS median -- despite
the median HAPLESS stellar mass being 4 times {\it lower} than that of
the HRS. Once again, the bias of the \cnp\ sample towards more massive
objects is manifest (Table~\ref{Table:Parameters}).

The \HI\ gas fractions (Equation~\ref{Equation:Gas_Fraction}) of the HAPLESS galaxies have a median value of 0.52 (Table~\ref{Table:Parameters}), and show a relatively flat distribution from 0.03 to 0.96, spanning a much wider range than those in HRS or Planck C13N13. Of the HAPLESS
galaxies with \HI\ detections, 58\,per\,cent (18) have baryonic masses which
are in fact dominated by their atomic gas component. This is
without any consideration of molecular gas, the inclusion of
which would only serve to drive up the gas fractions still further. In
contrast, the HRS and \cnp\ distributions are strongly skewed towards
lower gas fractions, with medians of 0.18 and 0.17 respectively. A
  K-S test suggests that HAPLESS galaxies are drawn from a different
  underlying population in terms of gas fraction (Table 6).

The right hand panel of Figure~\ref{Fig:Comparison_Gas_Hist_Grid}
shows the baryonic masses of the three samples, where $M_B = M_{\it
  HI} + M_{\star}$. This measure of galaxy mass may be more
appropriate for comparing samples where stars make up only a small
fraction of the total baryonic mass of some of the
galaxies.  Whilst HAPLESS and the HRS have very
different distributions of stellar mass and \HI\ mass
(Figures~\ref{Fig:Comparison_Dust_Stellar_Hist_Grid} and
\ref{Fig:Comparison_Gas_Hist_Grid}), the differences are far less pronounced once
we consider baryonic mass (see Table~\ref{Table:Parameters}). The\cnp\ sample is again limited by its high 550\,\micron\ flux limit, primarily sampling galaxies with high baryonic masses. In the local
Universe -- where the largest halos have already completed more of
their star formation -- this tends to populate the \cnp\ sample with
 a relatively high fraction of passive, high stellar mass and
  low gas fraction galaxies.

\begin{figure}
\begin{center}
\includegraphics[width=0.5\textwidth]{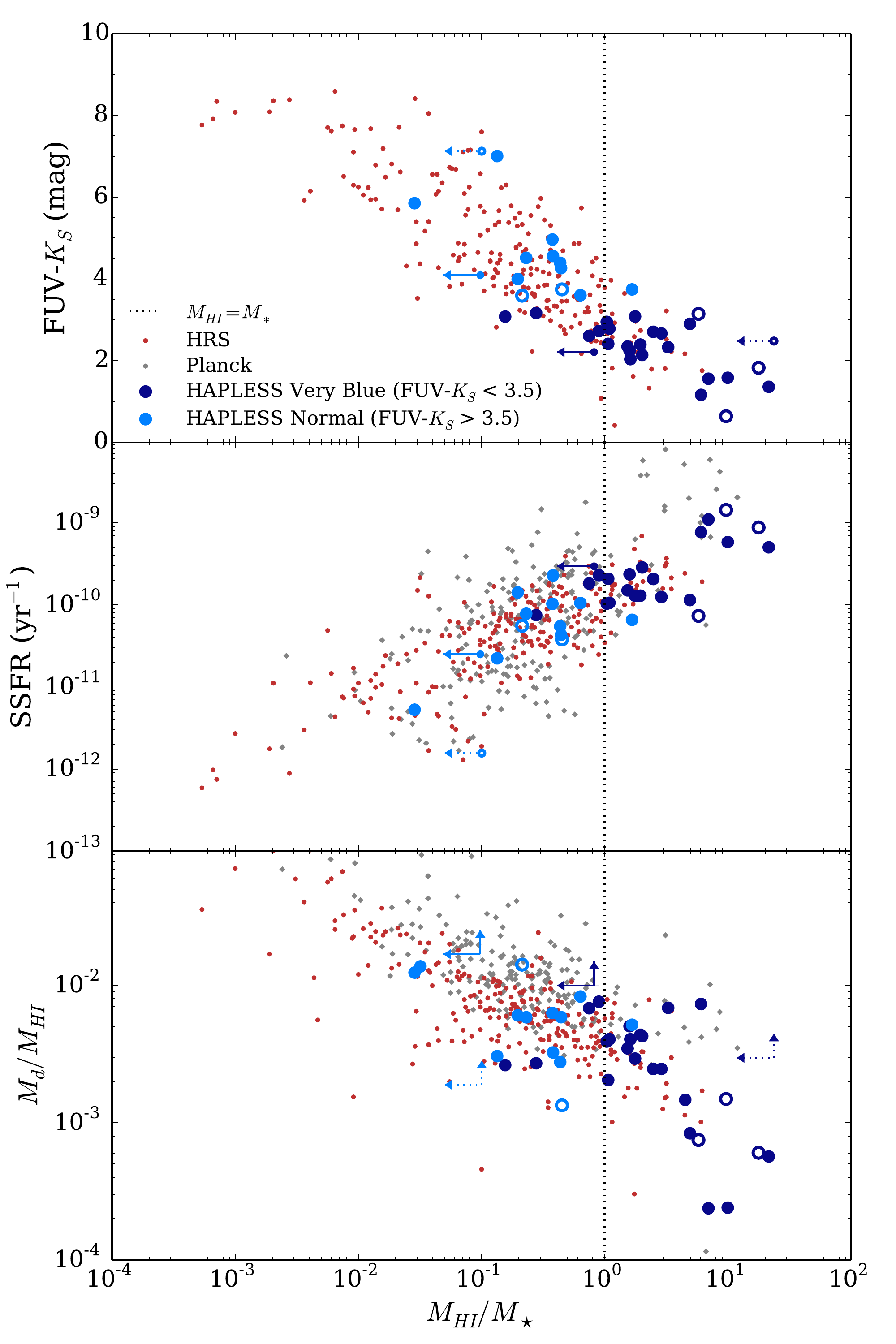}
\caption{Trends with $M_{\it HI}/M_{\star}$ (ie, gas richness) for the HAPLESS, HRS and \cnp\ samples.  {\it Upper:} \fK\ versus $M_{\it HI}/M_{\star}$. Bluer colours are strongly associated with higher gas-to-stellar mass fractions; The (\fK\,\textless\,3.5) colour-criterion we use to define the curious very blue galaxies transpires to correspond to $M_{\it HI} \approx M_{\star}$ (vertical dotted line). {\it Centre:} $M_{\it HI}/M_{\star}$ against SSFR. {\it Lower:} $M_{d}/M_{\it HI}$ against $M_{\it HI}/M_{\star}$.  Hollow circles indicate galaxies that are beneath the luminosity-limit of the sample. HIPASS 3\,$\sigma$ upper limits (Equation~\ref{Equation:HI_Mass_Upper}) are shown (dotted in the case of galaxies not in our 250\,\micron\ luminosity-limit sub-sample.)}
\label{Fig:Mg_per_Ms_Comparison_Grid_1}
\end{center}
\end{figure}

\begin{figure}
\begin{center}
\includegraphics[width=0.5\textwidth]{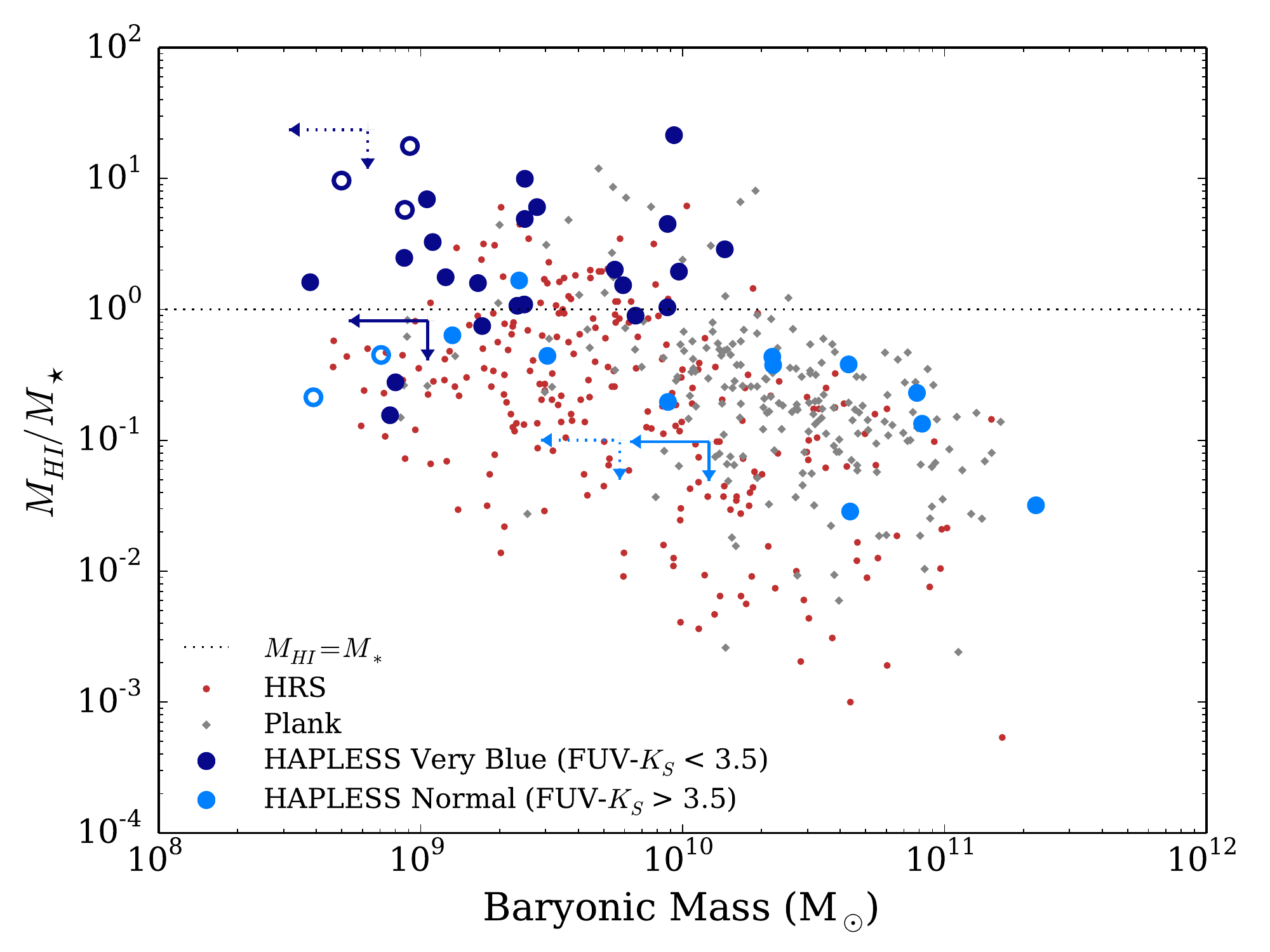}
\caption{$M_{\it HI}/M_{\star}$ against the baryonic mass of the HAPLESS, HRS and \cnp\ samples. The dotted line indicates $M_{\it HI}/M_{\star}=1$.  Symbols as in Figure~\ref{Fig:Mg_per_Ms_Comparison_Grid_1}. }
\label{Fig:Mb_vs_Mg_per_Ms}
\end{center}
\end{figure}

In Figure~\ref{Fig:Mg_per_Ms_Comparison_Grid_1}, we consider the 
  properties of the galaxies in relation to their {\it atomic gas richness};
ie, $M_{\it HI}/M_{\star}$. The top panel shows that bluer
\fK\ colour is strongly correlated with higher levels of gas richness;
we also note that the edges of this distribution appear to be quite
sharp; for a given \fK\ colour, only a small range of $M_{\it
  HI}/M_{\star}$ seems permissible. This is also seen in the
correlation between \fK\ colour and SSFR
(Figure~\ref{Fig:FUV-K_Comparison_Grid_2}), and together with the
middle panel, shows that SSFR is strongly related to gas fraction
in the local Universe. This relation will be explored further in
  a companion paper (De Vis et al., {\it in prep.}) The curious blue
HAPLESS galaxies (\fK\,\textless\,3.5) are the most gas-rich of all;
80\,per\,cent of those with \HI\ detections contain a greater mass of \HI\ than
of stars, and their median gas fraction is 0.66.  Conversely, all
  but one of the HAPLESS sources with $M_{\it HI}/M_{\star}>1$ are in
the curious blue category. It transpires that the
(\fK\,\textless\,3.5) colour criterion we adopted to identify the
curious very blue galaxies corresponds to the
divide between galaxies whose baryonic mass is gas-dominated, and
  those which are star-dominated.

The dust-to-gas ratio of the samples ($M_d/M_{\it HI}$) are compared
in the lower panel of
Figure~\ref{Fig:Mg_per_Ms_Comparison_Grid_1}. Until now we have
described the HAPLESS galaxies, especially the curious blue subset, as
being very dust-rich compared to other FIR surveys, in light of their
high values of $M_{d}/M_{\star}$. But the HAPLESS galaxies are in fact
{\it dust-poor} relative to their gas mass. The median value
(see Table~\ref{Table:Parameters}) of $M_d/M_{\it HI}$ for the HRS and
\cnp\ galaxies are $6.2 \times 10^{-3}$ and $1.2 \times 10^{-2}$ (ie,
gas-to-dust ratios of $\approx\,160$ and $\approx\,90$), whilst the
median for the HAPLESS galaxies is $3.9 \times 10^{-3}$ (gas-to-dust
$\approx\,260$). Furthermore, the median $M_{d}/M_{\it HI}$ of the
curious blue subset is only $2.7 \times 10^{-3}$ -- {\it a median
  gas-to-dust ratio of ${\it \approx\,370}$}.  In comparing
  dust-to-gas ratios of high and low gas fraction samples in this way,
  we do need to worry about our lack of molecular gas information. The
  \cnp\ and HRS may have higher $H_2$/\HI\ ratios than the higher gas
  fraction HAPLESS sources \citep{Saintonge2011S} and so the
  difference in dust-to-gas ratio may be less when this is taken into
  account.  The dust properties in relation to gas-richness will be explored further in De Vis et al. ({\it in
      prep}).


Figure~\ref{Fig:Mb_vs_Mg_per_Ms} compares baryonic mass to $M_{\it
  HI}/M_{\star}$ for the HAPLESS, HRS, and \cnp\ galaxies. Across all
three samples, we see a trend where galaxies with large baryonic
masses tend to have depleted more of their gas than smaller
objects, though the \cnp\ galaxies tend to have higher gas-to-stellar mass ratios for a given baryonic mass. As the HRS is essentially a stellar-mass-selected sample, it
is biased towards objects that have already converted a large fraction
of their gas into stars. The high flux limit of the \cnp\ sample means
that it is biased towards more massive galaxies; but being
selected by dust brightness, it nonetheless tends to select the more
ISM-rich examples of these massive systems. Our blind submm HAPLESS
sample favours ISM-rich objects and consistently features the most
gas-rich galaxies of a given baryonic mass.

\section{The Evolution of Gas and Dust} \label{Section:Chemical_Evolution}  

Here, we will attempt to explain the dust masses and high gas fractions of the HAPLESS sources using a chemical and dust evolution model to follow the build up of heavy elements and dust over time as gas is converted into stars. We assume a closed box model as the optimistic case for the build up of dust (that is, we do not consider inflows and outflows of gas) and instead simply follow the gas (and gas fraction $f_{g}$) as it is converted into stars using a star formation rate $\psi(t)$ and an IMF $\phi(m)$ (using the \citealp{Chabrier2003C} IMF consistent with Sections~\ref{Subsection:Stellar_Masses} and \ref{Subsection:Star_Formation_Rates}). More details of the model can be found in Appendix~\ref{AppendixSection:Chemev}; see also \citet{Morgan2003B} and \citet{Rowlands2014B}. 

We assume two possible scenarios for dust formation by stars (see \citealp{Rowlands2014B} for a more in depth discussion): firstly, where dust is only contributed via the stellar winds of evolved Low-to-Intermediate Mass Stars (LIMS); and secondly, where dust is contributed via both LIMS and SuperNovae (SNe). Whether the majority of dust in galaxies is contributed by LIMS or SNe is a long-standing question (see the review in \citealp{Gomez2013A}), though recent results \citep{Gall2014A} suggest not only do SNe form significant quantities of dust, but that also these grains are big enough (\textgreater 1 \micron) to survive their journey through the harsh reverse shock. We use the dust yields from LIMS consistent with FIR observations \citep{Ladjal2010C} and theoretical models \citep{Ventura2012E}. For supernova dust yields, we use those of \citet{Todini2001A}, which are consistent with the upper range of dust masses observed in historical SN remnants including the Crab Nebula (\citealp{Gomez2012B}; Owen \& Barlow, {\it submitted.}), Cassiopeia A \citep{Dunne2003D, Dunne2009A, Rho2009E, Barlow2010}, and SN1987A \citep{Matsuura2011E, Indebetouw2014A}. Type-Ia SNe are assumed to be negligible contributors to the dust budget \citep{Morgan2003C, Gomez2009A, Gomez2012A}. Note that we have no dust destruction in our model as we want to follow the maximum build up of dust mass at a given time\footnote{We also do not include models for grain growth \citep{Draine2009,Rowlands2014B,Mattsson2014} since this acts to counteract the effects of destruction \citep{Dunne2011,Asano2013}.} (see \citealp{Rowlands2014B}).

\begin{figure}
\begin{center}
\includegraphics[width=0.5\textwidth]{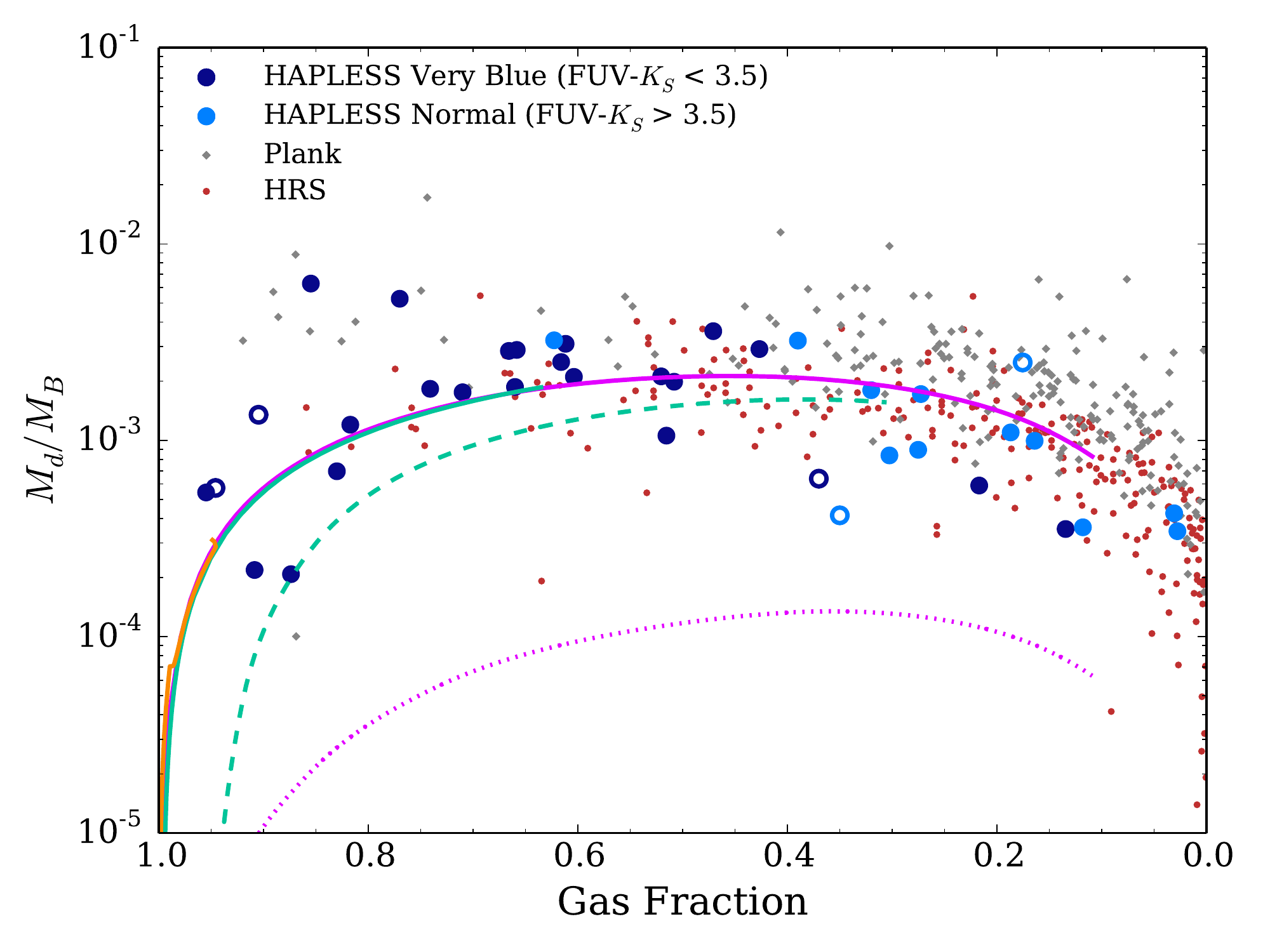}
\caption{$M_{d}/M_{B}$ against gas fraction for the three samples. Note that the x-axis of this plot goes from a gas fraction of 1 to 0. The curves show the results from the chemical evolution model for different SFHs (Appendix~\ref{Fig:App_t_vs_SFR}, Table~\ref{Table:SFR_Histories}) including SFR A - consistent with the Milky Way (purple, \citealp{Yin2009A}); SFR B - exponentially declining SFR with initial value of $0.06\,\rm M_{\odot}\,yr^{-1}$ and a burst (orange); SFH C - an exponentially declining rate, but with higher initial SFR of $2.4\,\rm M_{\odot}\,yr^{-1}$ (tourquoise); and finally SFR D - a scaled version of SFR C ($\times\,20$, tourquoise dashed). The dotted purple line is SFH A (MW) with dust from LIMS only.}
\label{Fig:fg_vs_Md_per_Mb}
\end{center}
\end{figure}

We use four fiducial Star Formation Histories (SFHs), shown in Figure~\ref{Fig:App_t_vs_SFR} and in Table~\ref{Table:SFR_Histories}, to model the HAPLESS galaxies. These SFHs are (i) SFH A - consistent with the Milky Way \citep{Yin2009A}; (ii) SFH B - an exponentially declining SFR with initial value of $\psi(t,0) = 0.06\,\rm M_{\odot}\,yr^{-1}$ and a short burst at $\sim 1$\,Gyr; (iii) SFH C - a faster exponentially declining SFR than B, with initial SFR of $2.4\,\rm M_{\odot}\,yr^{-1}$; (iv) SFH D - a scaled-up version of SFR C (multiplied by a factor of 20) to illustrate the evolution of a galaxy which is consuming its gas more rapidly. Using these fiducial SFHs, we follow the evolution of the dust mass relative to the baryonic mass as the gas fraction falls. The initial gas mass is set to $M_g(0) = 4\times 10^{10}\,\rm M_{\odot}$ for the Milky Way, and for the other models we use the observed gas masses and fractions (Table~\ref{Table:HI_Properties}) to derive the initial gas masses (these range from $M_g(0) = 3-5.5 \times 10^{9}\,{\rm M_{\odot}}$). 

The model results are shown in Figure~\ref{Fig:fg_vs_Md_per_Mb}. First, we compare the dust evolution with dust only from LIMS (dotted line). Second, including dust from LIMS and SNe in combination (solid and dashed lines). None of the models for the former scenario (ie, without SN dust) reach the high levels of $M_{d}/M_{B}$ observed in the HAPLESS, HRS and \cnp\ samples; this is in line with results from other studies, including \citet{Morgan2003B}, \citet{Matsuura2009B}, \citet{Dunne2011}, \citet{Gall2011B}, and \citet{Rowlands2014B}.  With dust from both SNe and LIMS, SFH models A--C all sit on the same evolutionary track in Figure~\ref{Fig:fg_vs_Md_per_Mb}, due to the models with lower star formation rates (SFHs B \& C) than the MW (SFH A) also having lower initial gas masses; ie the models lie on the same constant $SFR/M_{\it HI}$ tracks. These are in good agreement with the HAPLESS galaxies (at high gas fractions) and the HRS galaxies at lower gas fractions. The \cnp\ galaxies (clustered towards lower gas fractions) have somewhat elevated $M_d/M_B$ compared to the models presented here. When we multiply SFH C by a factor of 20 (SFH D) but keep the initial gas mass the same as SFH C, the evolutionary path is offset, due to the available gas reservoir being consumed faster and dust mass reduced due to astration. 

\begin{table*}
\begin{center}
\caption{The fiducial star formation histories (A--D) and initial gas masses used in this work to model the HAPLESS galaxies. SFH A is from \citealp{Yin2009A}.  Also given is the time ($t_{\rm end}$) where the SFH is truncated to match the present observed SFRs and gas fractions of the HAPLESS sample (Section~\ref{Section:HAPLESS_Properties}).}
\label{Table:SFR_Histories}
\begin{tabular}{lrrcrrr} 
\toprule \toprule
\multicolumn{1}{c}{SFH} &
\multicolumn{1}{c}{$g(0)$} &
\multicolumn{1}{c}{$\psi(0)$} &
\multicolumn{1}{c}{Burst?} &
\multicolumn{1}{c}{$t_{\rm end}$} &
\multicolumn{1}{c}{$\psi(t_{\rm end})$} &
\multicolumn{1}{c}{$f_g(t_{\rm end})$}  \\
\multicolumn{1}{c}{} &
\multicolumn{1}{c}{($\rm M_{\odot}$)} &
\multicolumn{1}{c}{($\rm M_{\odot}\,yr^{-1}$)} &
\multicolumn{1}{c}{(Y/N)} &
\multicolumn{1}{c}{(Gyr)} &
\multicolumn{1}{c}{($\rm M_{\odot}\,yr^{-1}$)}\\
\midrule
A (MW) & $4.0 \times 10^{10}$ & 10 & N & 20 & 0.7 & 0.11 \\  
B & $3.0 \times 10^{9}$ & 0.06 & Y & 1.35 & 0.029 & 0.95\\  
C & $5.5 \times 10^{9}$ & 2.5 & N & 2.8 & 0.5 & 0.64 \\  
D & $5.5 \times 10^{9}$ & 49 & N & 0.1 & 48 & 0.31 \\  
\bottomrule
\end{tabular}
\end{center}
\end{table*}

The evolutionary path suggested in Figure~\ref{Fig:fg_vs_Md_per_Mb} indicates that a galaxy's dust mass will peak when its gas fraction falls to $\sim$\,0.5, as predicted in \citet{Eales1996C}. Therefore this is the stage of a galaxy's development when it is most likely to meet the inclusion threshold of a dust-selected sample such as HAPLESS -- the median gas fraction of which is indeed 0.5. The stellar-mass selection of the HRS means that it is biased towards galaxies where most of the gas has already been converted into stars, hence it severely under-samples the gas-rich portion of this evolutionary path. Similarly, the tendency of the \cnp\ sample to mainly select more massive galaxies means that it too is biased towards systems with low gas fractions.


\section{Conclusions} 

We have presented a dust-selected sample of nearby galaxies drawn from the blind \HATLAS\ submm survey and introduced a pipeline used to derive photometry, dust masses and other properties for the sample. We have also studied correlations between the dust and other properties of the galaxies in our sample, and the HRS and Planck samples. We find the following results:

\begin{itemize}
\item A typical source seen by \hersc\ in this blind survey has a cold dust temperature of 14.6 K and dust mass of $5.6 \times 10^{6}\,{\rm M}_{\odot}$. 

\item HAPLESS galaxies have median $M_{d}/M_{\star}$ greater by a factor of $\sim$\,3.7 than the galaxies observed as part of the \hersc\ Reference Survey, and a factor of 1.8 than galaxies in the \planck\ Early Release Compact Source Catalogue. The median properties of this sample include: $\langle M_{\star}\rangle = 9.8 \times 10^8\,\rm M_{\odot}$, $\langle{\rm SFR}\rangle = 0.2\,\rm M_{\odot}\,yr^{-1}$, $\langle {\rm SSFR}\rangle = 1.3 \times 10^{-10}\,\rm \,yr^{-1}$ and are amongst the most actively star forming galaxies seen in local FIR and submm surveys.

\item This sample contains a high proportion of very blue galaxies (defined as \fK\,$<$\,3.5). These are generally irregular and/or highly flocculent; such galaxies tend to be UV-bright, NIR-faint, dust rich, and low stellar mass, with high specific star formation rates. The median dust-to-stellar mass ratio of the very blue subset is $\sim$\,3--5 times larger than the \planck\ and HRS samples. Whilst accounting for only 6\,per\,cent of the stellar mass in our sample, the bluest galaxies in our sample contain over 35\,per\,cent of the dust mass.

\item The dust mass volume density of our sample is $(3.7 \pm 0.7) \times 10^5 {\rm \,M_{\odot}\,Mpc^{-3}}$, which is higher than some other estimates, but consistent with the value found by \planck. Much of this difference seems to arise from the low dust temperatures of the galaxies in our sample, as the 250\,\micron\ luminosity function of our sample is in good agreement with surveys of larger volumes. Note however that our volume suffers from a high cosmic variance of $\sim$\,166\,per\,cent. 

\item The HAPLESS galaxies are extraordinarily gas rich, particularly the very blue sources. Of the 38 HAPLESS galaxies detected in \HI, 21 (55\,per\,cent) have atomic gas masses greater than or equivalent to their stellar mass. Their median gas fraction is 0.52, and 26\,per\,cent have gas fractions $>\,0.8$.  The median gas to dust ratios of these sources (\textgreater\,260) is 1.6--3.0 times greater than for the other local samples of dusty galaxies.   

\item The coldest dust seen in the local universe is consistently associated with galaxies that have lots of star formation relative to their older stellar population. Despite dust being so plentiful in these objects, UV photons apparently go unabsorbed -- giving rise to their very blue \fK\ colours, and colder dust temperatures. Comparing the star formation and stellar mass surface densities also shows that dust heating in galaxies selected by HRS is more strongly influenced by the old stellar population whereas galaxies selected in HAPLESS are more strongly heated by the young stellar population.

\item A chemical and dust evolution model confirms these galaxies are simply in an earlier stage of converting their gas into stars.  The bluest galaxies appear to be the most immature; they should therefore provide valuable insights into the chemical evolution of young galaxies. 
\end{itemize}

A blind dust-selected sample in the local universe reveals very blue, dusty, and gas-rich galaxies. Despite accounting for roughly half of all dusty galaxies, they have been severely under-represented in other FIR and submm surveys. We suggest the properties of these blue galaxies are in line with their `immaturity', and therefore may provide useful analogues to very young, high-$z$ galaxies, though we note that the interstellar medium in the HAPLESS sources is likely to be different. Resolved atomic and molecular gas maps of the bluest sources in this sample, combined with radiative transfer modelling, 850\,\micron\ observations, and integrated optical spectra, should be able to address this and test whether these blue-but-dusty galaxies have different grain properties or whether the dust is distributed in a `leaky' geometry (De Vis. et al., {\it in prep.}, Dunne et al., {\it in prep.}, Smith et al., {\it in prep.}).

\section{Acknowledgements}
CJRC acknowledges support from the Science and Technology Facilities Council (STFC) Doctoral Training Grant scheme and the European Research Council (ERC) FP7 project DustPedia (PI J Davies), and kindly thanks Marcel Clemens, Luca Cortese, Allison Kirkpatrick, Ivan Baldry, Lee Kelvin, and Michal Michalowski for helpful conversations. HLG and SAE acknowledge support from the STFC Consolidated Grant scheme. LD, SJM and RJI acknowledge support from the ERC in the form of the Advanced Investigator Program, {\sc COSMICISM}. KR acknowledges support from the ERC Starting Grant SEDmorph (PI V Wild). The \HATLAS\ is a project with {\it Herschel}, which is an ESA space observatory with science instruments provided by European-led Principal Investigator consortia and with important participation from NASA. The \HATLAS\ website is \url{http://www.h-atlas.org/}. GAMA is a joint European-Australasian project based around a spectroscopic campaign using the Anglo-Australian Telescope. The GAMA input catalogue is based on data taken from the Sloan Digital Sky Survey and the UKIRT Infrared Deep Sky Survey. Complementary imaging of the GAMA regions is being obtained by a number of independent survey programs including GALEX MIS, VST KIDS, VISTA VIKING, WISE, \hersc-ATLAS, GMRT, and ASKAP, providing UV to radio coverage. GAMA is funded by the STFC (UK), the ARC (Australia), the AAO, and the participating institutions. The GAMA website is: \url{http://www.gama-survey.org/}. The authors gratefully acknowledge Martha Haynes, Riccardo Giovanelli, and the ALFALFA team for supplying the latest ALFALFA survey data. 

This research has made use of Astropy\footnote{\url{http://www.astropy.org/}}, a community-developed core Python package for Astronomy \citep{Astropy2013}. This research has made use of TOPCAT\footnote{\url{http://www.star.bris.ac.uk/~mbt/topcat/}} \citep{Taylor2005A}, which was initially developed under the UK Starlink project, and has since been supported by PPARC, the VOTech project, the AstroGrid project, the AIDA project, the STFC, the GAVO project, the European Space Agency, and the GENIUS project. This research has made use of of APLpy\footnote{\url{http://aplpy.github.io/}}, an open-source astronomical image plotting package for Python. This research has made use of NumPy\footnote{\url{http://www.numpy.org/}} \citep{Walt2011B}, SciPy\footnote{\url{http://www.scipy.org/}}, and MatPlotLib\footnote{\url{http://matplotlib.org/}} \citep{Hunter2007A}. This research has made use of the SIMBAD\footnote{\url{http://simbad.u-strasbg.fr/simbad/}} database \citep{Wenger2000D} and the VizieR\footnote{\url{http://vizier.u-strasbg.fr/viz-bin/VizieR}} catalogue access tool \citep{Ochsenbein2000B}, both operated at CDS, Strasbourg, France. This research has made use of SAOImage DS9\footnote{\url{http://ds9.si.edu/site/Home.html}}, developed by the Smithsonian Astrophysical Observatory with support from the Chandra X-ray Science Center (CXC), the High Energy Astrophysics Science Archive Center (HEASARC), and the JWST Mission office at the Space Telescope Science Institute (STSI). This research has made use of the NASA/IPAC Extragalactic Database (NED\footnote{\url{http://ned.ipac.caltech.edu/}}) and the NASA/IPAC Infrared Science Archive (IRSA\footnote{\url{http://irsa.ipac.caltech.edu/frontpage/}}), both operated by the Jet Propulsion Laboratory, California Institute of Technology, under contract with the National Aeronautics and Space Administration.

\def\ref@jnl#1{{\rmfamily #1}}%
\newcommand\aj{\ref@jnl{AJ}}%
\newcommand\araa{\ref@jnl{ARA\&A}}%
\newcommand\apj{\ref@jnl{ApJ}}%
\newcommand\apjl{\ref@jnl{ApJ}}%
\newcommand\apjs{\ref@jnl{ApJS}}%
\newcommand\ao{\ref@jnl{Appl.~Opt.}}%
\newcommand\apss{\ref@jnl{Ap\&SS}}%
\newcommand\aap{\ref@jnl{A\&A}}%
\newcommand\aapr{\ref@jnl{A\&A~Rev.}}%
\newcommand\aaps{\ref@jnl{A\&AS}}%
\newcommand\azh{\ref@jnl{AZh}}%
\newcommand\baas{\ref@jnl{BAAS}}%
\newcommand\jrasc{\ref@jnl{JRASC}}%
\newcommand\memras{\ref@jnl{MmRAS}}%
\newcommand\mnras{\ref@jnl{MNRAS}}%
\newcommand\pra{\ref@jnl{Phys.~Rev.~A}}%
\newcommand\prb{\ref@jnl{Phys.~Rev.~B}}%
\newcommand\prc{\ref@jnl{Phys.~Rev.~C}}%
\newcommand\prd{\ref@jnl{Phys.~Rev.~D}}%
\newcommand\pre{\ref@jnl{Phys.~Rev.~E}}%
\newcommand\prl{\ref@jnl{Phys.~Rev.~Lett.}}%
\newcommand\pasp{\ref@jnl{PASP}}%
\newcommand\pasj{\ref@jnl{PASJ}}%
\newcommand\qjras{\ref@jnl{QJRAS}}%
\newcommand\skytel{\ref@jnl{S\&T}}%
\newcommand\solphys{\ref@jnl{Sol.~Phys.}}%
\newcommand\sovast{\ref@jnl{Soviet~Ast.}}%
\newcommand\ssr{\ref@jnl{Space~Sci.~Rev.}}%
\newcommand\zap{\ref@jnl{ZAp}}%
\newcommand\nat{\ref@jnl{Nature}}%
\newcommand\iaucirc{\ref@jnl{IAU~Circ.}}%
\newcommand\aplett{\ref@jnl{Astrophys.~Lett.}}%
\newcommand\apspr{\ref@jnl{Astrophys.~Space~Phys.~Res.}}%
\newcommand\bain{\ref@jnl{Bull.~Astron.~Inst.~Netherlands}}%
\newcommand\fcp{\ref@jnl{Fund.~Cosmic~Phys.}}%
\newcommand\gca{\ref@jnl{Geochim.~Cosmochim.~Acta}}%
\newcommand\grl{\ref@jnl{Geophys.~Res.~Lett.}}%
\newcommand\jcp{\ref@jnl{J.~Chem.~Phys.}}%
\newcommand\jgr{\ref@jnl{J.~Geophys.~Res.}}%
\newcommand\jqsrt{\ref@jnl{J.~Quant.~Spec.~Radiat.~Transf.}}%
\newcommand\memsai{\ref@jnl{Mem.~Soc.~Astron.~Italiana}}%
\newcommand\nphysa{\ref@jnl{Nucl.~Phys.~A}}%
\newcommand\physrep{\ref@jnl{Phys.~Rep.}}%
\newcommand\physscr{\ref@jnl{Phys.~Scr}}%
\newcommand\planss{\ref@jnl{Planet.~Space~Sci.}}%
\newcommand\procspie{\ref@jnl{Proc.~SPIE}}%

\bibliographystyle{mn2e}
\bibliography{ChrisBib}

\appendix
 
\section{Properties of the HAPLESS Galaxies} \label{AppendixSection:HAPLESS}

Multiwavelength imagery of the HAPLESS galaxies can be found in Figure~A1. Our UV to FIR photometry of the HAPLESS galaxies, with uncertainties, is given in Table~A1. Figure~A2 shows the spectral energy distributions of the HAPLESS galaxies.

\begin{center}
\begin{figure*}
\includegraphics[width=0.475\textwidth]{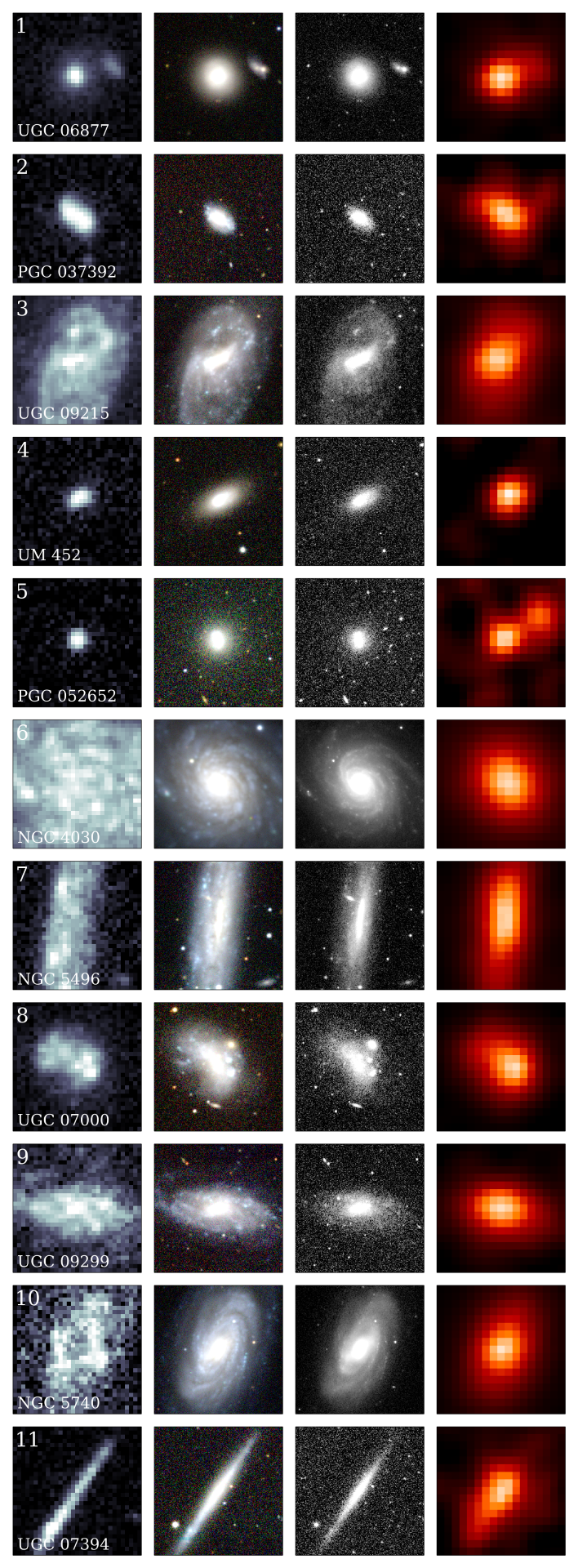}
\includegraphics[width=0.475\textwidth]{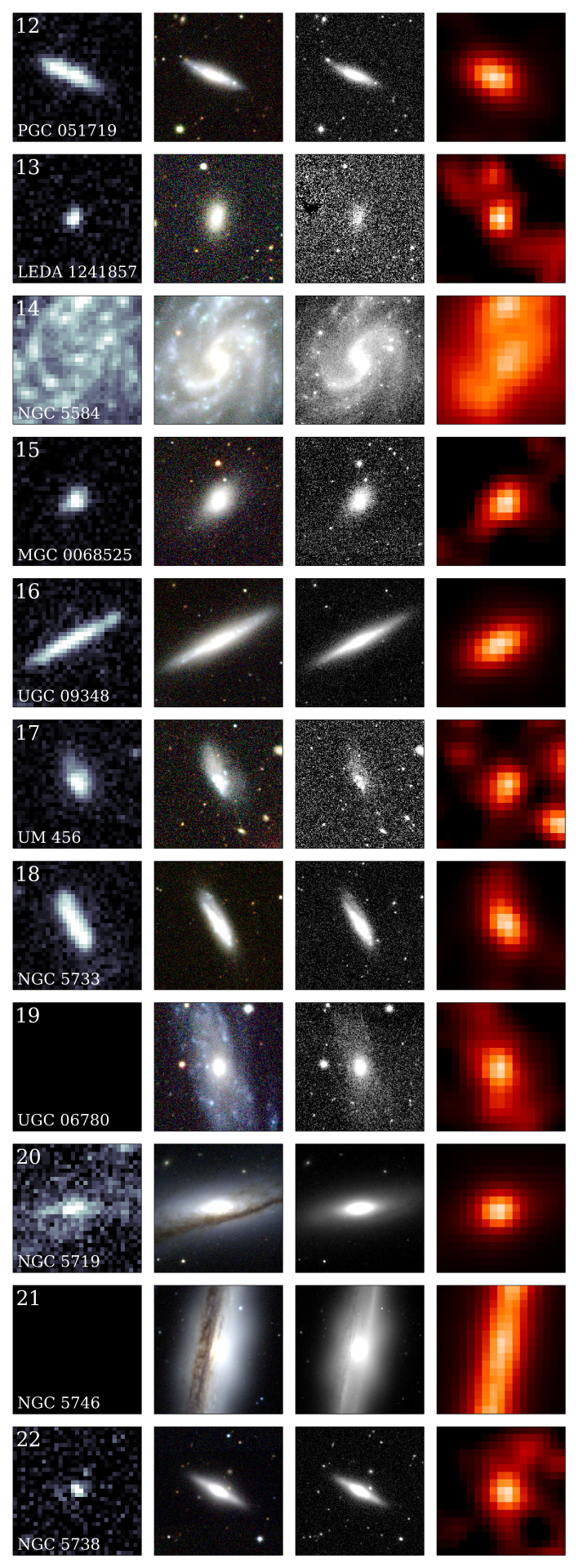}
\caption{Multiwavelength imagery of each of the HAPLESS galaxies. The bands displayed, from left-to-right, are: GALEX FUV, SDSS {\it gri} three-colour, VIKING \Kband, and \hersc\ 250 \micron. Each cutout is 100\arcsec\ on a side. HAPLESS 19 and 21 do not have GALEX coverage.} 
\label{AppendixFig:HAPLESS_Imagery}
\end{figure*}
\end{center}

\begin{center}
\begin{figure*}
\includegraphics[width=0.475\textwidth]{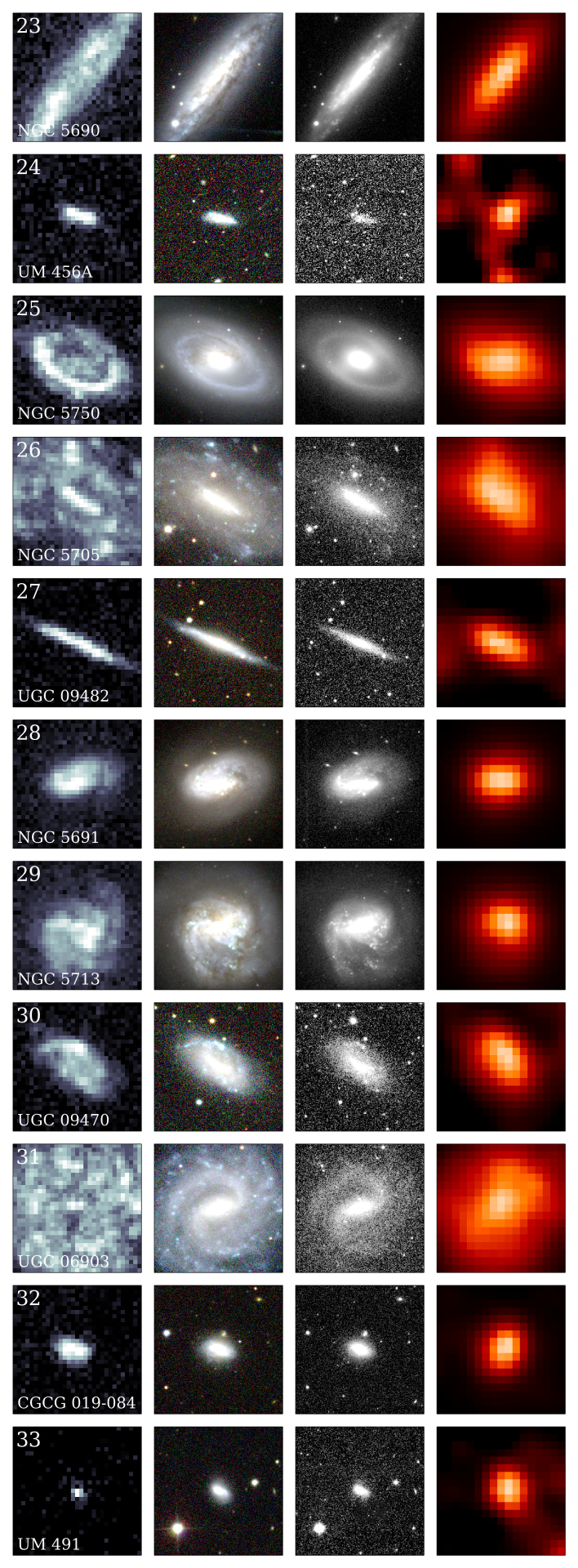}
\includegraphics[width=0.475\textwidth]{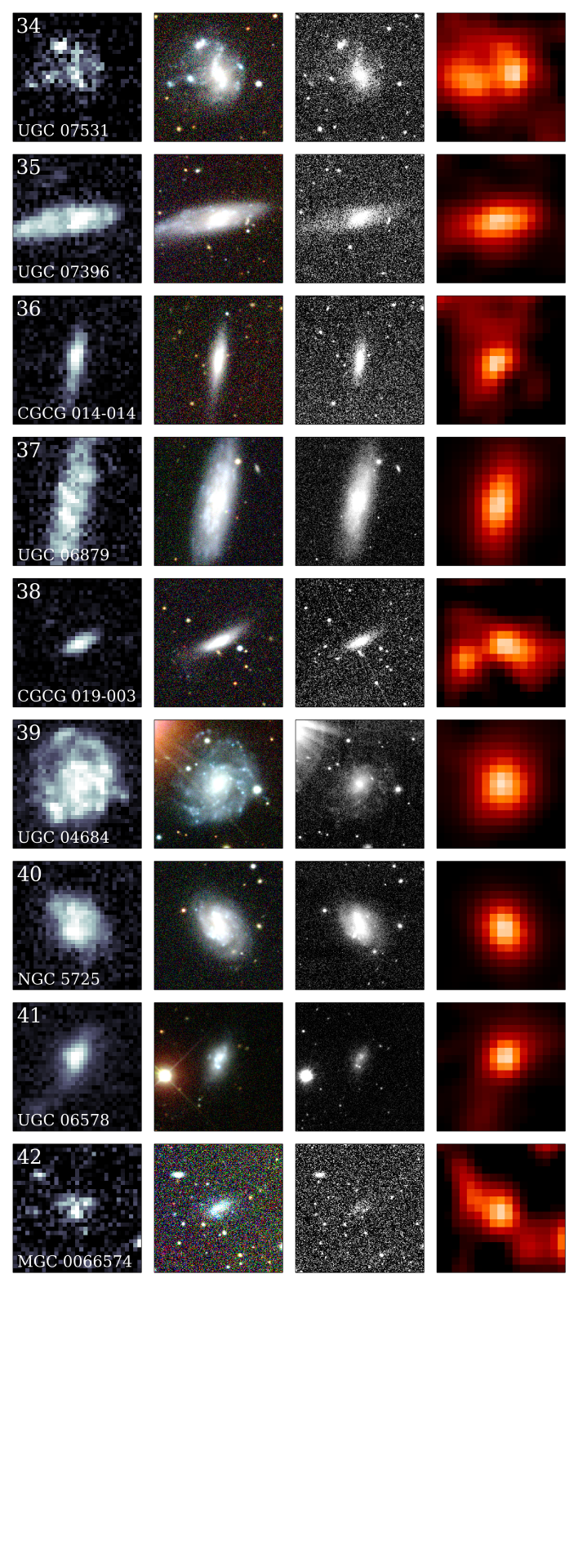}
\contcaption{}
\end{figure*}
\end{center}

\begin{figure*}
\begin{center}
\includegraphics[width=0.9625\textwidth]{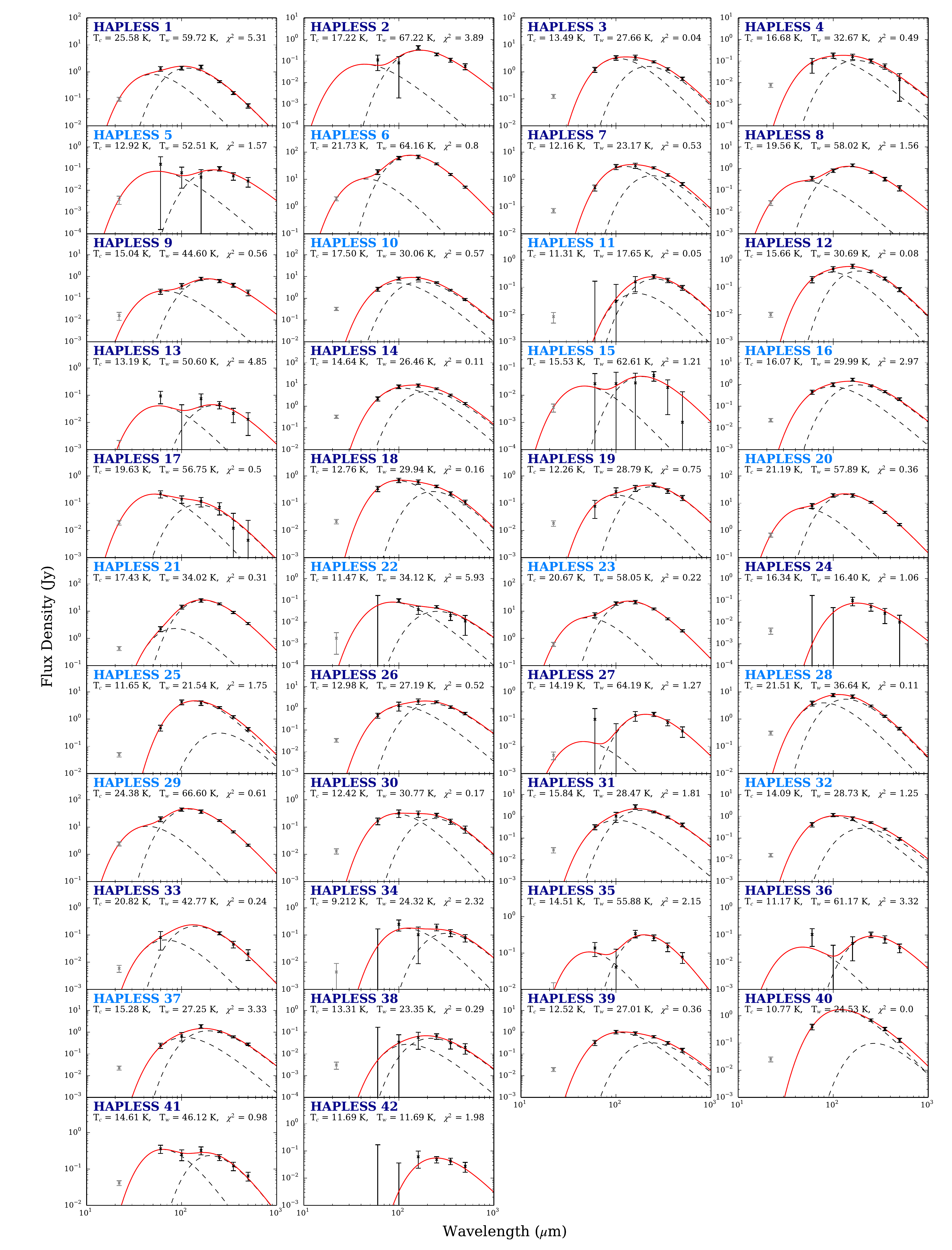}
\caption{Spectral energy distributions for the HAPLESS sample. The two-temperature modified blackbody fits are shown in red, with the contributions from the warm and cold dust components shown by the dashed curves. The grey 22\,\micron\ point was treated as an upper limit. Sources with dark blue names satisfied the \fK\,\textless\,3.5 colour criterion necessary to be counted amongst the curious blue sub-population; sources with light blue names did not.}
\label{AppendixFigure:HAPLESS_SED_Grid}
\end{center}
\end{figure*}

\begin{landscape}
\begin{table}
\begin{center}
\caption{Photometry of the HAPLESS galaxies, where '-' indicates cases where no observations were available. The semi-major axis is denoted by $a$, the position angle by $\theta$, and the axial ratio (the semi-major axis divided by the semi-minor axis) by $a/b$. \hersc-SPIRE fluxes were measured using maps reduced for extended-source photometry, but have not been colour-corrected. }
\label{AppendixTable:Photometry}
\begin{tabular}{lrrrrrrrrrrrrrrrrr}
\toprule \toprule
\multicolumn{1}{c}{HAPLESS} &
\multicolumn{3}{c}{Aperture dimensions} &
\multicolumn{4}{c}{GALEX (mag)} &
\multicolumn{8}{c}{SDSS (mag)} &
\multicolumn{2}{c}{VIKING (mag)} \\
\cmidrule(lr){2-4}
\cmidrule(lr){5-8} 
\cmidrule(lr){9-16}
\cmidrule(lr){17-18}
\multicolumn{1}{c}{} &
\multicolumn{1}{c}{$a$ (arcsec)} &
\multicolumn{1}{c}{$\theta$ (deg)} &
\multicolumn{1}{c}{$a/b$} &
\multicolumn{1}{c}{FUV} &
\multicolumn{1}{c}{$\Delta$\,FUV} &
\multicolumn{1}{c}{NUV} &
\multicolumn{1}{c}{$\Delta$\,NUV} &
\multicolumn{1}{c}{$u$} &
\multicolumn{1}{c}{$\Delta\,u$} &
\multicolumn{1}{c}{$g$} &
\multicolumn{1}{c}{$\Delta\,g$} &
\multicolumn{1}{c}{$r$} &
\multicolumn{1}{c}{$\Delta\,r$} &
\multicolumn{1}{c}{$i$} &
\multicolumn{1}{c}{$\Delta\,i$} &
\multicolumn{1}{c}{$Z$} &
\multicolumn{1}{c}{$\Delta\,Z$} \\
\midrule
1 & 28.4 & 158.2 & 1.075 & 15.78 & 0.05 & 15.23 & 0.03 & 14.57 & 0.10 & 13.72 & 0.04 & 13.26 & 0.05 & 13.06 & 0.06 & 12.83 & 0.07\\ 
2 & 36.5 & 23.8 & 1.064 & 16.67 & 0.05 & 16.36 & 0.03 & 15.84 & 0.33 & 15.08 & 0.06 & 14.83 & 0.07 & 14.72 & 0.10 & 14.59 & 0.08\\ 
3 & 122.0 & 72.8 & 1.477 & 14.48 & 0.05 & 14.18 & 0.03 & 13.53 & 0.22 & 12.72 & 0.06 & 12.42 & 0.06 & 12.23 & 0.08 & 12.14 & 0.07\\ 
4 & 36.6 & 28.0 & 1.437 & 17.31 & 0.05 & 16.94 & 0.03 & 15.96 & 0.13 & 15.00 & 0.05 & 14.51 & 0.06 & 14.28 & 0.07 & 14.18 & 0.07\\ 
5 & 36.5 & 0.7 & 1.381 & 18.08 & 0.10 & 17.41 & 0.03 & 16.41 & 0.37 & 15.29 & 0.06 & 14.84 & 0.07 & 14.61 & 0.09 & 14.47 & 0.07\\ 
6 & 178.9 & 117.2 & 1.455 & 13.69 & 0.05 & 12.99 & 0.03 & 11.99 & 0.11 & 10.68 & 0.04 & 10.10 & 0.05 & 9.76 & 0.06 & 9.54 & 0.07\\ 
7 & 178.9 & 85.3 & 3.424 & 14.58 & 0.05 & 14.08 & 0.03 & 13.54 & 0.30 & 12.41 & 0.04 & 12.04 & 0.05 & 11.82 & 0.06 & - & -\\ 
8 & 65.0 & 141.2 & 1.292 & 15.47 & 0.05 & 15.14 & 0.03 & 14.46 & 0.19 & 13.67 & 0.05 & 13.37 & 0.06 & 13.15 & 0.07 & 13.08 & 0.07\\ 
9 & 122.0 & 178.5 & 1.533 & 15.35 & 0.05 & 15.08 & 0.03 & 14.89 & 0.39 & 13.82 & 0.08 & 13.78 & 0.12 & 13.53 & 0.12 & 13.63 & 0.22\\ 
10 & 126.0 & 62.1 & 2.058 & 15.14 & 0.07 & 14.61 & 0.03 & 13.68 & 0.20 & 12.20 & 0.05 & 11.58 & 0.05 & 11.26 & 0.06 & 10.99 & 0.07\\ 
11 & 85.3 & 50.2 & 3.665 & 17.36 & 0.05 & 16.75 & 0.03 & 15.62 & 0.38 & 14.53 & 0.05 & 14.09 & 0.06 & 13.90 & 0.07 & 13.72 & 0.07\\ 
12 & 44.6 & 147.2 & 1.568 & 17.08 & 0.05 & 16.65 & 0.03 & 15.78 & 0.17 & 14.88 & 0.06 & 14.47 & 0.06 & 14.29 & 0.08 & 14.07 & 0.07\\ 
13 & 24.3 & 67.3 & 1.347 & 18.80 & 0.05 & 18.38 & 0.03 & 17.26 & 0.42 & 16.39 & 0.08 & 15.93 & 0.07 & 15.75 & 0.11 & 15.60 & 0.09\\ 
14 & 134.2 & 63.9 & 1.205 & 13.90 & 0.06 & 13.52 & 0.03 & 12.82 & 0.13 & 11.87 & 0.05 & 11.45 & 0.05 & 11.33 & 0.06 & 11.16 & 0.07\\ 
15 & 32.5 & 46.2 & 1.289 & 18.42 & 0.05 & 17.79 & 0.03 & 16.49 & 0.24 & 15.37 & 0.06 & 14.89 & 0.06 & 14.62 & 0.09 & 14.57 & 0.07\\ 
16 & 69.1 & 32.2 & 2.122 & 17.14 & 0.05 & 16.50 & 0.03 & 15.19 & 0.20 & 14.02 & 0.05 & 13.52 & 0.05 & 13.23 & 0.07 & 13.03 & 0.07\\ 
17 & 44.7 & 139.4 & 1.156 & 16.41 & 0.05 & 16.16 & 0.03 & 15.77 & 0.28 & 15.08 & 0.06 & 14.87 & 0.08 & 14.88 & 0.12 & 14.73 & 0.08\\ 
18 & 52.8 & 96.6 & 1.274 & 15.86 & 0.05 & 15.51 & 0.03 & 14.96 & 0.22 & 14.20 & 0.05 & 13.88 & 0.05 & 13.74 & 0.08 & 13.62 & 0.07\\ 
19 & 134.2 & 109.9 & 3.776 & - & - & - & - & 14.64 & 0.20 & 13.71 & 0.06 & 13.49 & 0.07 & 13.15 & 0.07 & 13.34 & 0.08\\ 
20 & 142.3 & 30.3 & 2.096 & 17.01 & 0.10 & 16.27 & 0.04 & 13.90 & 0.39 & 12.29 & 0.05 & 11.37 & 0.05 & 10.91 & 0.06 & 10.62 & 0.07\\ 
21 & 276.6 & 78.5 & 4.490 & - & - & - & - & 12.54 & 0.16 & 10.94 & 0.04 & 10.07 & 0.05 & 9.61 & 0.06 & 9.27 & 0.07\\ 
22 & 40.6 & 151.1 & 1.638 & 19.70 & 0.08 & 18.50 & 0.05 & 15.73 & 0.24 & 14.22 & 0.05 & 13.50 & 0.05 & 13.11 & 0.06 & 12.85 & 0.07\\ 
23 & 117.9 & 60.0 & 2.317 & 15.64 & 0.05 & 14.91 & 0.03 & 13.58 & 0.18 & 12.29 & 0.05 & 11.72 & 0.05 & 11.38 & 0.06 & 11.12 & 0.07\\ 
24 & 36.5 & 152.8 & 1.729 & 17.76 & 0.05 & 17.59 & 0.03 & 17.19 & 0.47 & 16.43 & 0.10 & 16.24 & 0.12 & 16.20 & 0.25 & 16.09 & 0.10\\ 
25 & 126.1 & 151.5 & 1.811 & 16.10 & 0.05 & 15.42 & 0.03 & 13.49 & 0.15 & 11.95 & 0.04 & 11.22 & 0.05 & 10.83 & 0.06 & 10.59 & 0.07\\ 
26 & 113.8 & 162.3 & 1.366 & 14.83 & 0.05 & 14.52 & 0.03 & 14.35 & 0.53 & 12.83 & 0.06 & 12.48 & 0.06 & 12.25 & 0.08 & 12.22 & 0.07\\ 
27 & 65.0 & 157.4 & 3.380 & 17.35 & 0.05 & 16.80 & 0.03 & 15.91 & 0.34 & 15.01 & 0.06 & 14.64 & 0.06 & 14.47 & 0.08 & 14.35 & 0.08\\ 
28 & 85.3 & 28.8 & 1.394 & 15.56 & 0.05 & 15.03 & 0.03 & 13.45 & 0.12 & 12.43 & 0.05 & 11.98 & 0.05 & 11.75 & 0.06 & 11.61 & 0.07\\ 
29 & 117.9 & 85.0 & 1.197 & 14.70 & 0.05 & 13.87 & 0.03 & 12.53 & 0.15 & 11.40 & 0.04 & 10.85 & 0.05 & 10.58 & 0.06 & 10.38 & 0.07\\ 
30 & 60.9 & 143.7 & 1.439 & 16.19 & 0.05 & 15.86 & 0.03 & 15.18 & 0.26 & 14.43 & 0.06 & 14.08 & 0.06 & 13.94 & 0.09 & 13.83 & 0.08\\ 
31 & 101.6 & 47.4 & 1.123 & 15.24 & 0.05 & 15.16 & 0.03 & 14.67 & 0.72 & 13.14 & 0.07 & 12.73 & 0.07 & 12.51 & 0.08 & 12.24 & 0.07\\ 
32 & 37.9 & 46.3 & 1.391 & 17.49 & 0.07 & 17.00 & 0.03 & 15.93 & 0.30 & 14.84 & 0.05 & 14.42 & 0.06 & 14.19 & 0.10 & 13.99 & 0.07\\ 
33 & 28.4 & 136.9 & 1.540 & 16.60 & 0.05 & 16.36 & 0.03 & 16.08 & 0.14 & 15.41 & 0.05 & 15.18 & 0.07 & 15.07 & 0.07 & 14.97 & 0.07\\ 
34 & 65.0 & 178.7 & 1.338 & 15.55 & 0.06 & 15.46 & 0.04 & 15.22 & 0.36 & 14.39 & 0.07 & 14.25 & 0.08 & 14.16 & 0.15 & 14.20 & 0.09\\ 
35 & 73.1 & 8.1 & 1.846 & 16.73 & 0.05 & 16.25 & 0.03 & 15.35 & 0.34 & 14.55 & 0.07 & 14.17 & 0.06 & 13.97 & 0.09 & 13.90 & 0.08\\ 
36 & 48.7 & 76.6 & 2.341 & 17.66 & 0.05 & 17.24 & 0.03 & 16.49 & 0.42 & 15.64 & 0.07 & 15.36 & 0.08 & 15.22 & 0.14 & 15.14 & 0.09\\ 
37 & 89.4 & 76.2 & 2.711 & 16.42 & 0.05 & 16.01 & 0.03 & 14.99 & 0.19 & 13.70 & 0.05 & 13.04 & 0.05 & 12.73 & 0.06 & 12.45 & 0.07\\ 
38 & 32.4 & 30.8 & 1.650 & 17.94 & 0.05 & 17.53 & 0.03 & 16.77 & 0.35 & 15.83 & 0.06 & 15.57 & 0.08 & 15.39 & 0.09 & 15.15 & 0.08\\ 
39 & 48.8 & 131.0 & 1.110 & 15.72 & 0.05 & 15.34 & 0.03 & 14.63 & 0.13 & 13.58 & 0.05 & 13.24 & 0.05 & 13.06 & 0.06 & 12.70 & 0.07\\ 
40 & 52.8 & 117.6 & 1.161 & 15.89 & 0.05 & 15.47 & 0.03 & 14.78 & 0.24 & 13.94 & 0.05 & 13.57 & 0.06 & 13.40 & 0.08 & 13.25 & 0.07\\ 
41 & 48.7 & 50.5 & 1.619 & 15.98 & 0.05 & 15.83 & 0.03 & 15.39 & 0.14 & 14.75 & 0.04 & 14.59 & 0.05 & 14.57 & 0.07 & 14.39 & 0.08\\ 
42 & 24.3 & 3.6 & 1.471 & 19.77 & 0.10 & 19.53 & 0.08 & 18.62 & 1.31 & 17.70 & 0.33 & 17.39 & 0.36 & 17.28 & 1.11 & 17.11 & 0.28\\ 
\bottomrule
\end{tabular}
\end{center}
\end{table}
\end{landscape}

\begin{landscape}
\begin{table}
\begin{center}
\contcaption{}
\begin{tabular}{lrrrrrrrrrrrrrrrr}
\toprule \toprule
\multicolumn{1}{c}{HAPLESS} &
\multicolumn{8}{c}{VIKING (mag)} &
\multicolumn{8}{c}{WISE (mJy)} \\
\cmidrule(lr){2-9} 
\cmidrule(lr){10-17}
\multicolumn{1}{c}{} &
\multicolumn{1}{c}{$Y$} &
\multicolumn{1}{c}{$\Delta\,Y$} &
\multicolumn{1}{c}{$J$} &
\multicolumn{1}{c}{$\Delta\,J$} &
\multicolumn{1}{c}{$H$} &
\multicolumn{1}{c}{$\Delta\,H$} &
\multicolumn{1}{c}{$K_{S}$} &
\multicolumn{1}{c}{$\Delta\,K_{S}$} &
\multicolumn{1}{c}{3.4 \micron} &
\multicolumn{1}{c}{$\Delta$\,3.4\,\micron} &
\multicolumn{1}{c}{4.6 \micron} &
\multicolumn{1}{c}{$\Delta$\,4.6\,\micron} &
\multicolumn{1}{c}{12 \micron} &
\multicolumn{1}{c}{$\Delta$\,12\,\micron} &
\multicolumn{1}{c}{22\,\micron} &
\multicolumn{1}{c}{$\Delta$\,22\,\micron} \\
\midrule
1 & 12.70 & 0.07 & 12.62 & 0.08 & 12.50 & 0.07 & 12.70 & 0.06 & 14.91 & 0.86 & 9.52 & 0.75 & 32.61 & 3.71 & 98.06 & 16.53\\ 
2 & 14.56 & 0.09 & 14.57 & 0.10 & 14.48 & 0.10 & 14.63 & 0.12 & 2.62 & 0.21 & 1.36 & 0.20 & 2.57 & 0.46 & - & -\\ 
3 & 12.05 & 0.08 & 12.00 & 0.10 & 12.03 & 0.10 & 12.34 & 0.10 & 30.03 & 1.93 & 18.25 & 1.62 & 52.02 & 6.26 & 124.94 & 21.41\\ 
4 & 14.05 & 0.08 & 13.97 & 0.09 & 13.93 & 0.16 & 14.15 & 0.09 & 3.54 & 0.24 & 2.26 & 0.24 & 2.55 & 0.46 & 7.68 & 1.75\\ 
5 & 14.36 & 0.08 & 14.32 & 0.09 & 14.24 & 0.09 & 14.49 & 0.10 & 2.80 & 0.19 & 1.65 & 0.18 & 1.65 & 0.33 & 3.81 & 1.57\\ 
6 & 9.32 & 0.07 & 9.18 & 0.08 & 8.96 & 0.07 & 9.18 & 0.06 & 463.11 & 26.73 & 287.13 & 22.43 & 1287.59 & 146.06 & 1945.55 & 326.94\\ 
7 & - & - & 11.66 & 0.09 & 11.48 & 0.08 & 11.92 & 0.06 & 39.32 & 2.36 & 24.25 & 2.01 & 43.01 & 5.18 & 71.70 & 13.16\\ 
8 & 13.00 & 0.08 & 12.93 & 0.10 & 12.82 & 0.08 & 13.06 & 0.10 & 11.70 & 0.78 & 7.12 & 0.71 & 15.21 & 2.10 & 26.63 & 6.63\\ 
9 & 13.53 & 0.14 & 13.49 & 0.25 & 14.26 & 6.85 & 13.99 & 0.35 & 7.59 & 0.86 & 4.26 & 0.93 & 9.68 & 1.99 & 15.94 & 6.29\\ 
10 & 10.79 & 0.07 & 10.65 & 0.08 & 10.49 & 0.07 & 10.74 & 0.06 & 102.86 & 5.96 & 59.48 & 4.67 & 182.60 & 20.72 & 327.67 & 55.13\\ 
11 & 13.63 & 0.08 & 13.56 & 0.10 & 12.88 & 0.27 & 13.62 & 0.17 & 6.01 & 0.41 & 3.39 & 0.41 & 4.97 & 2.05 & 8.35 & 3.51\\ 
12 & 13.98 & 0.08 & 13.90 & 0.10 & 13.80 & 0.10 & 14.00 & 0.09 & 4.86 & 0.32 & 2.66 & 0.26 & 7.34 & 0.89 & 9.95 & 2.00\\ 
13 & 15.45 & 0.09 & 15.53 & 0.12 & 15.54 & 0.19 & 15.66 & 0.16 & 0.88 & 0.08 & 0.54 & 0.08 & 0.60 & 0.19 & 0.94 & 1.23\\ 
14 & 11.09 & 0.07 & 11.09 & 0.08 & 11.08 & 0.09 & 11.18 & 0.07 & 77.59 & 4.57 & 46.95 & 3.79 & 158.51 & 18.05 & 331.32 & 55.87\\ 
15 & 14.48 & 0.08 & 14.46 & 0.09 & 14.40 & 0.11 & 14.68 & 0.12 & 3.02 & 0.21 & 1.66 & 0.20 & 0.85 & 0.62 & 3.57 & 1.17\\ 
16 & 12.89 & 0.07 & 12.82 & 0.08 & 12.72 & 0.08 & 12.87 & 0.06 & 13.89 & 0.82 & 8.28 & 0.69 & 17.15 & 1.99 & 23.12 & 4.25\\ 
17 & 14.66 & 0.11 & 14.64 & 0.15 & 14.65 & 0.20 & 14.86 & 0.35 & 2.23 & 0.25 & 1.60 & 0.27 & 2.14 & 0.67 & 19.54 & 3.86\\ 
18 & 13.54 & 0.08 & 13.49 & 0.09 & 13.43 & 0.09 & 13.65 & 0.16 & 6.85 & 0.45 & 3.93 & 0.40 & 7.02 & 0.89 & 21.33 & 4.03\\ 
19 & 13.46 & 0.11 & 13.15 & 0.12 & 13.18 & 0.15 & 13.62 & 0.17 & 10.21 & 0.70 & 5.20 & 0.57 & 4.54 & 1.02 & 18.33 & 4.18\\ 
20 & 10.33 & 0.07 & 10.10 & 0.08 & 9.84 & 0.07 & 10.01 & 0.06 & 203.06 & 11.75 & 122.99 & 9.62 & 378.25 & 42.99 & 683.62 & 114.96\\ 
21 & 8.95 & 0.07 & 8.73 & 0.08 & 8.50 & 0.07 & 8.66 & 0.06 & 641.00 & 36.97 & 358.08 & 27.95 & 390.83 & 44.48 & 429.46 & 72.33\\ 
22 & 12.65 & 0.07 & 12.53 & 0.08 & 12.36 & 0.07 & 12.58 & 0.06 & 15.94 & 0.92 & 8.86 & 0.70 & 3.15 & 0.48 & 1.81 & 1.48\\ 
23 & 10.90 & 0.07 & 10.75 & 0.08 & 10.49 & 0.07 & 10.68 & 0.06 & 122.38 & 7.07 & 78.87 & 6.17 & 397.79 & 45.15 & 609.66 & 102.47\\ 
24 & 16.13 & 0.16 & 16.01 & 0.31 & 16.01 & 0.36 & 15.94 & 0.36 & 0.60 & 0.11 & 0.19 & 0.26 & 0.25 & 0.06 & 4.03 & 1.29\\ 
25 & 10.36 & 0.07 & 10.22 & 0.08 & 10.12 & 0.07 & 10.25 & 0.06 & 143.77 & 8.32 & 78.92 & 6.21 & 64.70 & 7.74 & 49.72 & 9.06\\ 
26 & 12.18 & 0.08 & 12.20 & 0.09 & 12.09 & 0.08 & 12.44 & 0.17 & 28.30 & 1.75 & 16.15 & 1.41 & 26.79 & 3.19 & 34.04 & 6.42\\ 
27 & 14.27 & 0.09 & 14.24 & 0.11 & 14.19 & 0.19 & 14.45 & 0.12 & 2.72 & 0.22 & 1.57 & 0.21 & 1.58 & 0.44 & 4.67 & 1.46\\ 
28 & 11.48 & 0.07 & 11.40 & 0.08 & 11.28 & 0.07 & 11.56 & 0.07 & 52.03 & 3.04 & 33.82 & 2.68 & 134.25 & 15.25 & 308.95 & 52.01\\ 
29 & 10.19 & 0.07 & 10.07 & 0.08 & 9.94 & 0.07 & 10.14 & 0.08 & 200.25 & 11.59 & 134.95 & 10.57 & 947.13 & 107.46 & 2427.67 & 408.04\\ 
30 & 13.79 & 0.11 & 13.80 & 0.11 & 13.81 & 0.26 & 13.95 & 0.13 & 5.53 & 0.40 & 2.95 & 0.40 & 2.56 & 0.53 & 13.02 & 2.79\\ 
31 & 12.17 & 0.08 & 12.11 & 0.10 & 12.05 & 0.09 & 12.29 & 0.13 & 25.93 & 1.71 & 15.75 & 1.45 & 31.86 & 5.48 & 28.27 & 7.75\\ 
32 & 13.86 & 0.08 & 13.82 & 0.09 & 13.71 & 0.08 & 13.89 & 0.09 & 5.42 & 0.35 & 3.38 & 0.30 & 10.59 & 1.26 & 16.55 & 2.94\\ 
33 & 14.92 & 0.10 & 14.97 & 0.11 & 13.97 & 0.11 & 15.02 & 0.11 & 1.85 & 0.13 & 0.84 & 0.14 & 2.00 & 0.53 & 5.94 & 1.72\\ 
34 & 14.19 & 0.20 & 14.09 & 0.16 & 14.19 & 0.18 & 14.38 & 0.24 & 3.61 & 0.41 & 1.96 & 0.44 & 4.44 & 0.86 & 4.37 & 4.64\\ 
35 & 13.83 & 0.09 & 13.78 & 0.12 & - & - & 13.94 & 0.38 & 5.32 & 0.44 & 2.23 & 0.46 & 4.13 & 2.48 & 4.84 & 10.47\\ 
36 & 15.11 & 0.10 & 14.99 & 0.14 & 14.88 & 0.34 & 15.33 & 0.83 & 1.43 & 0.16 & 0.51 & 0.28 & 0.73 & 0.69 & - & -\\ 
37 & 12.26 & 0.08 & 12.25 & 0.08 & 12.22 & 0.09 & 12.33 & 0.08 & 21.91 & 1.28 & 12.13 & 1.02 & 21.03 & 2.47 & 22.77 & 4.68\\ 
38 & 15.09 & 0.11 & 15.08 & 0.14 & 15.19 & 0.20 & 15.23 & 0.19 & 1.15 & 0.15 & 0.81 & 0.13 & 0.71 & 0.60 & 3.08 & 1.11\\ 
39 & 12.71 & 0.08 & 12.83 & 0.09 & 12.84 & 0.08 & 13.38 & 0.13 & 19.89 & 1.18 & 12.19 & 1.00 & 11.35 & 1.59 & 19.43 & 3.88\\ 
40 & 13.15 & 0.08 & 13.08 & 0.08 & 12.92 & 0.09 & 13.29 & 0.08 & 10.86 & 0.70 & 6.41 & 0.62 & 17.66 & 2.13 & 25.36 & 5.13\\ 
41 & 14.51 & 0.12 & 14.46 & 0.11 & 15.09 & 0.23 & 15.34 & 0.26 & 2.80 & 0.32 & 1.98 & 0.29 & 7.57 & 1.02 & 41.26 & 6.42\\ 
42 & 17.06 & 0.59 & 17.07 & 0.35 & 16.91 & 1.05 & 17.29 & 0.59 & 0.20 & 0.12 & 0.13 & 0.38 & - & - & - & -\\ 
\bottomrule
\end{tabular}
\end{center}
\end{table}
\end{landscape}

\begin{landscape}
\begin{table}
\begin{center}
\contcaption{}
\begin{tabular}{lrrrrrrrrrrrr}
\toprule \toprule
\multicolumn{1}{c}{HAPLESS} &
\multicolumn{2}{c}{IRAS SCANPI (mJy)} &
\multicolumn{4}{c}{\hersc-PACS (mJy)} &
\multicolumn{6}{c}{\hersc-SPIRE (mJy)} \\
\cmidrule(lr){2-3} 
\cmidrule(lr){4-7} 
\cmidrule(lr){8-13}
\multicolumn{1}{c}{} &
\multicolumn{1}{c}{60 \micron} &
\multicolumn{1}{c}{$\Delta$\,60\,\micron} &
\multicolumn{1}{c}{100 \micron} &
\multicolumn{1}{c}{$\Delta$\,100\,\micron} &
\multicolumn{1}{c}{160 \micron} &
\multicolumn{1}{c}{$\Delta$\,160\,\micron} &
\multicolumn{1}{c}{250 \micron} &
\multicolumn{1}{c}{$\Delta$\,250\,\micron} &
\multicolumn{1}{c}{350 \micron} &
\multicolumn{1}{c}{$\Delta$\,350\,\micron} &
\multicolumn{1}{c}{500 \micron} &
\multicolumn{1}{c}{$\Delta$\,500\,\micron} \\
\midrule
1 & 1310.00 & 266.45 & 1387.15 & 219.81 & 1602.08 & 238.71 & 427.94 & 34.84 & 161.96 & 20.96 & 54.96 & 9.87\\ 
2 & 120.00 & 76.14 & 84.28 & 79.95 & 421.45 & 89.68 & 204.11 & 30.24 & 111.00 & 23.39 & 57.35 & 17.70\\ 
3 & 1270.00 & 256.58 & 3331.62 & 631.78 & 3502.96 & 705.72 & 2351.96 & 205.46 & 1279.07 & 136.93 & 560.28 & 79.40\\ 
4 & 80.00 & 52.70 & 184.36 & 49.68 & 164.46 & 45.74 & 102.47 & 21.44 & 56.53 & 16.18 & 13.81 & 12.36\\ 
5 & 170.00 & 191.88 & 64.41 & 51.71 & 40.66 & 48.09 & 99.70 & 23.01 & 46.14 & 17.04 & 26.92 & 12.48\\ 
6 & 18780.00 & 3756.40 & 61020.59 & 8519.13 & 69358.11 & 10198.06 & 36792.35 & 2775.40 & 14854.21 & 1276.14 & 5134.39 & 448.21\\ 
7 & 550.00 & 123.30 & 2900.78 & 616.31 & 3236.56 & 688.95 & 2657.91 & 243.22 & 1453.49 & 158.95 & 683.97 & 80.92\\ 
8 & 350.00 & 81.51 & 814.95 & 153.08 & 1433.18 & 214.11 & 675.04 & 82.80 & 327.31 & 62.87 & 128.52 & 34.82\\ 
9 & 210.00 & 55.48 & 374.73 & 103.74 & 784.69 & 138.80 & 625.65 & 108.22 & 402.94 & 80.91 & 184.27 & 54.10\\ 
10 & 2730.00 & 548.10 & 8166.43 & 1360.14 & 8316.84 & 1295.71 & 5199.65 & 395.95 & 2337.76 & 209.50 & 882.87 & 91.98\\ 
11 & 0.00 & 56.00 & 33.36 & 96.72 & 161.65 & 90.14 & 249.67 & 40.45 & 182.15 & 33.64 & 99.09 & 19.22\\ 
12 & 200.00 & 51.21 & 473.04 & 102.40 & 602.62 & 109.52 & 391.55 & 35.79 & 211.30 & 26.93 & 83.64 & 13.75\\ 
13 & 100.0 & 45.56 & -33.8 & 44.50 & 74.30 & 37.55 & 45.34 & 13.72 & 21.34 & 11.57 & 13.22 & 9.69\\ 
14 & 2340.00 & 469.96 & 7941.90 & 1494.84 & 9091.88 & 1587.11 & 6340.79 & 533.89 & 3256.95 & 309.99 & 1350.70 & 139.03\\ 
15 & 30.00 & 36.08 & 27.00 & 42.92 & 28.34 & 36.07 & 53.25 & 20.23 & 19.20 & 17.34 & 1.02 & 12.42\\ 
16 & 460.00 & 98.60 & 1011.81 & 185.56 & 1684.40 & 260.50 & 874.30 & 71.99 & 470.14 & 48.91 & 213.04 & 29.09\\ 
17 & 240.00 & 64.17 & 141.94 & 42.77 & 120.14 & 44.53 & 69.67 & 33.30 & 11.91 & 30.21 & 4.34 & 19.00\\ 
18 & 350.00 & 76.69 & 696.79 & 128.00 & 617.64 & 112.90 & 418.04 & 45.76 & 228.25 & 34.23 & 109.71 & 21.17\\ 
19 & 80.00 & 49.83 & 264.56 & 101.85 & 358.95 & 88.39 & 480.02 & 72.13 & 286.89 & 54.87 & 161.44 & 33.59\\ 
20 & 8090.00 & 1618.42 & 19549.90 & 2770.41 & 19867.75 & 2939.29 & 10656.58 & 807.11 & 4569.69 & 400.61 & 1629.31 & 147.40\\ 
21 & 2350.00 & 471.67 & 14624.22 & 2392.07 & 24916.23 & 3807.53 & 18567.37 & 1415.15 & 8892.77 & 773.69 & 3516.77 & 317.25\\ 
22 & 0.00 & 56.00 & 99.52 & 19.26 & 38.62 & 15.17 & 51.10 & 6.92 & 20.30 & 8.30 & 11.48 & 8.83\\ 
23 & 7250.00 & 1450.64 & 19264.58 & 2772.15 & 22023.09 & 3249.34 & 11932.25 & 887.39 & 5139.88 & 436.57 & 1896.64 & 164.87\\ 
24 & 0.00 & 56.00 & -2.0 & 46.68 & 97.51 & 41.24 & 51.84 & 19.90 & 25.41 & 16.99 & 10.15 & 10.59\\ 
25 & 550.00 & 117.20 & 4222.31 & 840.95 & 4005.41 & 766.60 & 2637.02 & 223.50 & 1198.41 & 128.49 & 428.35 & 61.84\\ 
26 & 500.00 & 122.91 & 1333.25 & 588.14 & 2116.58 & 594.26 & 2015.23 & 215.48 & 1147.87 & 151.78 & 589.46 & 82.07\\ 
27 & 110.00 & 145.24 & -8.7 & 68.14 & 137.46 & 53.35 & 151.16 & 26.04 & 75.29 & 18.44 & 37.06 & 15.27\\ 
28 & 3770.00 & 760.63 & 7735.75 & 1066.90 & 7023.54 & 972.91 & 3008.62 & 230.80 & 1254.65 & 115.44 & 447.29 & 52.58\\ 
29 & 19560.00 & 3912.25 & 44031.71 & 6054.90 & 38567.82 & 5528.44 & 17142.52 & 1272.60 & 6536.57 & 559.94 & 2158.08 & 191.51\\ 
30 & 170.00 & 45.89 & 323.81 & 98.58 & 307.54 & 81.30 & 274.88 & 44.08 & 160.90 & 34.82 & 85.20 & 24.36\\ 
31 & 340.00 & 82.77 & 1028.02 & 499.16 & 2740.92 & 638.74 & 1588.00 & 157.54 & 922.73 & 101.27 & 413.94 & 76.84\\ 
32 & 430.00 & 93.31 & 1154.40 & 192.59 & 817.59 & 152.62 & 516.41 & 43.37 & 253.84 & 25.62 & 93.09 & 13.21\\ 
33 & 80.00 & 52.75 & - & - & - & - & 115.18 & 15.51 & 45.39 & 12.74 & 20.12 & 8.59\\ 
34 & 0.00 & 56.00 & 247.96 & 108.46 & 103.19 & 93.55 & 198.34 & 54.04 & 122.85 & 34.92 & 81.44 & 24.24\\ 
35 & 150.00 & 57.75 & 41.46 & 86.88 & 341.33 & 79.16 & 267.19 & 47.46 & 148.84 & 39.72 & 78.26 & 25.40\\ 
36 & 120.00 & 66.55 & -36.4 & 42.24 & 51.08 & 37.62 & 104.39 & 23.32 & 72.53 & 21.47 & 35.06 & 12.11\\ 
37 & 260.00 & 66.55 & 686.97 & 289.21 & 1864.80 & 348.26 & 1060.72 & 93.62 & 608.12 & 63.30 & 283.41 & 35.29\\ 
38 & 0.00 & 56.00 & 35.23 & 41.20 & 57.68 & 41.36 & 66.09 & 18.69 & 32.65 & 15.85 & 19.91 & 9.95\\ 
39 & 350.00 & 88.46 & 1026.31 & 189.93 & 896.63 & 160.63 & 619.56 & 74.65 & 321.32 & 46.29 & 151.88 & 30.18\\ 
40 & 430.00 & 92.28 & - & - & - & - & 691.09 & 64.53 & 324.21 & 45.35 & 127.06 & 22.43\\ 
41 & 370.00 & 90.46 & 251.76 & 82.52 & 324.24 & 79.87 & 209.03 & 39.02 & 121.31 & 31.74 & 64.60 & 17.37\\ 
42 & 0.00 & 56.00 & -36.1 & 36.09 & 63.73 & 37.65 & 48.98 & 12.64 & 42.76 & 11.38 & 27.75 & 10.86\\ 
\bottomrule
\end{tabular}
\end{center}
\end{table}
\end{landscape}

\section{The Chemical Evolution Model} \label{AppendixSection:Chemev}

\begin{figure}
\begin{center}
\includegraphics[width=0.5\textwidth]{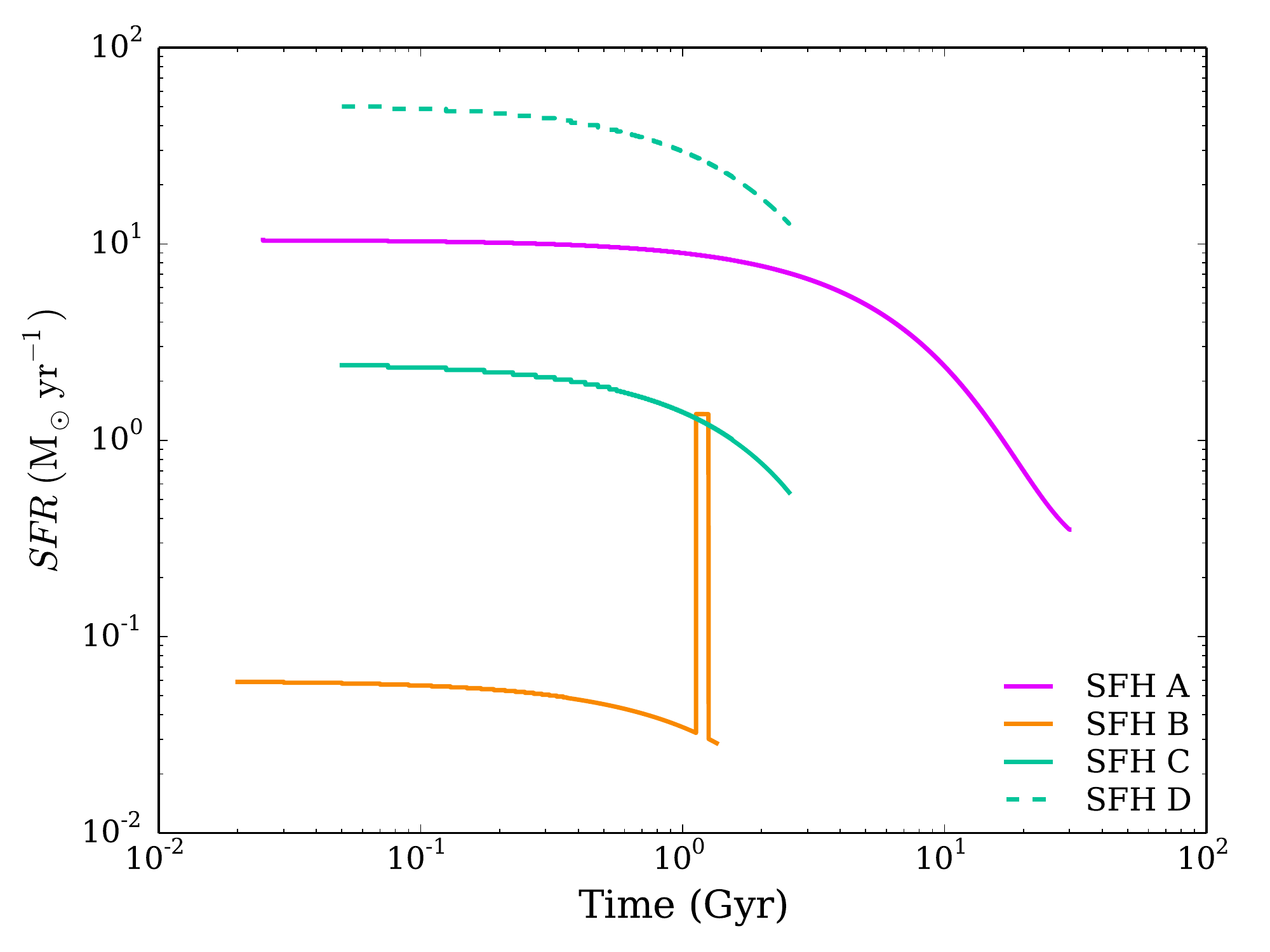}
\caption{The star formation histories used to model the HAPLESS galaxies. SFH A follows the evolution of the MW (pink, uppermost solid track, Yin et al. 2009) with initial gas mass $M_g(0) = 4\times 10^{10}\,\rm M_{\odot}$ and initial SFR $\psi(0) = \rm 10\,M_{\odot}\,yr^{-1}$. SFH B (orange, lowermost solid track) has initial gas mass $M_g(0) \sim 3\times 10^{9}\,\rm M_{\odot}$ and initial SFR $\rm 0.058\,M_{\odot}\,yr^{-1}$ with a burst at 1 Gyr superimposed on top of the exponentially declining rate. SFH C (orange, middle solid track) has $M_g(0) = 5.5\times 10^{9}\,\rm M_{\odot}$ and initial SFR $\rm 2.4\,M_{\odot}\,yr^{-1}$ and exponentially declines until reaching a gas fraction of $f_g \sim 0.6$ at 2.5 Gyr. SFR D (turquoise, dashed track) is a scaled version of SFR C ($\times 20$).}
\label{Fig:App_t_vs_SFR}
\end{center}
\end{figure}

Briefly, the equations to follow the evolution of gas and dust in the HAPLESS galaxies are:

\begin{equation}
M_{\it tot}=M_{g} + M_{\star},
\label{Equation:Gomez_1}
\end{equation}

\noindent where $M_{g}$ is the gas mass and $M_{\star}$ is the stellar mass. The gas mass evolution with time is described by:

\begin{equation}
\frac{dM_{g}}{dt} = -\psi(t) + e(t).
\label{Equation:Gomez_2}
\end{equation}

\noindent where $\psi(t)$ is the rate at which gas is depleted by the SFR, and $e(t)$ is the rate at which it is returned as stars die.

Assuming that mass loss occurs suddenly at the end of stellar evolution at time $\tau_{m}(m)$ \citep{Schaller1992A}, the ejected mass, $e(t)$, from stars is:

\begin{equation}
e(t)=\int_{m_{\tau_m}}^{m_U}{\left[m-m_{R}(m)\right]\psi(t-\tau_m)\phi(m) dm}
\label{Equation:Gomez_3}
\end{equation}

\noindent where $m_{R(m)}$ (from \citealp{Prantzos1993C}) is the remnant mass and $m_{\tau_{m}}$ is the mass of a star whose age is that of a system where a star formed at $(t-\tau_{m})$ has died at $\tau_{m}$. The evolution over time of the mass of metals in the ISM, $M_{Z}$, is described by:

\begin{equation}
\frac{d(M_Z)}{dt} = -Z(t)\psi(t) + e_z(t)
\label{Equation:Gomez_4}
\end{equation}

\noindent where $Z$ is the fraction of heavy elements by mass in the gas phase. The mass of heavy elements ejected by stars at the end of their lives is denoted by $e_{z(t)}$:

\begin{eqnarray}
 e_z(t)&=& \int_{m_{\tau_m}}^{m_U}\bigl({\left[m-m_{R}(m)\right] Z(t-\tau_m)+mp_z}\bigr)  \nonumber \\
 & &\mbox{} \times \psi(t-\tau_m)\phi(m)dm
\label{Equation:Gomez_5}
\end{eqnarray}

Yields from stars ($mp_z$) are taken from the theoretical models of \citet{Maeder1992C} and \citet{vandenHoek1997D}. The development of the mass of dust with time is described by:

\begin{eqnarray}
{dM_d\over{dt}}&=&\int_{m_{\tau_m}}^{m_U}\bigl({\left[m-m_{R}(m)\right] Z(t-\tau_m) \delta_{\it old}+mp_z\delta_{\it new}}\bigr) \nonumber \\
 & &\mbox{} \times \psi(t-\tau_m)\phi(m)dm - (M_d/M_g)\psi(t)
\label{Equation:Gomez_6}
\end{eqnarray}

\noindent where dust is built up from two sources, the fraction of the heavy elements that are recycled through star formation and ejected in stellar winds ($\delta_{\it old}$), and the fraction of new elements freshly synthesised in stars and ejected in both supernovae and stellar winds ($\delta_{\it new}$); the final term describes dust removed from the interstellar medium due to astration.

We use different star formation histories including one consistent with the Milky Way, and others consistent with galaxies with low star formation rates throughout their evolution (see Figure~B1 and Table~8). The initial gas masses are derived from the observed properties of the HAPLESS galaxies (ie $M_g(0) = M_g/f_g$, Table~4), we truncate the star formation histories when they reach the observed gas fraction. E.g. for HAPLESS 3, we start the model with an initial gas mass of $5.53\times 10^9\,\rm M_{\odot}$, and with the star formation history SFR C we reach a gas fraction of $f_g \sim 0.6$ at 2.5\,Gyrs consistent with the observations. Note that the lower SFHs are consistent with the current SFRs of the HAPLESS galaxies as derived from their UV and MIR fluxes; the SFHs are also compatible with the range of SFHs derived from more complex multiwavelength modelling of their SEDs (De Vis et al., {\it in prep.}). 

\end{document}